\documentclass[11pt,a4paper]{article}
\usepackage{authblk}
\usepackage{lmodern}
\usepackage{jstyle}
\usepackage{amsmath}
\usepackage{mathtools}

\usepackage{tikz}
\usetikzlibrary{snakes}
\usetikzlibrary{positioning}
\usetikzlibrary{arrows}

\tikzstyle{spring}=[line width=0.8,blue!7!black!80,snake=coil,segment amplitude=5,segment length=5,line cap=round]
\usepackage{tikz-cd}
\usetikzlibrary{shapes,arrows,cd,chains,decorations.markings,decorations.pathmorphing,calc,positioning,patterns}

\tikzset{
	->-/.style args={#1rotate#2}{decoration={markings, mark=at position #1 with {\arrow[scale=1.5,rotate = #2 ]{stealth}}}, postaction={decorate}}
}
\tikzset{
	-r-/.style args={#1rotate#2}{decoration={markings, mark=at position #1 with {\arrow[scale=1,rotate = #2 ]{>}}}, postaction={decorate}}
}

\usepackage{empheq}
\usepackage{enumitem}
\usepackage{graphics}
\usepackage{wrapfig}
\usepackage{caption}
\usepackage{natbib}
\usepackage{xcolor}
\usepackage{framed}
\usepackage{array}
\definecolor{shadecolor}{gray}{0.925}

\usepackage[cbgreek]{textgreek}

\def\sideremark#1{\ifvmode\leavevmode\fi\vadjust{\vbox to0pt{\vss
 \hbox to 0pt{\hskip\hsize\hskip1em
 \vbox{\hsize3cm\tiny\raggedright\pretolerance10000
 \noindent #1\hfill}\hss}\vbox to8pt{\vfil}\vss}}}%

\newcommand{\bi}{\begin{itemize}}
\newcommand{\ei}{\end{itemize}}
\newcommand{\bea}{\begin{align}}
\newcommand{\eea}{\end{align}}
\newcommand{\be}{\begin{equation}}
\newcommand{\ee}{\end{equation}}

\makeatletter
\renewcommand*\env@matrix[1][\arraystretch]{%
  \edef\arraystretch{#1}%
  \hskip -\arraycolsep
  \let\@ifnextchar\new@ifnextchar
  \array{*\c@MaxMatrixCols c}}
\makeatother

\author{Md. Abhishek}
\author{\, Charlotte Sleight}
\author{\, Massimo Taronna}

\affiliation{Dipartimento di Fisica ``Ettore Pancini'', Universit\`a degli Studi di Napoli Federico II, \\Monte S. Angelo, Via Cintia, 80126 Napoli, Italy}

\affiliation{INFN, Sezione di Napoli, Monte S. Angelo, Via Cintia, 80126 Napoli, Italy}

\emailAdd{abhishek.mohammad@na.infn.it, charlotte.sleight@na.infn.it, massimo.taronna@unina.it}

\title{\centering \huge Cosmological Correlators in \\ Gauge Theory and Gravity from EAdS}

\abstract{In this work we examine in more detail the map between late-time correlators in de Sitter space and boundary correlators in Euclidean anti-de Sitter space, elaborating on the general construction presented in \cite{Sleight:2020obc,Sleight:2021plv} for EFTs of bosonic spinning fields. This map may be phrased as an equivalence between the generating functional of late-time correlators in the Schwinger--Keldysh formalism and the generating functional for boundary correlators in the corresponding EAdS theory. We extend the construction to gauge bosons and gravitons, and clarify additional subtleties that appear in even boundary dimensions. We give the resulting EAdS Feynman rules for scalar QED, pure Yang--Mills theory and Einstein gravity in dS space, illustrating them with contact and tree-level exchange diagrams. The sinusoidal factors relating dS and EAdS diagrams lead to selection rules for late-time falloffs, with possible residual local terms only in
IR-divergent cases.  Finally, we emphasise that the relation between dS and EAdS propagators is manifest in Mellin space, and we provide new expressions for gauge-boson and graviton propagators in axial/temporal gauge, including their longitudinal components. These results provide a practical framework for studying cosmological correlators involving gauge fields and gravitons.}

\begin{document}

\begin{flushright}    
\texttt{}
\end{flushright}

\maketitle

\section{Introduction}\label{sec::Intro}

Cosmological correlators provide a key window into the dynamics of the early universe. The spatial correlations in the large-scale structure in our universe can be traced back to the spacelike boundary at the end of a postulated period of quasi-de Sitter expansion. The
detailed structure of these boundary correlations encodes information about both the dynamics
and particle content of inflation. 

\vskip 4pt
Our understanding of correlators on the future boundary of de Sitter (dS) space however remains far more rudimentary than for their negative–curvature counterparts on the boundary of anti–de Sitter (AdS) space. In AdS, the gravitational field is frozen at a boundary lying at spatial infinity, while time flows in the same way it does in the interior. The boundary system is then a non-gravitational Conformal Field Theory (CFT) in Minkowski space, rigorously defined at the non-perturbative level by conformal symmetry, unitarity, and an associative operator product expansion. In dS, by contrast, the boundary is purely spatial, with no notion of boundary time—obscuring how cosmological correlators encode a consistent picture of unitary time evolution in the interior. These differences make boundary correlators in de Sitter space more elusive, and motivate the search for frameworks that connect them to the well-developed tools of AdS/CFT---with foundational works on the subject including \cite{Strominger:2001pn,Maldacena:2002vr,McFadden:2009fg,McFadden:2010vh,McFadden:2011kk,Bzowski:2011ab}. 

\vskip 4pt
Despite this gap in understanding, the structural similarities between dS and Euclidean AdS (EAdS) space facilitate connections between the two. Both share the same isometry group, which for 
$(d+1)$-dimensional (EA)dS is $SO(1,d+1)$. In each case the isometries act on the $\mathbb{R}^d$ boundaries as the conformal group, with the upshot that boundary correlators in (EA)dS are constrained in the same way by conformal symmetry \cite{Antoniadis:2011ib,Maldacena:2011nz,Creminelli:2011mw,Bzowski:2011ab,Kehagias:2012pd,Kehagias:2012td,Schalm:2012pi,Bzowski:2012ih,Mata:2012bx,Bzowski:2013sza,Ghosh:2014kba,Kundu:2014gxa,Arkani-Hamed:2015bza,Shukla:2016bnu,Arkani-Hamed:2018kmz}. These similarities are made even more striking by the fact that dS and EAdS are related by analytic continuation. The analytic structure of de Sitter correlators was clarified in early field-theoretic studies \cite{Bros:1994dn,Bros:1995js,Moschella:2024kvk}, and the relation to EAdS was first exploited in a holographic context through the Bunch–Davies/Hartle–Hawking wavefunction $\Psi_{\text{dS}}$, whose form closely resembles that of the partition function in a Euclidean AdS background upon analytic continuation \cite{Maldacena:2002vr,McFadden:2009fg,Harlow:2011ke,Anninos:2014lwa}:
\begin{equation}\label{wfwick}
    \Psi_{\text{dS}}[J] = Z_{\text{EAdS}}[J].
\end{equation}
More recently this has been extended to late-time correlators in the Schwinger-Keldysh (S-K) formalism, whose doubled field structure \cite{Feynman:1963fq} inherited from the in-in contour can, under the appropriate analytic continuation \eqref{wickzeta}, be recast as boundary correlators of a theory in EAdS with doubled field content \cite{Sleight:2019mgd,Sleight:2019hfp,Sleight:2020obc,Sleight:2021plv}:
 \begin{equation}
    Z_{\text{S-K}}[J_+,J_-] = Z_{\text{EAdS}}\left[J_{\Delta_+},J_{\Delta_-}\right].
\end{equation} 
This analytic relation manifests in a close correspondence between bulk Feynman rules in dS and EAdS. Under analytic continuation, propagators in dS can be traded for linear combinations of propagators in EAdS corresponding to pairs of fields $\phi_{\Delta_\pm}$, where the free theory scaling dimensions $\Delta_\pm$ satisfy the shadow relation $\Delta_++\Delta_-=d$ \cite{Sleight:2020obc,Sleight:2021plv}. In this way, Feynman rules for late–time correlators in the in-in/Schwinger-Keldysh formalism can be recast as a set of Feynman rules for boundary correlators of a corresponding theory in EAdS. This construction was carried out in \cite{Sleight:2020obc,Sleight:2021plv} for generic dS EFTs of scalar and integer–spin fields in the Bunch–Davies vacuum, and has since been extended to fermions \cite{Schaub:2023scu} and to more general Bogoliubov initial states \cite{Chopping:2024oiu}. It is important to stress, however, that these are not the standard EAdS theories that would arise from the Wick rotation of AdS theories,\footnote{In other words, they are not bound to satisfy the Osterwalder–Schrader axioms that provide a Euclidean AdS formulation of Lorentzian AdS theories under Wick rotation to Euclidean time.} but instead represent a reformulation of dS dynamics. Expressing dS late–time correlators in terms of EAdS boundary correlators has provided new insights into their analytic structure and allows one to import techniques that have proven highly effective in AdS/CFT and the conformal bootstrap. This has already enabled the application of techniques in harmonic analysis and conformal partial wave expansions \cite{Sleight:2020obc,Sleight:2021plv,Hogervorst:2021uvp,DiPietro:2021sjt,Bissi:2023bhv,Loparco:2023rug,Werth:2024mjg}, Mellin amplitudes \cite{Sleight:2020obc,Dey:2025kci}, and methods for loop calculations \cite{Sleight:2021plv,DiPietro:2021sjt,Heckelbacher:2022hbq,Chowdhury:2023arc,Chowdhury:2025ohm,Nowinski:2025cvw}, to the study and calculation of late-time correlators in dS space.

\vskip 4pt
In this paper we revisit the general framework \cite{Sleight:2020obc,Sleight:2021plv} in detail for the cases of gauge bosons and gravitons. Massless fields play an important role during inflation, with their quantum fluctuations amplified by the expansion and seeding structure formation in the late universe. These massless representations of the de Sitter isometry group present subtleties, particularly in even boundary dimensions, which naively lead to divergences in the general EAdS reformulation of the Feynman rules for in–in correlators. We clarify how such cases can be consistently accommodated within the framework, and provide a streamlined reformulation for their cosmological correlators in EAdS. In contrast to gauge bosons and gravitons in AdS/CFT, where their boundary conditions at spatial infinity are reflective, we emphasise that late-time correlators in dS space also receive contributions from boundary gauge bosons and gravitons corresponding to Neumann boundary conditions---that codify outgoing radiation.

\vskip 4pt
The relationship between dS and EAdS propagators is made manifest in Mellin space \cite{Sleight:2019mgd,Sleight:2019hfp}, which diagonalises the action of dilatations, much as Fourier space diagonalises the action of translations. It also provides a convenient representation of the (EA)dS propagators themselves, which for gauge bosons and gravitons we use to package all components (transverse and longitudinal) in the axial/temporal gauge.

\vskip 4pt
In this paper we present the general framework for reformulating gauge boson and graviton theories in EAdS. A detailed evaluation and renormalisation of explicit spinning boundary correlators will
be presented elsewhere.

\vskip 4pt
The paper is organised as follows:
\begin{itemize}
    \item In Section~\ref{sec::SKEAdS} we review the Schwinger-Keldysh formalism for late-time correlators in de Sitter space and its reformulation \cite{Sleight:2020obc,Sleight:2021plv} in terms of boundary correlators of a corresponding theory in EAdS.
    \item In Section~\ref{sec::SKPI} we provide further details behind the EAdS formulation of late-time correlators \cite{Sleight:2020obc,Sleight:2021plv}, showing how to derive the generating functional of boundary correlators in the corresponding EAdS theory starting from the Schwinger--Keldysh generating functional in the path integral representation.
    \item In Section~\ref{sec::(EA)dSprops} we derive the Mellin-space representation of gauge boson and graviton propagators in (EA)dS in axial/temporal gauge, drawing on similarities with the analysis for scalar fields. In Mellin space it becomes transparent that dS propagators are linear combinations of their EAdS counterparts under analytic continuation. We further clarify subtleties in this relation that arise for massless representations in even boundary dimensions.
\item In Sections~\ref{sec::scalarqed}, \ref{sec::pureYM}, and \ref{sec::EHGr} we present the complete EAdS reformulation of the Schwinger-Keldysh formalism for late-time correlators in scalar QED, pure Yang–Mills theory, and Einstein gravity. We illustrate this in perturbation theory with examples of contact and tree-level exchange diagrams, and note that certain late-time falloffs yield vanishing boundary correlators at separated points. Any residual term in IR-divergent cases is a local coincident-point contribution and can be absorbed into boundary counterterms (as in \cite{Bzowski:2023nef}).

\item In Appendix \ref{app::props} we compile various technical details regarding the Mellin space representation of bulk-to-bulk propagators and their relation to other representations available in the literature.
    
\end{itemize}

\subsection{Notation and conventions.}
\label{subsec::notation and conventions}

We work in Poincar\'e coordinates for EAdS$_{d+1}$ and dS$_{d+1}$:
\begin{align}\label{poincare}
    {\rm d}s^2_{\text{EAdS}}=R^2_{\text{AdS}}\frac{{\rm d}z^2+{\rm d}{\bf x}^2}{z^2}\,, \qquad {\rm d}s^2_{\text{dS}}=R^2_{\text{dS}}\frac{-{\rm d}\eta^2+{\rm d}{\bf x}^2}{\eta^2}\,,
\end{align}
where $z\in[0,\infty)$ and $\eta\in(-\infty,0]$, where the latter parametrises the dS expanding patch. The boundary limit corresponds to $z \to 0$ and $ \eta \to 0$ respectively. We will take $R_{\text{(A)dS}}=1$ unless stated otherwise. We use Greek letters for spacetime indices, $\mu = 0, 1, \ldots, d$, and Latin letters for spatial indices, $i=1,\ldots d$. 

\vskip 4pt
The boundary directions are parametrised by the spatial vector ${\bf x}$. In these directions it is convenient to switch to Fourier space, introducing spatial momenta ${\bf k}$ so that translation invariance becomes explicit. For a function $f({\bf x})$ and its Fourier transform $\hat f({\bf k})$, the relation is
\begin{equation}\label{FT}
f({\bf x}) = \int \frac{{\rm d}^d {\bf k}}{(2\pi)^d}, e^{i {\bf k}\cdot{\bf x}}, \hat f({\bf k}),
\qquad
\hat f({\bf k}) = \int {\rm d}^d{\bf x}, e^{-i {\bf k}\cdot{\bf x}}, f({\bf x}),.
\end{equation}

\vskip 4pt
For the bulk directions $z$ (or $\eta$ in dS) it is often useful to work in a basis that diagonalises the dilatation generator. This is achieved by working in Mellin space (see \cite{Sleight:2021plv} and references therein), where $z$ (or $\eta$) are replaced by a Mellin variable $s$. For a function $f\left(z\right)$ and its Mellin transform ${\tilde f}(s)$ we have 
\begin{equation}\label{Mellin Transform}
    f\left(z\right) = \int^{+i\infty}_{-i\infty} \frac{{\rm d}s}{2\pi i}\, 2 {\tilde f}(s) z^{-(2s-\frac{d}{2})}, \qquad {\tilde f}(s) = \int^\infty_0 \frac{{\rm d}z}{z} f\left(z\right) z^{2s-\frac{d}{2}}.
\end{equation}
The integration contour is chosen to separate $\Gamma$ function poles. 

\vskip 4pt

We often employ the following shorthand notation for products of $\Gamma$ functions
\begin{equation}\label{shorthand Gamma}
    \Gamma\left(a\pm b\right) = \Gamma\left(a+b\right)\Gamma\left(a-b\right).
\end{equation}

Various parallels between Mellin space and Fourier space are summarised in the table below.

\begin{center}
\centering
\begin{tabular}{ | m{18em} | m{18em}|} 
  \hline
 \hfil {\bf Fourier space}  & \hfil {\bf Mellin space}  \\ 
  \hline
 \hfil ${\bf k}$ & \hfil $s$ \\ 
  \hline
\hfil  $e^{\pm i{\bf k}\cdot {\bf x}}$ & \hfil $z^{\mp\left(2s-\tfrac{d}{2}\right)}$  \\ 
\hline
\hfil  $\partial_{{\bf x}}\; \to \; i{\bf k}$ & \hfil $z\partial_z\;\to\;-\left(2s-\frac{d}{2}\right)$\\
\hline
  \hfil $\int\,{\rm d}^d{\bf x}\, e^{i{\bf x}\cdot {\bf k}}e^{-i{\bf x}\cdot {\bar {\bf k}}}=\left(2\pi\right)^d\delta^{(d)}\left({\bf k}-{\bar {\bf k}}\right)$& \vspace*{0.25cm} \hfil $\int_0^\infty\frac{{\rm d}z}{z^{d+1}}\,z^{-2s+\tfrac{d}{2}}z^{2{\bar s}+\tfrac{d}{2}}=\pi i\,\delta(s-{\bar s})$\vspace*{0.25cm}\\
  \hline
    \hfil $\int\,\frac{{\rm d}^d{\bf p}}{\left(2\pi\right)^d}\, e^{i{\bf x}\cdot {\bf p}}e^{-i{\bar {\bf x}}\cdot {\bf p} }=\delta^{(d)}\left({\bf x}-{\bar {\bf x}}\right)$& \vspace*{0.25cm} \hfil $2\int_{-i\infty}^{+i\infty}\frac{{\rm d}s}{2\pi i} \,z^{2s+\frac{d}{2}} {\bar z}^{-2s+\frac{d}{2}}=z^{d+1}\,\delta(z-{\bar z})$\vspace*{0.25cm}\\
  \hline
  \hfil  $\left(2\pi \right)^d\, \delta^{(d)}\left(\sum\limits^n_{i=1} {\bf k}_i\right)$ &  \hfil $2 \pi i\, \delta\left(\sum\limits^n_{i=1}\left(2s_i-\frac{d}{2}\right)\right)$\\
  \hline
\end{tabular}
\end{center}

We will often make use of the Mellin representation of Bessel  functions: 
\begin{align}\label{Bessels}
    &J_{i\nu}(kz):=\int_{-i\infty}^{
    +i\infty}\frac{{\rm d}s}{2\pi i}\left(\frac{zk}{2}\right)^{-2s}\frac{\Gamma(s+\frac{i\nu}{2})}{\Gamma(1-s+\frac{i\nu}{2})},\nonumber\\
    &I_{i\nu}(kz):=\int_{-i\infty}^{
    +i\infty}\frac{{\rm d}s}{2\pi i}e^{-i{\pi}(s+\frac{i\nu}{2})}\left(\frac{zk}{2}\right)^{-2s}\frac{\Gamma(s+\frac{i\nu}{2})}{\Gamma(1-s+\frac{i\nu}{2})},\nonumber\\
    &K_{i\nu}(kz):=\int_{-i\infty}^{
    +i\infty}\frac{{\rm d}s}{2\pi i}\left(\frac{zk}{2}\right)^{-2s}\frac{\Gamma(s+\frac{i\nu}{2})\Gamma(s-\frac{i\nu}{2})}{2}.
\end{align}

\vskip 4pt
When dealing with tensorial expressions we will often employ index-free notation. For a symmetric tensor $T_{i_i \ldots i_J}$ we introduce constant auxiliary vectors $w^i$ and write:
\begin{equation}\label{indexfree}
    T(w) = T_{i_i \ldots i_J} w^{i_1} \ldots w^{i_J}.
\end{equation}

\newpage

\section{Review: Schwinger-Keldysh formalism and rotation to EAdS}
\label{sec::SKEAdS}

\begin{figure}[t]
    \centering
    \includegraphics[width=0.75\textwidth]{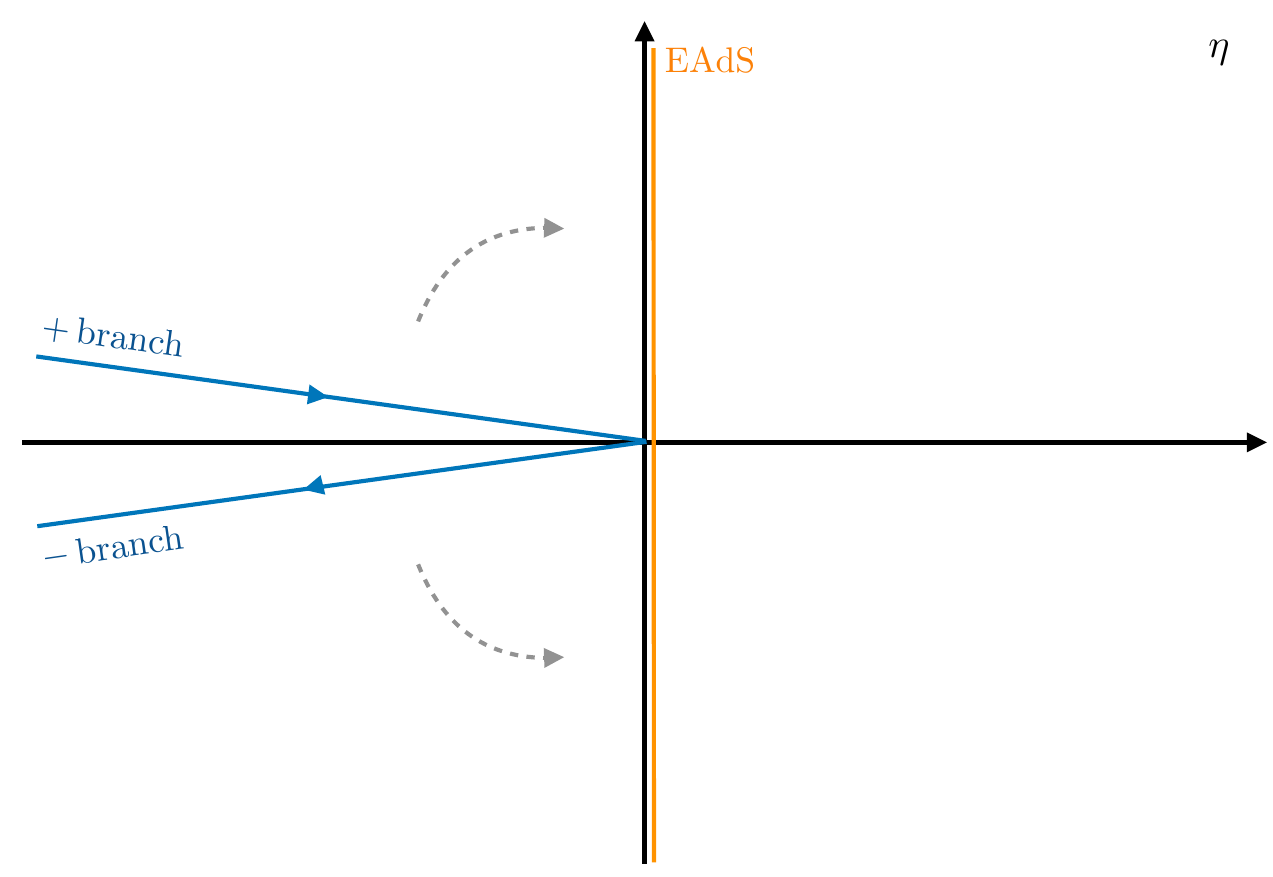}
    \caption{This figure illustrates the Schwinger-Keldysh contour (blue line) and the rotation of each branch to EAdS (yellow line) under the Wick rotations \eqref{wickzeta}.}
    \label{fig::Wick}
\end{figure}

Correlators on the future boundary of de Sitter space are expectation values 
\begin{equation}\label{LTcorr}
    \langle \phi_1(\eta_0, {\bf x}_1) \ldots \phi_n(\eta_0,{\bf x}_n) \rangle \equiv  \langle  \Omega | \phi_1(\eta_0,{\bf x}_1) \ldots \phi_n(\eta_0,{\bf x}_n) | \Omega \rangle,
\end{equation}
of operators $\phi_i$ inserted at various spatial points ${\bf x}_i$ on the future boundary $\eta_0 \to 0$. These can be computed using the Schwinger-Keldysh formalism \cite{Maldacena:2002vr,Bernardeau:2003nx,Weinberg:2005vy} (for a review see e.g. \cite{Chen:2017ryl}), which prescribes\footnote{Late-time correlators \eqref{LTcorr} can also be obtained by applying the Born rule to the cosmological wavefunction \cite{Maldacena:2002vr}.}
\begin{multline}
   \langle \phi_1(\eta_0, {\bf x}_1) \ldots \phi_n(\eta_0,{\bf x}_n) \rangle \\ =  \langle  \Omega | {\bar T} e^{+i \int^{\eta_0}_{-\infty}{\rm d}\eta^\prime\,H^{I}_{\text{int}}(\eta^\prime)}\phi_1(\eta_0,{\bf x}_1) \ldots \phi_n(\eta_0,{\bf x}_n) T e^{-i \int^{\eta_0}_{-\infty}{\rm d}\eta^\prime\,H^{I}_{\text{int}}(\eta^\prime)} | \Omega \rangle,
\end{multline}
where $({\bar T})T$ denotes (anti-)time ordering and $H^{I}_{\text{int}}$ is the interaction Hamiltonian in the interacting picture. The state $| \Omega \rangle$ is the early time vacuum of the fully interacting theory, which in the interaction picture can be expressed in terms of the free (Fock) vacuum $| 0 \rangle $. In this work we take the free theory vacuum to be the Bunch-Davies vacuum \cite{Chernikov:1968zm,Schomblond:1976xc,Gibbons:1977mu,Bunch:1978yq}, which can be implemented by introducing the following $i\epsilon$ prescription:
\begin{align}\label{Skcorr}
   & \langle \phi_1(\eta_0, {\bf x}_1) \ldots \phi_n(\eta_0,{\bf x}_n) \rangle \\ &  \hspace*{0.75cm}=  \langle  0 | {\bar T} e^{+i \int^{\eta_0}_{-\infty(1+i\epsilon)}{\rm d}\eta^\prime\,H^{I}_{\text{int}}(\eta^\prime)}\phi_1(\eta_0,{\bf x}_1) \ldots \phi_n(\eta_0,{\bf x}_n) T e^{-i \int^{\eta_0}_{-\infty(1-i\epsilon)}{\rm d}\eta^\prime\,H^{I}_{\text{int}}(\eta^\prime)} | 0 \rangle. \nonumber 
\end{align}
The integration contour in the complex $\eta$ plane (known as the Schwinger-Keldysh or in-in contour) is illustrated in figure \ref{fig::Wick}. It consists of two branches: the $+$ branch corresponding to time ordering and the $-$ branch anti-time-ordering.

\vskip 4pt
The correlators \eqref{Skcorr} can be computed perturbatively in the Schwinger-Keldysh formalism by expanding in powers of $H^{I}_{\text{int}}$ and applying Wick's theorem. This gives rise to four bulk-to-bulk propagators:
\begin{subequations}\label{SK propagators}
  \begin{align}
    G^{++}\left(x_1;x_2\right) &= \langle  0 | T \phi\left(x_1\right)\phi\left(x_2\right) | 0 \rangle,\\
    G^{--}\left(x_1;x_2\right) &= \langle  0 | {\bar T} \phi\left(x_1\right)\phi\left(x_2\right) | 0 \rangle,\\
    G^{+-}\left(x_1;x_2\right) &= \langle  0 |  \phi\left(x_2\right)\phi\left(x_1\right) | 0 \rangle,\\
    G^{-+}\left(x_1;x_2\right) &= \langle  0 |  \phi\left(x_1\right)\phi\left(x_2\right) | 0 \rangle.
\end{align}  
\end{subequations}

In Fourier space the mode expansion of each field operator $\phi$ in terms of creation and annihilation operators takes the form
\begin{equation}\label{modeexp}
\phi_{{\bf k}}\left(\eta\right) = f_{{\bf k}}\left(\eta\right) a^\dagger_{{\bf k}}\,+{\bar f}_{{\bf k}}\left(\eta\right)a_{-{\bf k}}\,, 
\end{equation}
where
\begin{equation}
    \phi_{{\bf k}}\left(\eta\right)=\int {\rm d}^d{\bf x}\,e^{-i\bf k\cdot x}\phi(\eta,{\bf x}).
\end{equation}
In terms of mode functions the Schwinger-Keldysh propagators read :
\begin{subequations}\label{SK mode functions}
\begin{align} G^{++}(\eta,\bar{\eta};{\bf k})&=\theta\left(\eta - {\bar \eta}\right){\bar f}_{{\bf k}}\left(\eta\right)f_{{\bf k}}\left({\bar \eta}\right)+\theta\left({\bar \eta} - \eta\right){\bar f}_{{\bf k}}\left({\bar \eta}\right)f_{{\bf k}}\left(\eta\right),\\   G^{--}(\eta,\bar{\eta};{\bf k})&=\theta\left({\bar \eta}-\eta\right){\bar f}_{{\bf k}}\left(\eta\right)f_{{\bf k}}\left({\bar \eta}\right)+\theta\left(\eta - {\bar \eta}\right){\bar f}_{{\bf k}}\left({\bar \eta}\right)f_{{\bf k}}\left(\eta\right),
    \\  G^{+-}(\eta,\bar{\eta};{\bf k})&={\bar f}_{{\bf k}}\left({\bar \eta}\right)f_{{\bf k}}\left(\eta\right),\\
 \hspace*{-0.5cm}   G^{-+}(\eta,\bar{\eta};{\bf k})&={\bar f}_{{\bf k}}\left( \eta\right)f_{{\bf k}}\left({\bar \eta}\right).
\end{align}
\end{subequations}

For correlators on the future boundary it is useful to introduce bulk-to-boundary propagators. At late times a free field $\phi_J$ of spin-$J$ behaves as\footnote{Here we have employed index-free notation as in \eqref{indexfree}.}
\begin{equation}\label{bdrybehaviour}
    \phi_{J}\left(\eta \to 0,{\bf x}\right) = \left(-\eta\right)^{\Delta_+-J}{\cal O}_{\Delta_+,J}\left({\bf x}\right)+\left(-\eta\right)^{\Delta_--J}{\cal O}_{\Delta_-,J}\left({\bf x}\right),
\end{equation}
where the two fall-offs $\Delta_\pm$ are fixed in terms of the mass as:
\begin{equation}
    m^2 = \Delta_+\Delta_-+J, \qquad \Delta_++\Delta_-=d.
\end{equation}
The boundary operators ${\cal O}_{\Delta_\pm,J}({\bf x})$ are spin-$J$ conformal primaries with shadow scaling dimension $\Delta_\pm$. The corresponding Schwinger-Keldysh bulk-to-boundary propagators are defined as 
\begin{equation}\label{SKbubo}
  \lim_{{\bar \eta}\to 0}   G^{\pm\hat{\pm}}(\eta,\bar{\eta};{\bf k}) = \left(-{\bar \eta}\right)^{\Delta_+-J}K^{\pm}_{\Delta_+}(\eta;{\bf k})+\left(-{\bar \eta}\right)^{\Delta_--J}K^{\pm}_{\Delta_-}(\eta;{\bf k}).
\end{equation}
We will often parameterise the scaling dimensions as $\Delta_\pm =  \frac{d}{2}\pm i \nu$, which are related under $\nu \to -\nu$.

\vskip 4pt
In QFT it is often useful to pass to Euclidean signature by analytically continuing the time coordinate. In Poincar\'e coordinates \eqref{poincare}, this is implemented by the continuation\footnote{With our mostly-plus convention, this continuation sends the dS line element to minus the standard positive-definite EAdS line element. This is the usual overall sign encountered in Wick rotation and does not require an analytic continuation of the curvature radius.}
\begin{equation}
\eta \to \pm i z .
\end{equation}
The sign is fixed by the Schwinger-Keldysh branch, or equivalently by the Bunch-Davies $i\epsilon$ prescription. Under this rotation, de Sitter two-point functions map to two-point functions in Euclidean AdS. This was exploited in \cite{Sleight:2019mgd,Sleight:2019hfp,Sleight:2020obc,Sleight:2021plv} to map the Schwinger-Keldysh propagators \eqref{SK propagators} to linear combinations of bulk-to-bulk propagators for EAdS Witten diagrams. In the Bunch-Davies vacuum, one opens up the Schwinger-Keldysh contour so that it runs parallel to the imaginary axis in the complex $\eta$ plane, rotating the $+$ and $-$ branches by $90^\circ$ clockwise and anticlockwise respectively \cite{Sleight:2019mgd,Sleight:2019hfp}\footnote{In the embedding-space description of (EA)dS$_{d+1}$, the two Schwinger-Keldysh branches in dS Poincar\'e patch continue to the two-sheets of the EAdS hyperboloid $X^2 = -1$ where $X \in \mathbb{R}^{1,d+1}$. For Bunch-Davies correlators, the absence of short-distance singularities between the two sheets allows both contour integrals to be parameterised by a single EAdS Poincar\'e coordinate $z\in(0,\infty)$. This simplification is not generally available for $\alpha$-states, where antipodal singularities require keeping the two sheets distinct \cite{Chopping:2024oiu}.}
\begin{shaded}
 \begin{equation}\label{wickzeta}
   \pm \,\text{branch:}\quad \eta_\pm \to e^{\mp i\frac{\pi}2}\eta_{\pm}= e^{\pm i\frac{\pi}{2}} z\quad z\in(0,\infty).
\end{equation}   
\end{shaded}
\noindent This is illustrated in figure \ref{fig::Wick}. These paths follow the prescriptions for going around the light-cone singularity of propagators in the Bunch-Davies vacuum, which are the same as their Minkowski counterparts.

\vskip 4pt
For fields of (integer) spin-$J$ it was shown in \cite{Sleight:2020obc,Sleight:2021plv} that under the Wick rotations \eqref{wickzeta} the Schwinger-Keldysh propagators \eqref{SK propagators} for late-time correlators are identified with the following linear combinations of bulk-to-bulk propagators for the $\Delta_\pm$ boundary conditions in EAdS:\footnote{Note that in \cite{Sleight:2020obc,Sleight:2021plv} the space-time indices were assumed to be contracted with constant auxiliary vectors $u^\alpha$ according to (see e.g. \cite{Taronna:2012gb}):
\begin{equation}
    \varphi(x,u) = \varphi_{\mu_1 \ldots \mu_J}(x) u \cdot e^{\mu_1}(x) \ldots  u \cdot e^{\mu_J}(x), 
\end{equation}
where, under the Wick rotation \eqref{wickzeta}, the inverse Vielbein cancels the spin $J$ dependence in the phases that appear \eqref{bubuwick} and \eqref{bubowick}. In terms of spacetime indices, as in \eqref{bubuwick} and \eqref{bubowick}, extra care should be taken when raising indices. This should be carried out with respect to the inverse metric in dS, where in Poincar\'e coordinates we have $g^{\mu \nu} = \eta^2 \delta^{\mu \nu}$. This introduces additional powers of $\eta$ and hence additional phases when Wick rotated according to \eqref{wickzeta}.}

\begin{shaded}
\begin{multline}\label{bubuwick}
    G^{\pm\hat{\pm}}_{\mu_1 \ldots \mu_J;\nu_1 \ldots \nu_J}(\eta,\bar{\eta})\to c^{\text{dS-AdS}}_{\Delta_+}e^{\mp \left(\Delta_+-J\right)\frac{\pi i}{2}}e^{\hat{\mp} \left(\Delta_+-J\right)\frac{\pi i}{2}}G^{\text{AdS}\,\Delta_+}_{\mu_1 \ldots \mu_J;\nu_1 \ldots \nu_J}(z,\bar{z}) \\+ c^{\text{dS-AdS}}_{\Delta_-}e^{\mp \left(\Delta_--J\right)\frac{\pi i}{2}}e^{\hat{\mp} \left(\Delta_--J\right)\frac{\pi i}{2}}G^{\text{AdS}\,\Delta_-}_{\mu_1 \ldots \mu_J;\nu_1 \ldots \nu_J}(z,\bar{z}).
\end{multline}    
\end{shaded}
\noindent This is equivalent to the following field redefinition:
\begin{multline}\label{fieldredef}
    \left(\phi_\pm\right)_{\mu_1 \ldots \mu_J} \to e^{\mp \left(\Delta_+-J\right)\frac{\pi i}{2}}\sqrt{|c^{\text{dS-AdS}}_{\Delta_+}|}   \left(\phi_{\Delta_+}\right)_{\mu_1 \ldots \mu_J}\\ + e^{\mp \left(\Delta_--J\right)\frac{\pi i}{2}}  \sqrt{|c^{\text{dS-AdS}}_{\Delta_-}|} \left(\phi_{\Delta_-}\right)_{\mu_1 \ldots \mu_J},
\end{multline}
where the fields $\phi_{\Delta_\pm}$ are subject to the $\Delta_\pm$ boundary conditions in EAdS. The coefficients $c^{\text{dS-AdS}}_{\Delta_\pm}$ take into account the change in two-point function normalisation from EAdS to dS, and are given explicitly by:
\begin{align}\label{cseadstods}
   c^{\text{dS-AdS}}_{\Delta} =\frac{1}{2}\text{csc}\left(\left(\frac{d}{2}-\Delta\right)\pi \right).
\end{align}
The bulk-to-boundary propagators \eqref{SKbubo} are related via: 
\begin{shaded}
   \begin{equation}\label{bubowick}
    K^{\pm\, \Delta}_{\mu_1 \ldots \mu_J;j_1 \ldots j_J}(\eta;{\bf k}) \to e^{\mp \left(\Delta-J\right)\frac{\pi i}{2}} c^{\text{dS-AdS}}_{\Delta}K^{\text{AdS}\,\Delta}_{\mu_1 \ldots \mu_J;j_1 \ldots j_J}(z;{\bf k}).
\end{equation}  
\end{shaded}

\vskip 4pt
Under the rotation \eqref{wickzeta} the integrals over the $\pm$ branches of the in-in contour can be re-cast as integrals over EAdS: 
\begin{shaded}
 \begin{align}\label{measurewick}
  \pm \,\text{branch}: &\quad  \pm i \int_{-\infty}^0 \frac{{\rm d}\eta}{\left(-\eta\right)^{d+1}} (...) = \pm i e^{\pm \frac{d\pi i}{2} }\int^\infty_0 \frac{{\rm d}z}{z^{d+1}} (...),
\end{align}   
\end{shaded}
\noindent while Lagrangian vertices ${\cal V}$ acquire a phase
\begin{shaded}
 \begin{equation}\label{vertexwick}
   \pm \,\text{branch}: \quad  {\cal V}\left(\eta\right) \to e^{\mp\frac{N \pi i}{2}} {\cal V}\left(z\right),
\end{equation}   
\end{shaded}
\noindent where $N$ is an integer determined by the number of inverse metric contractions in the vertex.

\begin{figure}[t]
    \centering
    \includegraphics[width=\textwidth]{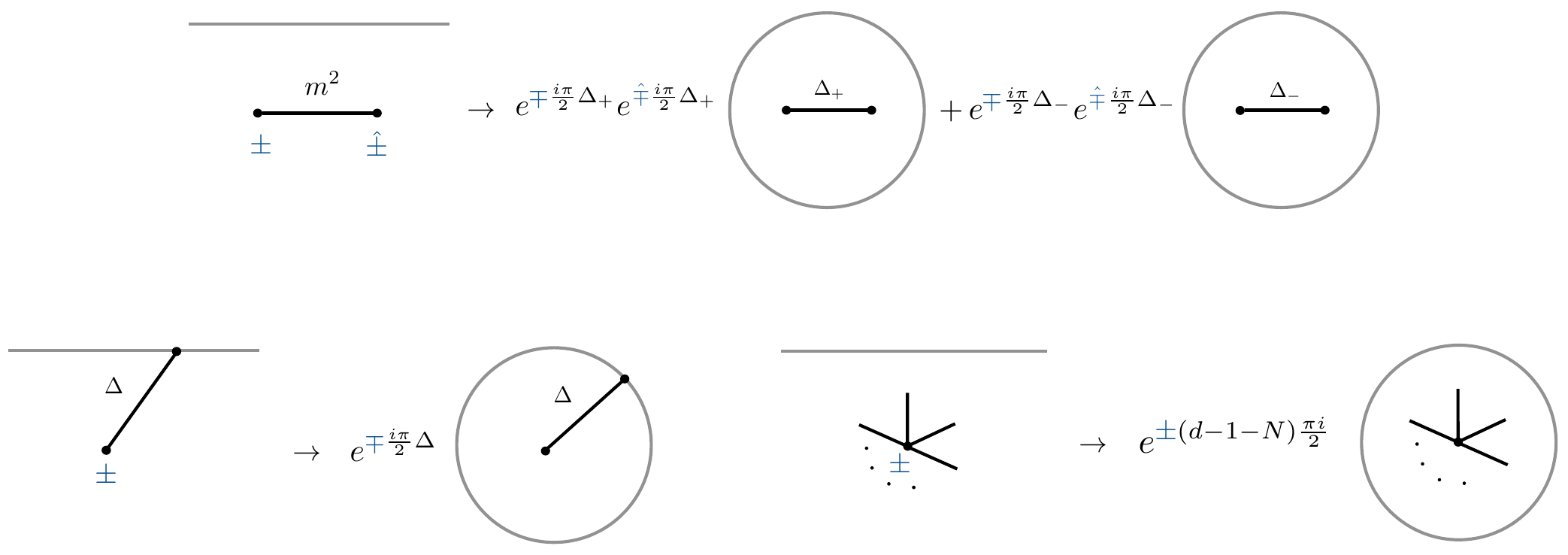}
    \caption{Graphical summary of the rules \eqref{bubuwick}, \eqref{bubowick}, \eqref{measurewick} and \eqref{vertexwick}, derived in \cite{Sleight:2020obc,Sleight:2021plv}, recasting perturbation theory for late-time correlators in the Schwinger-Keldysh formalism in terms of Witten diagrams in EAdS. The dS late time boundary is the horizontal grey line and the EAdS boundary the grey circle.}
    \label{fig::FeyndSAdS}
\end{figure}

\vskip 4pt
Together, the identities \eqref{bubuwick}, \eqref{bubowick}, \eqref{measurewick} and \eqref{vertexwick} recast the Feynman rules for late-time correlators into Feynman rules for the boundary correlators of a corresponding theory in EAdS \cite{Sleight:2020obc,Sleight:2021plv}, which are summarised graphically in figure \ref{fig::FeyndSAdS}. This result is an equivalence between the generating functional of Schwinger-Keldysh (S-K) correlators and the generating functional for boundary correlators in the EAdS theory:
\begin{shaded}
 \begin{equation}
    Z_{\text{S-K}}\left[J_+,J_-\right] = Z_{\text{EAdS}}\left[J_{\Delta_+},J_{\Delta_-}\right],
\end{equation}   
\end{shaded}
\noindent where $J_\pm$ source $\phi_\pm$ and $J_{\Delta_\pm}$ source $\phi_{\Delta_\pm}$. This is constructed explicitly from the path integral formalism in the next section.

\vskip 4pt
Using this framework, late-time correlators in dS can be written, to all orders in perturbation theory, as boundary correlators in EAdS. For example, for contact interactions we have \cite{Sleight:2020obc,Sleight:2021plv}:
\begin{multline}
    \langle {\cal O}_{\Delta_1, J_1}\left({\bf x}_1\right)  \ldots {\cal O}_{\Delta_n, J_n}\left({\bf x}_n\right) \rangle_{\text{dS}\,\text{contact}} = \sum_{\pm} \langle {\cal O}_{\Delta_1, J_1}\left({\bf x}_1\right)  \ldots {\cal O}_{\Delta_n, J_n}\left({\bf x}_n\right) \rangle^\pm_{\text{dS}\,\text{contact}},
\end{multline}
where applying the rules \eqref{bubowick}, \eqref{measurewick} and \eqref{vertexwick} the $\pm$ branch contributions are given in terms of the corresponding contact diagram in EAdS:
\begin{multline}\label{pmcontact}
   \langle {\cal O}_{\Delta_1, J_1}\left({\bf x}_1\right)  \ldots {\cal O}_{\Delta_n, J_n}\left({\bf x}_n\right) \rangle^\pm_{\text{dS}\,\text{contact}} =  \mp i \left(\prod^n_{i=1} c^{\text{dS-AdS}}_{\Delta_i} \right) e^{\mp \left(N-d+\sum_i (\Delta_i- J_i)\right)\frac{\pi i}{2}}\\ \times \langle {\cal O}_{\Delta_1, J_1}\left({\bf x}_1\right)  \ldots {\cal O}_{\Delta_n, J_n}\left({\bf x}_n\right) \rangle_{\text{EAdS}\,\text{contact}}.
\end{multline}
These combine to give:
\begin{multline}\label{sinecontact}
    \langle {\cal O}_{\Delta_1, J_1}\left({\bf x}_1\right)  \ldots {\cal O}_{\Delta_n, J_n}\left({\bf x}_n\right) \rangle_{\text{dS}\,\text{contact}} = \underbrace{\left(\prod^n_{i=1} c^{\text{dS-AdS}}_{\Delta_i} \right) 2\sin\left(\left(-d+N+\sum_i (\Delta_i- J_i)\right)\frac{\pi}{2}\right)}_{c^{\text{dS-AdS}}_{\Delta_i;J_i;N}}\\
   \times \langle {\cal O}_{\Delta_1, J_1}\left({\bf x}_1\right)  \ldots {\cal O}_{\Delta_n, J_n}\left({\bf x}_n\right) \rangle_{\text{EAdS}\,\text{contact}},
\end{multline}
where the sinusoidal factor arises from combining the contributions from each branch of the in-in contour. For exchange processes, the rule \eqref{bubuwick} for internal legs implies that these decompose as a sum of corresponding particle exchanges in EAdS---where the sum is taken over the $\Delta_\pm$ boundary conditions on the internal legs. These are multiplied by products of the sinusoidal factors \eqref{sinecontact}, relating each EAdS contact subdiagram to their dS counterpart. For example, the ${\sf s}$-channel exchange of a spin-$J$ particle of mass $m^2 = \Delta_+\Delta_-+J$ decomposes in terms of corresponding EAdS exchange Witten diagrams as follows \cite{Sleight:2020obc,Sleight:2021plv}:
\begin{align}
   & \langle {\cal O}_{\Delta_1, J_1}\left({\bf x}_1\right)  \ldots {\cal O}_{\Delta_4, J_4}\left({\bf x}_n\right) \rangle_{\text{dS}\,\text{exchange}} \\ \nonumber  & \hspace*{1cm} = \frac{c^{\text{dS-AdS}}_{\Delta_1\Delta_2\Delta_+;J_1,J_2,J;N_{12}}c^{\text{dS-AdS}}_{\Delta_+\Delta_3\Delta_4;J,J_3,J_4;N_{34}}}{c^{\text{dS-AdS}}_{\Delta_+} }  \langle {\cal O}_{\Delta_1, J_1}\left({\bf x}_1\right)  \ldots {\cal O}_{\Delta_4, J_4}\left({\bf x}_n\right) \rangle_{\text{EAdS}\,\text{exchange}\,\Delta_+}\\ \nonumber 
     & \hspace*{1cm} +\frac{c^{\text{dS-AdS}}_{\Delta_1\Delta_2\Delta_-;J_1,J_2,J;N_{12}}c^{\text{dS-AdS}}_{\Delta_-\Delta_3\Delta_4;J,J_3,J_4;N_{34}}}{c^{\text{dS-AdS}}_{\Delta_-} }  \langle {\cal O}_{\Delta_1, J_1}\left({\bf x}_1\right)  \ldots {\cal O}_{\Delta_4, J_4}\left({\bf x}_n\right) \rangle_{\text{EAdS}\,\text{exchange}\,\Delta_-},
\end{align}
where $N_{12}$ and $N_{34}$ are the phases \eqref{vertexwick} assigned to the vertices connected to ${\bf x}_{1,2}$ and ${\bf x}_{3,4}$ respectively. The constants $c^{\text{dS-AdS}}_{\Delta_1\Delta_2\Delta_\pm;J_1,J_2,J;N_{12}}$ and $c^{\text{dS-AdS}}_{\Delta_\pm\Delta_3\Delta_4;J,J_3,J_4;N_{34}}$ are the coefficients \eqref{sinecontact} relating the dS and EAdS contact subdiagrams. For further examples beyond tree level, see \cite{Sleight:2021plv}.

\vskip 4pt
The above EAdS reformulation of late-time correlators in the Bunch-Davies vacuum was presented in \cite{Sleight:2020obc,Sleight:2021plv} for generic dS EFTs of scalar and (integer) spinning fields, and extended to Fermions in \cite{Schaub:2023scu}. In the present work we analyse the map for theories of gauge bosons and gravitons, which correspond to the following specific values of $\Delta_\pm$:
\begin{align}
    \text{gauge boson}:& \quad \Delta_+ = d-1,\quad \Delta_- = 1, \\
    \text{graviton}:& \quad \Delta_+ = d,\quad \Delta_- = 0.
\end{align}
In these cases, the $\Delta_+$ falloff is the standard AdS/CFT Dirichlet boundary condition corresponding to the boundary current/ stress tensor. The $\Delta_-$ falloff instead corresponds to the Neumann boundary condition, where the gauge bosons/ gravitons are propagating on the boundary and codify outgoing radiation. For even boundary dimensions $d$, the general relations \eqref{bubuwick} between dS and EAdS propagators break down for these values of $\Delta_\pm$ since the coefficient \eqref{cseadstods} diverges. This occurs because the discriminant of the propagator equation in EAdS has a degeneracy in these cases and requires a modification of the relations \eqref{bubuwick} and \eqref{bubowick}, which we present in section \ref{subsec::nuimint}.

\newpage
\section{Schwinger-Keldysh Path Integral and EAdS reformulation}
\label{sec::SKPI}

In section \ref{sec::SKEAdS} we reviewed the Schwinger-Keldysh computation of late-time correlators in the operator formalism \eqref{Skcorr}. The Schwinger-Keldysh formalism can also be implemented via a path integral with a doubled field content \cite{Feynman:1963fq,Weinberg:2005vy}, with generating functional:
\begin{multline}\label{SkPI}
   Z_{\text{S-K}}\left[J_+,J_-\right] =   \underset{\phi_+(0,{\bf x})=\phi_-(0,{\bf x})}{\int {\cal D}\phi_+\,{\cal D}\phi_-}  \,\\ \times e^{i \int^{\eta_0}_{-(\infty-i\epsilon)}\frac{{\rm d}\eta^\prime{\rm d}^{d}{\bf x}}{\left(-\eta^\prime\right)^{d+1}}\left( {\cal L}\left[\phi_+\right]+J_+ \phi_+\right)}e^{-i \int^{\eta_0}_{-(\infty+i\epsilon)}\frac{{\rm d}\eta^\prime{\rm d}^{d}{\bf x}}{\left(-\eta^\prime\right)^{d+1}}\left({\cal L}\left[\phi_-\right]+J_- \phi_- \right)}.
\end{multline}
In the following we consider the Lagrangian density for a theory in dS of the generic form
\begin{equation}\label{Langrangian}
    {\cal L}\left[\phi\right] = -\left[\frac{1}{2} \nabla_\mu \phi_{\mu_1 \ldots \mu_J}\nabla^\mu \phi^{\mu_1 \ldots \mu_J} +\frac{1}{2}m^2 \phi_{\mu_1 \ldots \mu_J} \phi^{\mu_1 \ldots \mu_J}+ \ldots + {\cal V}\left(\phi\right)\right],
\end{equation}
where the $\ldots$ denote additional (off-shell) contributions to the kinetic term which are present for spin $J>0$ and required for consistency of the action. The field content is collectively denoted by $\phi$.

\vskip 4pt
In section \ref{sec::SKEAdS} we saw that the Feynman rules for the late-time correlators, encoded by the generating functional \eqref{SkPI}, can be recast under Wick rotations \eqref{wickzeta}, as Feynman rules for boundary correlators of a theory in EAdS \cite{Sleight:2020obc,Sleight:2021plv}. This result is an equivalence between the Schwinger-Keldysh generating functional \eqref{SkPI} and the generating functional for correlators in the EAdS theory:\footnote{The sources $J_\pm$ on the rotated contours can be
written in terms of the EAdS sources $J_{\Delta_\pm}$. For a spin-$J$ field, under $\eta_a=e^{a\pi i/2}z$ we have 
    \begin{equation}
    J_a(\eta_a,{\bf x})
    =
    \frac{
        c^{\text{dS-AdS}}_{\Delta_+}
    }{
        \sqrt{
            \left|
            c^{\text{dS-AdS}}_{\Delta_+}
            \right|
        }
    }
        e^{-a\left(\Delta_+ +J\right)\frac{\pi i}{2}}
        J_{\Delta_+}(z,{\bf x})+ (\Delta_+ \to \Delta_-).
\end{equation}}
\begin{equation}
    Z_{\text{S-K}}\left[J_+,J_-\right] = Z_{\text{EAdS}}\left[J_{\Delta_+},J_{\Delta_-}\right].
\end{equation}
Indeed, applying the field redefinition \eqref{fieldredef} gives the following EAdS Lagrangians:
\begin{multline}\label{Lphipm}
  {\cal L}_{\text{EAdS}}\left[\phi_\pm\right] = \pm i e^{\pm \frac{d \pi i}{2} } \left[e^{\mp 2\Delta_+\frac{\pi i}{2}} e^{i\theta[c^{\text{dS-AdS}}_{\Delta_+}]}|c^{\text{dS-AdS}}_{\Delta_+}|\left(\frac{1}{2} \nabla_\mu \left(\phi_{\Delta_+}\right)_{\mu_1 \ldots \mu_J}\nabla^\mu \left(\phi_{\Delta_+}\right)^{\mu_1 \ldots \mu_J} \right. \right.\\ \left. \left. \hspace*{3cm}- \frac{1}{2}m^2 \left(\phi_{\Delta_+}\right)_{\mu_1 \ldots \mu_J} \left(\phi_{\Delta_+}\right)^{\mu_1 \ldots \mu_J}+ \ldots \right) + (\Delta_+ \to \Delta_-)  \right. \\ \left. + e^{\mp\frac{N \pi i}{2}} {\cal V}\left(e^{\mp \left(\Delta_+-J\right)\frac{\pi i}{2}}e^{i\theta[c^{\text{dS-AdS}}_{\Delta_+}]} \sqrt{|c^{\text{dS-AdS}}_{\Delta_+}|}   \left(\phi_{\Delta_+}\right)_{\mu_1 \ldots \mu_J}+ \Delta_+ \to \Delta_-\right)\right],
\end{multline}
The phase is defined by
\begin{equation}
    e^{i\theta[c]}=\frac{c}{|c|}.
\end{equation}
$N$ is the same as in \eqref{vertexwick} and determined by the number of derivatives and space-time index contractions in the vertex ${\cal V}$. For a vertex of order $n$ in fields $\phi_i$ of mass $m^2_i$ we have in particular:
\begin{multline}
{\cal V}\left(\phi_{i\,\pm}\right) = \sum\limits_{(\Delta_1)_{\pm_1} \ldots (\Delta_n)_{\pm_n}} e^{\mp \left(N-d+\sum_i ((\Delta_i)_{\pm_i}- J_i)\right)\frac{\pi i}{2}} \left(\prod^n_{i=1}  \sqrt{|c^{\text{dS-AdS}}_{(\Delta_i)_{\pm_i}}|} \right) \\ \times {\cal V}\left(\phi_{(\Delta_1)_{\pm_1}}, \ldots, \phi_{(\Delta_n)_{\pm_n}}\right),
\end{multline}  
which is equivalent to \eqref{pmcontact}. The Schwinger-Keldysh generating functional \eqref{SkPI} is then recast as the EAdS generating functional: 
\begin{shaded}
\noindent \emph{EAdS generating functional for dS late-time correlators.}
 \begin{equation}\label{EAdSPI}
Z_{\text{EAdS}}\left[J_{\Delta_+},J_{\Delta_-}\right] =  \underset{\phi_{\Delta_\pm}(z\to 0,{\bf x})\sim \# z^{\Delta_{\mp}}}{\int {\cal D}\phi_{\Delta_+}\,{\cal D}\phi_{\Delta_-}} \,e^{-S_{\text{EAdS}}\left[\phi_{\Delta_+},\phi_{\Delta_-}\right]}e^{-\int_{z_0}^{\infty}\frac{{\rm d}z}{z^{d+1}}\,\left( J_{\Delta_+} \phi_{\Delta_+}+J_{\Delta_-} \phi_{\Delta_-}\right)},
\end{equation}   
\end{shaded}
\noindent with 
\begin{equation}
    S_{\text{EAdS}}\left[\phi_{\Delta_+},\phi_{\Delta_-}\right]  = \int {\rm d}z {\rm d}^d{\bf x}\,\sqrt{g}\,{\cal L}_{\text{EAdS}}\left[\phi_{\Delta_+},\phi_{\Delta_-}\right], 
\end{equation}
and Lagrangian
\begin{align}\label{SEAdScanon}
   & {\cal L}_{\text{EAdS}}\left[\phi_{\Delta_+},\phi_{\Delta_-}\right]=  \\ \nonumber
    & \hspace*{0.15cm} \times   \left[\sum_ie^{i\theta[c^{\text{dS-AdS}}_{\Delta_{i\,,+}}]}\left(  \frac{1}{2}\nabla_\mu \left(\phi_{\Delta_{i\,+}}\right)_{\mu_1 \ldots \mu_J}\nabla^\mu \left(\phi_{\Delta_{i\,+}}\right)^{\mu_1 \ldots \mu_J} - \frac{1}{2}m^2_i \left(\phi_{\Delta_{i\,+}}\right)_{\mu_1 \ldots \mu_J} \left(\phi_{\Delta_{i\,+}}\right)^{\mu_1 \ldots \mu_J}+ \ldots \right) \right. \\ \nonumber &  \left. \hspace*{0.25cm}+\, \sum_ie^{i\theta[c^{\text{dS-AdS}}_{\Delta_{i\,,-}}]}\left( \frac{1}{2} \nabla_\mu \left(\phi_{\Delta_{i\,-}}\right)_{\mu_1 \ldots \mu_J}\nabla^\mu \left(\phi_{\Delta_{i\,-}}\right)^{\mu_1 \ldots \mu_J} - \frac{1}{2}m^2_i \left(\phi_{\Delta_{i\,-}}\right)_{\mu_1 \ldots \mu_J} \left(\phi_{\Delta_{i\,-}}\right)^{\mu_1 \ldots \mu_J}+ \ldots \right) \right. \\ \nonumber & \hspace*{0.65cm} \left. + 2 \sum\limits_{(\Delta_1)_{\pm_1} \ldots (\Delta_n)_{\pm_n}} \left(\prod^n_{i=1}  \sqrt{|c^{\text{dS-AdS}}_{(\Delta_i)_{\pm_i}}|} \right) \sin\left(\left(-d+N+\sum_i ((\Delta_i)_{\pm_i}- J_i)\right)\frac{\pi}{2}\right) \right. \\ & \left. \hspace*{10cm}\times {\cal V}\left(\phi_{(\Delta_1)_{\pm_1}}, \ldots, \phi_{(\Delta_n)_{\pm_n}}\right)\right], \nonumber
\end{align}
where the phases in \eqref{Lphipm} combine to give the sinusoidal factor \eqref{sinecontact} and canonically normalised kinetic term (up to constant phases). Since $c^{\text{dS-AdS}}_{\Delta_+}
    =
    -c^{\text{dS-AdS}}_{\Delta_-}$, the two shadow sectors always have opposite phase factors,\footnote{Consequently, one of the two fields $\phi_{\Delta_\pm}$ necessarily carries a kinetic term with the opposite sign. This is simply the Euclidean counterpart of the sign-indefinite structure inherited from the doubled Schwinger--Keldysh (in-in) formulation.}
\begin{equation}
    e^{i\theta[c^{\text{dS-AdS}}_{\Delta_{i,-}}]}
    =
    - e^{i\theta[c^{\text{dS-AdS}}_{\Delta_{i,+}}]} .
\end{equation}
For real scaling dimensions, the coefficients
$c^{\text{dS-AdS}}_{\Delta_\pm}$ are real, and hence
\begin{equation}
    e^{i\theta[c^{\text{dS-AdS}}_{\Delta_{i,+}}]}=\pm 1,
    \qquad
    e^{i\theta[c^{\text{dS-AdS}}_{\Delta_{i,-}}]}=\mp 1 .
\end{equation}
For principal-series weights $\Delta_\pm=\frac d2\pm i\nu,\,\nu\in\mathbb{R}$ the coefficients $c^{\text{dS-AdS}}_{\Delta_\pm}$ are purely imaginary, and therefore
\begin{equation}
    e^{i\theta[c^{\text{dS-AdS}}_{\Delta_{i,+}}]}=\pm i,
    \qquad
    e^{i\theta[c^{\text{dS-AdS}}_{\Delta_{i,-}}]}=\mp i .
\end{equation}
Special
integer values for which the coefficients $c^{\text{dS-AdS}}_{\Delta}$ are singular require
a limiting prescription---these cases will be discussed separately in section
\ref{subsec::nuimint}.

\vskip 4pt
Although the explicit correspondence \eqref{EAdSPI} has so far been verified in perturbation theory, the EAdS generating functional \eqref{SEAdScanon} is naturally suggested by the Wick rotation of the Schwinger–Keldysh contour (figure \ref{fig::Wick}) and may be viewed as a formal Euclidean representation of dS late-time correlators at the level of the functional integral, analogous to similar EAdS reformulations \eqref{wfwick} at the level of the wavefunction of the universe \cite{Maldacena:2002vr,McFadden:2009fg} and the Wick rotation in standard path integral formulations of quantum field theory.

\vskip 4pt
The sinusoidal factors \eqref{sinecontact} can lead to substantial simplifications of late-time correlators relative to their wavefunction/EAdS Witten-diagram counterparts, as noted in \cite{Sleight:2019mgd,Sleight:2019hfp,Goodhew:2020hob,Chowdhury:2023arc,Chowdhury:2025ohm}. In particular, at the zeros of the sine function---which occur for special values of the scaling dimensions---the non-local part of the late-time correlator vanishes at separated points. A non-vanishing contribution can arise only when the corresponding EAdS diagram is itself IR divergent, in which case any residual term is local and supported at coincident points. This is reminiscent of the mechanism familiar from extremal correlators in AdS/CFT, where a vanishing prefactor multiplies a singular AdS integral, leaving only a local contribution after regularisation \cite{DHoker:1999jke,DHoker:2000xhf}. In later sections we will see that this leads to significant simplifications in correlators of gauge bosons and gravitons in certain dimensions, where their scaling dimensions lie at zeros of some of the corresponding sine factors. 

\vskip 4pt
The same sinusoidal factors can also render finite a late-time correlator whose EAdS counterpart is IR divergent, through cancellations between would-be divergent contributions. This was realised in \cite{Sleight:2019mgd,Sleight:2019hfp} using dimensional regularisation. The tools developed in the present work make it possible to extend these analyses of IR divergences to theories of gauge bosons and gravitons, which we leave to future work.

\section{(EA)dS Propagators}
\label{sec::(EA)dSprops}

In this section we derive the relations \eqref{bubuwick} and \eqref{bubowick} between Schwinger-Keldysh propagators and EAdS propagators for gauge bosons and gravitons. To this end we work in Mellin space \cite{Sleight:2019mgd,Sleight:2019hfp,Sleight:2021plv}, where such relations are made manifest. Various properties of the Mellin transform are summarised in section \ref{subsec::notation and conventions}. We begin in section \ref{subsec::massivescalar} by revisiting the case of a massive scalar field, presenting an approach which straightforwardly extends to gauge bosons (section \ref{subsec::gaugebosonprop}) and gravitons (section \ref{subsec::gravitonprop}). In section \ref{subsec::nuimint} we discuss some subtleties that arise for even boundary dimensions $d$.

\subsection{Massive scalar}
\label{subsec::massivescalar}

Consider the free theory of a massive scalar field $\phi$, 
\begin{equation}\label{action free scalar}
    {\cal L} = -\frac{1}{2} \left( \partial^\mu 
\phi \partial_\mu \phi +m^2 \phi^2\right).
\end{equation}
In (EA)dS$_{d+1}$ the mass is related to the scaling dimensions $\Delta_\pm$:
\begin{align}
    \sigma_{\text{(A)dS}} m^2 =  \Delta_+\Delta_-,\qquad \sigma_{\text{(A)dS}}=(-)1,
\end{align}
which label the representation of the isometry group $SO\left(1,d+1\right)$. We will often parameterise the scaling dimensions as $\Delta_\pm =  \frac{d}{2}\pm i \nu$, which are related under $\nu \to -\nu$. 

\vskip 4pt
\subsection*{Euclidean anti-de Sitter.} 
The free action in EAdS takes the following form in Poincar\'e coordinates \eqref{poincare}:
  \begin{align}
     S=\frac{1}{2}\int {\rm d}z{\rm d}^d{\bf x}\phi \left[\partial_z(z^{1-d}\partial_{z})+z^{1-d}\partial^i\partial_{i}+z^{-d-1}\Delta_+\Delta_-\right]\phi+\mathcal{B},
 \end{align}
where $\mathcal{B}$ is the total derivative term. The equation of motion for $\phi$ is,
\begin{align}\label{Massive scalar EOM}
   \left[z^2 \partial^2_z+(1-d)z\partial_z+z^2\partial^{i}\partial_{i}+\Delta_+\Delta_-\right]\phi=0,
 \end{align}
which in Fourier space \eqref{FT} reads
\begin{align}\label{Massive scalar EOM Fourier}
   &\left[z^2 \partial^2_z+(1-d)z\partial_z-(z^2k^2-\Delta_+\Delta_-)\right]\phi_{{\bf k}}=0.
 \end{align}

\vskip 4pt
The bulk-to-bulk propagator is a solution to the equation of motion with a Dirac delta unit source term:
\begin{align}\label{scalar bubu eom eads}
    \left[z^2 \partial^2_z+(1-d)z\partial_z-(z^2k^2-\Delta_+\Delta_-)\right]G^{\text{AdS}}(z,\bar{z};{\bf  k})=-z^{d+1}\delta(z-\bar{z})\,,
\end{align}
where the two independent solutions can be expressed in the well known form \cite{Liu:1998ty}:
\begin{equation}\label{standard bubu sc eads}
   G^{\text{AdS}}_{\frac{d}{2}+i\nu}(z,\bar{z};{\bf  k}) = z^{\frac{d}{2}}{\bar z}^{\frac{d}{2}}\left[\theta\left({\bar z}-z\right)K_{i\nu}\left(k z\right)I_{i\nu}\left(k {\bar z}\right)+\theta\left(z-{\bar z}\right)I_{i\nu}\left(k z\right)K_{i\nu}\left(k {\bar z}\right)\right],
\end{equation}
in terms of modified Bessel functions $I_{i\nu}$ and $K_{i\nu}$ of the first and second kind. Recall that $\Delta_\pm = \frac{d}{2}\pm i\nu$.

\begin{figure}[t]
    \centering
    \includegraphics[width=0.8\textwidth]{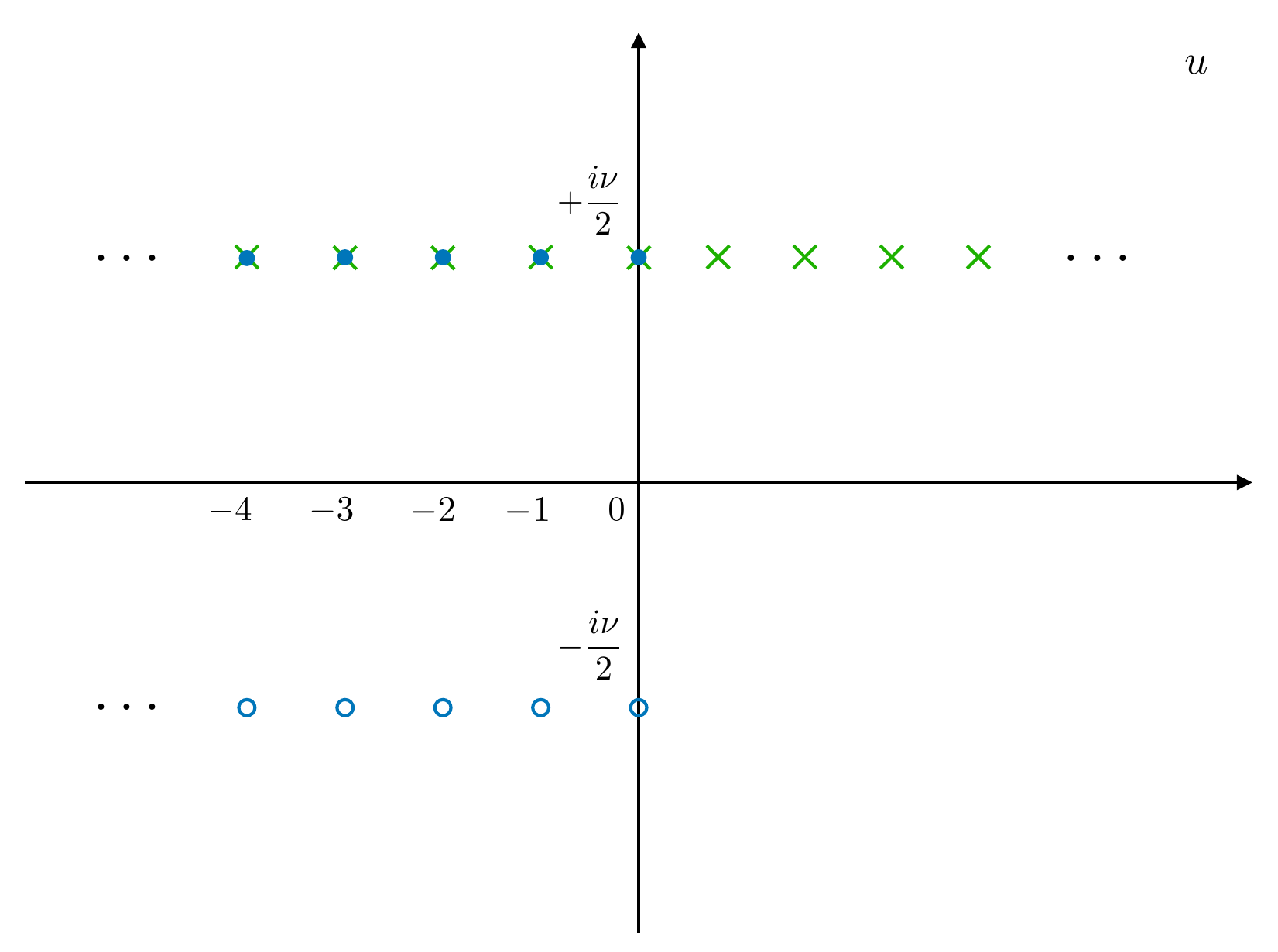}
    \caption{Plot of the $u$ poles from the factors $\Gamma\left(u\pm \frac{i\nu}{2}\right)$ in the Mellin space representation \eqref{EAdS scalar prop Mellin bubu} of the bulk-to-bulk propagator. The poles from $\Gamma\left(u - \frac{i\nu}{2}\right)$, which generate the falloff $z^{\frac{d}{2}-i\nu+2n}$ as $z \to 0$, are denoted by solid blue circles and the poles from $\Gamma\left(u + \frac{i\nu}{2}\right)$, generating the falloff $z^{\frac{d}{2}+i\nu+2n}$, are the hollow blue circles. The zeros from the factor $\omega_{\nu}(u,\bar{u})$ are the green crosses, which cancel the poles from $\Gamma\left(u - \frac{i\nu}{2}\right)$. Likewise, the zeros of $\omega_{-\nu}(u,\bar{u})$ would cancel the poles from $\Gamma\left(u + \frac{i\nu}{2}\right)$. To plot the poles we assumed that $\nu \in \mathbb{R}$, which corresponds to unitary Principal Series representations of the dS isometry group.}\label{fig::upoles}
\end{figure}

\vskip 4pt
In Mellin space this takes the form \cite{Sleight:2020obc,Sleight:2021plv}:
\begin{subequations}\label{EAdS scalar prop Mellin bubu}
\begin{align}
    G^{\text{AdS}}_{\frac{d}{2}+i\nu}(u,\bar{u};{\bf  k})&= \int^\infty_0 \frac{{\rm d}z}{z} \frac{{\rm d}{\bar z}}{\bar z}\, G^{\text{AdS}}_{\frac{d}{2}+i\nu}(z,\bar{z};{\bf  k}) z^{2u-\frac{d}{2}} {\bar z}^{2{\bar u}-\frac{d}{2}},\\
    &= \frac1{16\pi} \csc(\pi(u+\bar{u})) \omega_{\nu}(u,\bar{u})\Gamma(u\pm \tfrac{i\nu}{2})\Gamma(\bar{u}\pm \tfrac{i\nu}{2})\left(\frac{k}{2}\right)^{-2 u-2 \bar{u}},
\end{align}    
\end{subequations}
and we used the shorthand notation \eqref{shorthand Gamma}. The derivation is reviewed in appendix \ref{app::MTprops}. To define the Mellin integration contour we use the following convention to separate the poles of the cosecant function: 
\begin{align}\label{csc}
    \csc(\pi z)\equiv \frac{\Gamma (1-z) \Gamma (z)}{\pi }.
\end{align}
The function $\omega_{\nu}(u,\bar{u})$ is given by\footnote{Equivalently, writing $\Delta=\frac d2+i\nu$, we define
\begin{equation}
    \omega_{\Delta}(u,\bar u)
    =
    2\sin\left[
        \pi\left(
            u-\frac12\left(\Delta-\frac d2\right)
        \right)
    \right]
    \sin\left[
        \pi\left(
            \bar u-\frac12\left(\Delta-\frac d2\right)
        \right)
    \right].
\end{equation}}
\begin{align}\label{projector bc}
    \omega_\nu(u,\bar{u})=2\sin \left(\pi  \left(u-\tfrac{i \nu }{2}\right)\right) \sin \left(\pi  \left(\bar{u}-\tfrac{i \nu }{2}\right)\right)\,.
\end{align}
The sine factors in $\omega_\nu(u,\bar u)$ serve to cancel the poles of
$\Gamma\left(u-\frac{i\nu}{2}\right)$, and likewise for $\bar u$. See figure \ref{fig::upoles}. The effect is that $\omega_{\pm \nu}(u,\bar{u})$ acts as a projector onto the $\Delta_\pm = \frac{d}{2}\pm i\nu$ boundary behaviour \eqref{bdrybehaviour}. This can be seen from the $z \to 0$ expansion, where the leading $z^{\frac{d}{2}\pm i\nu}$ terms are generated by the residue of the poles $u = \mp \tfrac{i\nu}{2}$ in $\Gamma(u\pm \tfrac{i\nu}{2})$. Likewise, the leading ${\bar z}^{\frac{d}{2}\pm i\nu}$ terms in the ${\bar z}\to0$ expansion are generated by the residue of the poles ${\bar u} = \mp \tfrac{i\nu}{2}$ in $\Gamma({\bar u}\pm \tfrac{i\nu}{2})$.

\vskip 4pt
The bulk-to-boundary propagators can be derived by considering an asymptotic expansion in $\frac{{\bar z}}{z} \ll 1$. At the level of the Mellin representation \eqref{EAdS scalar prop Mellin bubu}, this is achieved by making the change of variables $u \to u - {\bar u}$ and evaluating the integral in ${\bar u}$ by closing the contour to the left: 
\begin{align}\nonumber
  \hspace*{-0.3cm} G^{\text{AdS}}_{\frac{d}{2}+i\nu}(z,\bar{z};{\bf  k})  &=\int^{+i\infty}_{-i\infty} \frac{{\rm d}u\,{\rm d}\bar{u}}{(2\pi i)^2}\,4G^{\text{AdS}}_{\frac{d}{2}+i\nu}(u,\bar{u};{\bf  k})\,z^{\frac{d}{2}-2 u} \bar{z}^{\frac{d}{2}-2 \bar{u}} \\ \nonumber 
  &= \bar{z}^{\frac{d}{2}+i\nu} \left[ \left(\frac{k}{2}\right)^{+i
    \nu} \frac{z^{\frac{d}{2}}}{2\Gamma\left(1+i\nu\right)}  \int^{+i\infty}_{-i\infty} \frac{{\rm d}u}{2\pi i}\,\Gamma\left(u-\tfrac{i\nu}{2}\right)\Gamma\left(u+\tfrac{i\nu}{2}\right)
    \left(\frac{kz}{2}\right)^{-2 u} \right. \\ & \left. \hspace*{10cm} +O({\bar z})\, \right],
\end{align}
where one recognises the Mellin representation of the modified Bessel function of the second kind. The bulk-to-boundary propagator is then given by:
\begin{subequations}\label{EAdS bubo scalar}
   \begin{align}
   K^{\text{AdS}}_{\frac{d}{2}+i\nu}\left(z;{\bf k}\right)&=  \lim_{{\bar z}\to 0} \left[{\bar z}^{-\left(\frac{d}{2}+i\nu\right)} G^{\text{AdS}}_{\frac{d}{2}+i\nu}(z,\bar{z};{\bf  k})\right]\\
    &= \left(\frac{k}{2}\right)^{i\nu}\frac{z^{\frac{d}{2}}}{\Gamma(i\nu+1)}K_{i\nu}(kz),
\end{align}  
\end{subequations}
which matches the standard result \cite{Gubser:1998bc}. The bulk-to-boundary propagator is a solution of the homogeneous equation of motion \eqref{Massive scalar EOM Fourier} with boundary condition:
\begin{equation}
    \lim_{z\to 0}\left[z^{-\left(\frac{d}{2}-i\nu\right)}K^{\text{AdS}}_{\frac{d}{2}+i\nu}\left(z;{\bf k}\right)\right]=\frac{1}{2i\nu}.
\end{equation}

\vskip 4pt
It is instructive to analyse the equations of motion in Mellin space \eqref{Massive scalar EOM Fourier}. Taking the Mellin transform of the scalar $\phi$
\begin{equation}
    {\tilde \phi}_{{\bf k}}(s) = \int^{\infty}_0 \frac{{\rm d}z}{z} \phi_{{\bf k}}(z)z^{2s-\frac{d}{2}},
\end{equation}
its equation of motion \eqref{Massive scalar EOM Fourier} reduces to a recursion relation:
\begin{equation}\label{EOM scalar mellin}
    \left[(1-d)(\tfrac{d}{2}-2s)+(\tfrac{d}{2}-2s)(\tfrac{d}{2}-1-2s)+\Delta_+\Delta_-\right]{\tilde \phi}_{{\bf k}}\left(s\right)-k^2{\tilde \phi}_{{\bf k}}\left(s+1\right)=0.
\end{equation}
Setting $\Delta_\pm = \frac{d}{2}\pm i \nu$ one can verify that this is solved by the Mellin transform \eqref{Bessels} of the modified Bessel function $K_{i\nu}(kz)$ of the second kind, and hence also the bulk-to-boundary propagator \eqref{EAdS bubo scalar} (as expected). 

\vskip 4pt
In Mellin space the differential equation \eqref{scalar bubu eom eads} for the bulk-to-bulk propagator reduces to:
\begin{multline}\label{EOM Mellin scalar bubu}
    \left[(1-d)(\tfrac{d}{2}-2u)+(\tfrac{d}{2}-2u)(\tfrac{d}{2}-1-2u)+\Delta_+\Delta_-\right]G^{\text{AdS}}(u,\bar{u};{\bf  k})\\-k^2G^{\text{AdS}}(u+1,\bar{u};{\bf  k})= - i \pi  \delta\left(u+{\bar u}\right),
\end{multline}
To show that the Mellin space expression \eqref{EAdS scalar prop Mellin bubu} for the bulk-to-bulk propagator satisfies this equation requires careful treatment of the integration contour in $u$; plugging the expression into the lhs of \eqref{EOM Mellin scalar bubu} without taking into account the integration contour of each term would give zero. In fact, $G^{\text{AdS}}_{\frac{d}{2}+i\nu}(u,\bar{z};{\bf  k})$ and $G^{\text{AdS}}_{\frac{d}{2}+i\nu}(u+1,\bar{z};{\bf  k})$ do not share the same integration contour owing to the cosecant function \eqref{csc}, since the shift $u \to u+1$ moves a pole from one $\Gamma$ function in  \eqref{csc} to the other. To bring both terms to the same integration contour one must cross this pole, whose residue generates the source term on the rhs of \eqref{EOM Mellin scalar bubu} and the remaining terms cancel each other. This mechanism is discussed in more detail in appendix \ref{app::mellincontour}. The role of the cosecant function in the Mellin space form \eqref{EAdS scalar prop Mellin bubu} of the propagator is therefore to generate the source term on the rhs of the propagator equation \eqref{scalar bubu eom eads}. 

\vskip 4pt
The above method to analyse solutions to the equations of motion in Mellin space is especially useful for the case of spinning fields, which will be considered in later sections.

\subsection*{de Sitter.} 

In dS the equation of motion in Poincar\'e coordinates reads in Fourier space
\begin{align}
   &\left[\eta^2 \partial^2_\eta+(1-d)\eta\partial_\eta-(\Delta_+\Delta_--\eta^2k^2)\right]\phi_{{\bf k}}=0.
 \end{align}
As reviewed in section \ref{sec::SKEAdS}, in de Sitter space the Schwinger-Keldysh propagators can be expressed in terms of the corresponding mode functions \eqref{SK mode functions}. For a massive scalar field these are Hankel functions of the first and second kind:
\begin{equation}\label{Scalar mode functions}
   f_{{}\bf k}\left(\eta\right) = \left(-\eta\right)^{\frac{d}{2}}\frac{\sqrt{\pi}}{2} e^{+\frac{\pi 
    \nu}{2}}H^{(2)}_{i\nu}\left(-k\eta\right), \qquad {\bar f}_{{}\bf k}\left(\eta\right) = \left(-\eta\right)^{\frac{d}{2}}\frac{\sqrt{\pi}}{2} e^{-\frac{\pi 
    \nu}{2}}H^{(1)}_{i\nu}\left(-k\eta\right).
\end{equation}

As in EAdS we analyse the propagators in Mellin space, where in \cite{Sleight:2020obc,Sleight:2021plv} the Mellin transform of the Schwinger-Keldysh propagators \eqref{SK mode functions} with mode functions \eqref{Scalar mode functions} was found to take the form:
\begin{align}\nonumber
  G^{\pm {\hat \pm}}(u,\bar{u};{\bf k}) &=\int_{-\infty}^0 \frac{{\rm d}\eta}{\eta} \frac{{\rm d}{\bar \eta}}{\bar \eta}\, G^{\pm {\hat \pm}}(\eta,\bar{\eta};{\bf  k}) \left(-\eta\right)^{2u-\frac{d}{2}} \left(-{\bar \eta}\right)^{2{\bar u}-\frac{d}{2}}\\ \nonumber
  &= \frac{1}{16\pi} c^{\text{dS-AdS}}_{\frac{d}{2}+i\nu}c^{\text{dS-AdS}}_{\frac{d}{2}-i\nu} e^{\mp\left(u+\tfrac{i\nu}{2}\right)\pi i}e^{{\hat \mp}\left({\bar u}-\tfrac{i\nu}{2}\right)\pi i} \csc\left(\pi\left(u+{\bar u}\right)\right)\\ & \times \left[ \alpha^{\pm {\hat \pm}}\,\omega_{\nu}\left(u,{\bar u}\right)+\beta^{\pm {\hat \pm}}\,\omega_{-\nu}\left(u,{\bar u}\right)\right] \Gamma(u\pm \tfrac{i\nu}{2})\Gamma(\bar{u}\pm \tfrac{i\nu}{2})\left(\frac{k}{2}\right)^{-2 u-2 \bar{u}},  \label{spin0bubudSMellin}
\end{align}
with
\begin{subequations}\label{alpha beta}
    \begin{align}
    \alpha^{\pm\pm}&=\frac1{c^{\text{dS-AdS}}_{\tfrac{d}{2}-i\nu}}\,e^{\pm \pi \nu}\,,& \beta^{\pm\pm}&=\frac1{c^{\text{dS-AdS}}_{\tfrac{d}{2}+i\nu}}\,e^{\mp \pi \nu}\,,\\
    \alpha^{\pm\mp}&=\frac1{c^{\text{dS-AdS}}_{\tfrac{d}{2}-i\nu}}\,e^{\mp \pi \nu}\,,& \beta^{\pm\mp}&=\frac1{c^{\text{dS-AdS}}_{\tfrac{d}{2}+i\nu}}\,e^{\mp \pi \nu}\,.
\end{align}
\end{subequations}
For a derivation see appendix A.2 of \cite{Sleight:2021plv}. As for the Mellin space representation of EAdS bulk-to-bulk propagators \eqref{EAdS scalar prop Mellin bubu}, the poles in $\Gamma(u\pm \tfrac{i\nu}{2})$ and $\Gamma(\bar{u}\pm \tfrac{i\nu}{2})$ generate the late-time expansions in $\eta \to 0$ and ${\bar \eta} \to 0$ respectively. Since $\alpha^{\pm{\hat \pm}}$ and $\beta^{\pm {\hat \pm}}$ are non vanishing, both $\Delta_\pm$ late-time behaviours \eqref{bdrybehaviour} are present in dS space. This is to be contrasted with the story in AdS space, where the behaviours \eqref{bdrybehaviour} correspond to a choice of boundary condition (Dirichlet or Neumann) at spatial infinity.

\vskip 4pt
Following the discussion above in EAdS it is useful to analyse the propagators at the level of the equation of motion, which read:
\begin{subequations}
  \begin{align}
   \left[\eta^2 \partial^2_\eta+(1-d)\eta\partial_\eta-(\Delta_+\Delta_--\eta^2k^2)\right]G^{\pm \pm}(\eta,\bar{\eta};{\bf k})&=\pm i \left(-\eta\right)^{d+1}\delta\left(\eta-{\bar \eta}\right),\\
   \left[\eta^2 \partial^2_\eta+(1-d)\eta\partial_\eta-(\Delta_+\Delta_--\eta^2k^2)\right]G^{\pm \mp}(\eta,\bar{\eta};{\bf k})&=0.
 \end{align}  
\end{subequations}
In Mellin space these become 
\begin{multline}
  \left[(\tfrac{d}{2}-2u)(\tfrac{d}{2}-1-2u)+(1-d)(\tfrac{d}{2}-2u)-\Delta_+\Delta_-\right]G^{\pm \pm}(u,\bar{u};{\bf k})\\  +k^2 G^{\pm \pm}(u+1,\bar{u};{\bf k})=\mp  \pi \delta\left(u+{\bar u}\right),
 \end{multline}
and
\begin{multline}\label{pmmpspin0bubumellin}
    \left[(\tfrac{d}{2}-2u)(\tfrac{d}{2}-1-2u)+(1-d)(\tfrac{d}{2}-2u)-\Delta_+\Delta_-\right]G^{\pm \mp}(u,\bar{u};{\bf k})\\+k^2 G^{\pm \mp}(u+1,\bar{u};{\bf k})=0.
\end{multline}
It is straightforward to verify that the Mellin space representation \eqref{spin0bubudSMellin} of the Schwinger-Keldysh propagators satisfy these equations, following the same steps as in the EAdS case. As before, the cosecant function \eqref{csc} generates the source term. Note that for the $G^{\pm \mp}(u,\bar{u};{\bf k})$ propagator the coefficients \eqref{alpha beta} conspire to cancel the cosecant function \eqref{csc} due to the identity
\begin{equation}
    \omega_\nu\left(u,{\bar u}\right)-\omega_{-\nu}\left(u,{\bar u}\right) = -2i\nu \sin(i\pi \nu ) \sin(\pi(u+{\bar u})),
\end{equation}
so that they satisfy the homogeneous equation \eqref{pmmpspin0bubumellin}.

\vskip 4pt
 We see that dS and EAdS propagators take a universal form in Mellin space, which was exploited in \cite{Sleight:2020obc,Sleight:2021plv} to map Schwinger-Keldysh propagators to a linear combination of EAdS ones under Wick rotation. This is reviewed in the following section.

\vskip 4pt
\subsection*{From dS to EAdS.} 
By comparing the Mellin space form of the dS Schwinger-Keldysh propagators \eqref{spin0bubudSMellin} and EAdS bulk-to-bulk propagators \eqref{EAdS scalar prop Mellin bubu}, under the Wick rotations \eqref{wickzeta} one finds 
\begin{shaded} 
\begin{multline}\label{dS scalar prop to EAdS}
    G^{\pm\hat{\pm}}(\eta,\bar{\eta};{\bf k})\to c^{\text{dS-AdS}}_{\Delta_+}e^{\mp \Delta_+ \frac{\pi i}{2}}e^{\hat{\mp} \Delta_+\frac{\pi i}{2}}G^{\text{AdS}}_{\Delta_+}(z,\bar{z};{\bf k}) \\+ c^{\text{dS-AdS}}_{\Delta_-}e^{\mp \Delta_-\frac{\pi i}{2}}e^{\hat{\mp} \Delta_-\frac{\pi i}{2}}G^{\text{AdS}}_{\Delta_-}(z,\bar{z};{\bf k}).
\end{multline}
\end{shaded}

\vskip 4pt
A similar relationship can be obtained between bulk-to-boundary propagators by performing the asymptotic expansion in $\frac{{\bar \eta}}{\eta} \ll 1$ directly at the level of the relationship \eqref{dS scalar prop to EAdS} between bulk-to-bulk propagators. We have
\begin{equation}
  \lim_{{\bar \eta}\to 0}   G^{\pm\hat{\pm}}(\eta,\bar{\eta};{\bf k}) = \left(-{\bar \eta}\right)^{\Delta_+}K^{\pm}_{\Delta_+}(\eta;{\bf k})+\left(-{\bar \eta}\right)^{\Delta_-}K^{\pm}_{\Delta_-}(\eta;{\bf k}),
\end{equation}
where, in terms of the EAdS bulk-to-boundary propagator \eqref{EAdS bubo scalar}:
\begin{shaded}
 \begin{equation}\label{bubuscdstoeads}
    K^{\pm}_{\Delta}(\eta;{\bf k}) \to e^{\mp \Delta \frac{\pi i}{2}} c^{\text{dS-AdS}}_{\Delta}K^{\text{AdS}}_{\Delta}(z;{\bf k}).
\end{equation}   
\end{shaded}

\subsection{Gauge Boson}
\label{subsec::gaugebosonprop}

Consider the free theory for a spin-1 gauge field $A_\mu$,
\begin{equation}
    {\cal L}
    =
    -\frac{1}{4}F_{\mu\nu}F^{\mu\nu},
    \qquad
    F_{\mu\nu}\left[A\right]
    =
    \partial_\mu A_\nu-\partial_\nu A_\mu,
\end{equation}
where for the moment we leave any colour indices implicit. For a massless spin-one field
the two boundary falloffs are
\begin{equation}
    \Delta_A^+=d-1,
    \qquad
    \Delta_A^-=1,
    \qquad
    \Delta_A^++\Delta_A^-=d .
\end{equation}
We will label the two contributions directly by their scaling dimensions.

\vskip 4pt
We follow the same steps as the scalar case reviewed in the previous section to establish
the Mellin-space representation of gauge-boson propagators in (EA)dS, which are then used
to establish the relationships \eqref{bubuwick} and \eqref{bubowick} under Wick rotation.

\subsection*{Euclidean anti-de Sitter.}

In Poincar\'e coordinates \eqref{poincare} the action on EAdS in terms of
$A_\mu=(A_z,A_i)$ reads
\begin{align}
    S
 =&
 \frac{1}{2}\int {\rm d}z\,{\rm d}^d{\bf x}
 \Big[
 \delta^{ij}
 \Big\{
 A_z\left(z^{3-d}\partial_j\partial_i A_z\right)
 -2A_z\left(\partial_j z^{3-d}\partial_zA_i\right)
 \Big\}
 \nonumber
 \\
 &+
 A_i\delta^{ij}
 \Big\{
 \partial_z z^{3-d}\partial_zA_j
 +
 \delta^{kl}\partial_kz^{3-d}\partial_lA_j
 \Big\}
 -
 A_i
 \Big\{
 z^{3-d}\delta^{ik}\delta^{jl}
 \partial_k\partial_l A_j
 \Big\}
 \Big]
 +
 \mathcal{B},
\end{align}
where $\mathcal{B}$ is the total derivative term. We will work in the axial gauge
$A_z=0$, where the equation of motion for $A_i$ is,\footnote{We are keeping
${\bf k}\cdot{\bf A}\neq0$ to keep track of the longitudinal modes of the gauge bosons.}
\begin{equation}
    \delta^{ij}
    \Big\{
    \partial_z z^{3-d}\partial_zA_j
    +
    \delta^{kl}\partial_k z^{3-d}\partial_lA_j
    \Big\}
    -
    z^{3-d}\delta^{ik}\delta^{jl}
    \partial_k\partial_lA_j
    =
    0.
\end{equation}
In Fourier space
$A_j(z,{\bf x})=\int {\rm d}^d{\bf k}\,e^{-i{\bf k}\cdot{\bf x}}A_j(z,{\bf k})$,
this reads
\begin{equation}\label{EOM YM}
    \left[
        \delta^{ij}z^2\partial_z^2
        +
        \delta^{ij}(3-d)z\partial_z
        -
        z^2k^2\pi^{ij}
    \right]A_j(z,{\bf k})
    =
    0,
\end{equation}
where $\pi_{ij}$ is the transverse projector:
\begin{equation}
    \pi_{ij}
    =
    \delta_{ij}
    -
    \frac{k_i k_j}{k^2}.
\end{equation}

\vskip 4pt
The bulk-to-bulk propagator is a solution to the equation of motion with a Dirac delta
unit source term,
\begin{equation}
    \left[
        \delta^{ij}z^2\partial_z^2
        +
        \delta^{ij}(3-d)z\partial_z
        -
        z^2k^2\pi^{ij}
    \right]
    G^{\text{AdS}}_{j;k}(z,\bar{z};{\bf k})
    =
    -z^{d-1}\delta^i_k\delta(z-\bar{z}) .
\end{equation}
Going to Mellin space,
\begin{equation}\label{EAdS GB prop}
    G^{\text{AdS}}_{i;j}(u,\bar{u};{\bf k})
    =
    \int^\infty_0
    \frac{{\rm d}z}{z}
    \frac{{\rm d}{\bar z}}{\bar z}\,
    G^{\text{AdS}}_{i;j}(z,\bar{z};{\bf k})\,
    z^{2u-\frac{d}{2}+1}
    {\bar z}^{2{\bar u}-\frac{d}{2}+1},
\end{equation}
this reads
\begin{multline}
    \left[
        \delta^{ij}
        \left(\frac{d}{2}-2u-1\right)
        \left(\frac{d}{2}-2u-2\right)
        +
        \delta^{ij}(3-d)
        \left(\frac{d}{2}-2u-1\right)
    \right]
    G^{\text{AdS}}_{j;k}(u,\bar{u};{\bf k})
    \\
    -
    k^2\pi^{ij}G^{\text{AdS}}_{j;k}(u+1,\bar{u};{\bf k})
    =
    -\delta^i_k\,i\pi\,\delta(u+\bar{u}) .
\end{multline}
Given the lessons learned from the scalar case in the previous section, one can write down
the solution as\footnote{Note that the two $\csc$ terms in the square bracket are defined
with respect to different integration contours, following the convention \eqref{csc}.
This is discussed in more detail in appendix \ref{app::mellincontour}.}
\begin{multline}\label{Spin 1 bubu Mellin}
    G^{\text{AdS}\,\Delta_A}_{i;j}(u,\bar{u};{\bf k})
    =
    \frac1{16\pi}
    \left[
        \delta_{ij}\csc(\pi(u+\bar{u}))
        +
        \frac{k_i k_j}{k^2}\,
        \csc(\pi(u+\bar{u}+1))
    \right]
    \\
    \times
    \omega_{\Delta_A}(u,\bar{u})
    \Gamma\left(u\pm \frac{1}{2}\left(\Delta_A-\frac d2\right)\right)
    \Gamma\left({\bar u}\pm \frac{1}{2}\left(\Delta_A-\frac d2\right)\right)
    \left(\frac{k}{2}\right)^{-2u-2\bar{u}} .
\end{multline}
Here $\Delta_A$ is to be set equal to either $\Delta_A^+=d-1$ or
$\Delta_A^-=1$. The factor $\omega_{\Delta_A}(u,\bar u)$ is the same spectral weight \eqref{projector bc} as
in the scalar case, written as a function of the scaling dimension. As a further check,
in appendix \ref{app::raju} we compare with other expressions available in the
literature.

\vskip 4pt
As in the scalar case we obtain bulk-to-boundary propagators by performing an asymptotic
expansion of the bulk-to-bulk propagator for $\bar{z}/z\ll 1$, treating separately terms
that do not share the same integration contour. This gives
\begin{subequations}\label{EAdS bubo GB}
\begin{align}
   K^{\text{AdS}\,\Delta_A}_{i;j}\left(z;{\bf k}\right)
   &=
   \lim_{\bar z\to 0}
   \left[
        {\bar z}^{-\Delta_A+1}
        G^{\text{AdS}\,\Delta_A}_{i;j}(z,\bar{z};{\bf k})
   \right]
   \\
   &=
   \pi_{ij}
   \frac{1}{\Gamma\left(\Delta_A-\frac d2+1\right)}
   \left(\frac{k}{2}\right)^{\Delta_A-\frac d2}
   z^{\frac d2-1}
   K_{\Delta_A-\frac d2}(kz)
   +
   \frac{k_i k_j}{2\left(\Delta_A-\frac d2\right)k^2}
   z^{d-\Delta_A-1}.
\end{align}
\end{subequations}
This matches with the expression derived in appendix D of \cite{Marotta:2024sce}. The
term proportional to the projector $\pi_{ij}$ is the transverse component, and the
remaining term is the longitudinal component. It is a straightforward exercise to show
that this satisfies the homogeneous equation \eqref{EOM YM} with boundary condition
\begin{equation}
    \lim_{z\to 0}
    \left[
        z^{-(d-\Delta_A)+1}
        K^{\text{AdS}\,\Delta_A}_{i;j}\left(z;{\bf k}\right)
    \right]
    =
    \frac{1}{2\left(\Delta_A-\frac d2\right)}
    \delta_{ij}.
\end{equation}

\subsection*{de Sitter.}

In de Sitter space the free equation of motion for the spin-1 gauge field
$A_\mu=(A_\eta,A_i)$ in the temporal gauge $A_\eta=0$ is
\begin{equation}\label{EOM YM dS}
    \left[
        \delta^{ij}\eta^2\partial_\eta^2
        +
        \delta^{ij}(3-d)\eta\partial_\eta
        +
        \eta^2k^2\pi^{ij}
    \right]A_j(\eta,{\bf k})
    =
    0.
\end{equation}
As before we proceed in Mellin space,
\begin{equation}\label{GB SK prop}
 G^{\pm{\hat \pm}}_{i;j}(u,\bar{u};{\bf k})
 =
 \int_{-\infty}^0
 \frac{{\rm d}\eta}{\eta}
 \frac{{\rm d}{\bar \eta}}{\bar \eta}\,
 G^{\pm{\hat \pm}}_{i;j}(\eta,\bar{\eta};{\bf k})\,
 \left(-\eta\right)^{2 u-\frac{d}{2}+1}
 \left(-{\bar \eta}\right)^{2 {\bar u}-\frac{d}{2}+1}.
\end{equation}
The differential equations for the Schwinger-Keldysh propagators take the form
\begin{multline}
    \left[
        \delta^{ij}
        \left(\frac{d}{2}-2u-1\right)
        \left(\frac{d}{2}-2u-2\right)
        +
        \delta^{ij}(3-d)
        \left(\frac{d}{2}-2u-1\right)
    \right]
    G^{\pm\pm}_{j;k}(u,\bar{u};{\bf k})
    \\
    +
    k^2\pi^{ij}G^{\pm \pm}_{j;k}(u+1,\bar{u};{\bf k})
    =
    \mp \delta^i_k \pi \delta(u+\bar{u}),
\end{multline}
and
\begin{multline}
    \left[
        \delta^{ij}
        \left(\frac{d}{2}-2u-1\right)
        \left(\frac{d}{2}-2u-2\right)
        +
        \delta^{ij}(3-d)
        \left(\frac{d}{2}-2u-1\right)
    \right]
    G^{\pm\mp}_{j;k}(u,\bar{u};{\bf k})
    \\
    +
    k^2\pi^{ij}G^{\pm \mp}_{j;k}(u+1,\bar{u};{\bf k})
    =
    0.
\end{multline}
Drawing lessons from the scalar case one can immediately write down the solution as
\begin{multline}\label{spin1 SK bubu mellin}
    G^{\pm\hat{\pm}}_{i;j}(u,\bar{u};{\bf k})
    =
    c^{\text{dS-AdS}}_{\Delta_A}
    c^{\text{dS-AdS}}_{d-\Delta_A}
    e^{\mp \left(u+\frac{\Delta_A}{2}-\frac d4\right)\pi i}
    e^{\hat{\mp} \left(\bar{u}-\frac{\Delta_A}{2}+\frac d4\right)\pi i}
    \\
    \times
    \left[
        \alpha^{\pm\hat{\pm}}\,
        G^{\text{AdS}\,\Delta_A}_{i;j}(u,\bar{u};{\bf k})
        +
        \beta^{\pm\hat{\pm}}\,
        G^{\text{AdS}\,d-\Delta_A}_{i;j}(u,\bar{u};{\bf k})
    \right].
\end{multline}
Here $\Delta_A$ may be taken to be either of the two falloffs. The coefficients
$\alpha^{\pm\hat{\pm}}$ and $\beta^{\pm\hat{\pm}}$ are the same as those in
\eqref{alpha beta} for the scalar case.

\subsection*{From dS to EAdS.}

By comparing the Mellin-space form of the dS Schwinger-Keldysh propagators
\eqref{spin1 SK bubu mellin} with the EAdS bulk-to-bulk propagators
\eqref{Spin 1 bubu Mellin}, under the Wick rotations \eqref{wickzeta} one finds
\begin{shaded}
\begin{multline}\label{dS Gauge Boson prop to EAdS}
    G^{\pm\hat{\pm}}_{i;j}(\eta,\bar{\eta};{\bf k})
    \to
    c^{\text{dS-AdS}}_{\Delta_A^+}
    e^{\mp \left(\Delta_A^+-1\right)\frac{\pi i}{2}}
    e^{\hat{\mp} \left(\Delta_A^+-1\right)\frac{\pi i}{2}}
    G^{\text{AdS}\,\Delta_A^+}_{i;j}(z,\bar{z};{\bf k})
    \\
    +
    c^{\text{dS-AdS}}_{\Delta_A^-}
    e^{\mp \left(\Delta_A^--1\right)\frac{\pi i}{2}}
    e^{\hat{\mp} \left(\Delta_A^--1\right)\frac{\pi i}{2}}
    G^{\text{AdS}\,\Delta_A^-}_{i;j}(z,\bar{z};{\bf k}) .
\end{multline}
\end{shaded}
\noindent Here
\begin{equation}
    \Delta_A^+=d-1,
    \qquad
    \Delta_A^-=1 .
\end{equation}
The $\Delta_A^+=d-1$ contribution corresponds to the boundary spin-one conserved current,
while $\Delta_A^-=1$ corresponds to the shadow falloff.

\vskip 4pt
As in the scalar case, one obtains the corresponding relation between bulk-to-boundary
propagators by taking the asymptotic expansion of \eqref{dS Gauge Boson prop to EAdS}.
We have
\begin{equation}
  \lim_{{\bar \eta}\to 0}
  G^{\pm\hat{\pm}}_{i;j}(\eta,\bar{\eta};{\bf k})
  =
  \left(-{\bar \eta}\right)^{\Delta_A^+-1}
  K^{\pm\,\Delta_A^+}_{i;j}(\eta;{\bf k})
  +
  \left(-{\bar \eta}\right)^{\Delta_A^--1}
  K^{\pm\,\Delta_A^-}_{i;j}(\eta;{\bf k}) .
\end{equation}
In terms of the EAdS bulk-to-boundary propagator \eqref{EAdS bubo GB}, this gives
\begin{shaded}
\begin{equation}
    K^{\pm\,\Delta_A}_{i;j}(\eta;{\bf k})
    \to
    e^{\mp \left(\Delta_A-1\right)\frac{\pi i}{2}}
    c^{\text{dS-AdS}}_{\Delta_A}
    K^{\text{AdS}\,\Delta_A}_{i;j}(z;{\bf k}),
    \qquad
    \Delta_A=\Delta_A^+,\Delta_A^- .
\end{equation}
\end{shaded}

\vskip 4pt
The results presented in this section show that the general relations \eqref{bubuwick}
and \eqref{bubowick} between spin-$J$ propagators for late-time correlators and
propagators for EAdS Witten diagrams presented in
\cite{Sleight:2020obc,Sleight:2021plv} extend to gauge theories as well.

\subsection{Graviton}
\label{subsec::gravitonprop}

In this section we consider the free theory for gravity fluctuations $h_{\mu \nu}$ about
an (EA)dS$_{d+1}$ background $g_{\mu \nu}$. Linearising the Lagrangian of
Einstein-Hilbert gravity, the quadratic Lagrangian is given in Lorentzian signature by
\begin{equation}\label{linEH}
    {\cal L}_{\text{lin EH}}
    =
    - \frac{1}{2}
    \left(
        {\tilde h}^{\mu \nu} \Box {\tilde h}_{\mu \nu}
        +
        2 {\tilde h}^{\mu \nu} R_{\mu \rho \nu \sigma} h^{\rho \sigma}
        +
        2 \nabla^\rho {\tilde h}_{\rho \mu}
        \nabla^\sigma {\tilde h}^\mu{}_{\sigma}
    \right),
\end{equation}
where
\begin{equation}
    {\tilde h}^{\mu \nu}
    =
    h^{\mu \nu}
    -
    \frac{1}{2}g^{\mu \nu}h^{\rho \sigma}g_{\rho \sigma},
\end{equation}
and all covariant derivatives are taken with respect to the background metric.

For a massless spin-two field the two boundary falloffs are
\begin{equation}
    \Delta_h^+=d,
    \qquad
    \Delta_h^-=0,
    \qquad
    \Delta_h^++\Delta_h^-=d .
\end{equation}
We will label the two contributions directly by their scaling dimensions.

\vskip 4pt
We follow the same steps as in the scalar case reviewed in section
\ref{subsec::massivescalar} to establish the Mellin-space representation of graviton
propagators in (EA)dS, which are then used to establish the relationships
\eqref{bubuwick} and \eqref{bubowick} under Wick rotation.

\subsection*{Euclidean anti-de Sitter.}

In EAdS the free theory action in Poincar\'e coordinates \eqref{poincare} is given in the
axial gauge $h_{z\mu}=0$ by
\begin{multline}
    S
    =
    \frac{1}{2}
    \int {\rm d}z\,{\rm d}^d{\bf x}\,
    h^{ij}
    \Big[
    z^2 \delta^{\gamma\rho}
    \partial_\rho z^{1-d}
    \partial_\gamma(z^2 h_{ij})
    -
    \delta_{ij}\delta^{kl} z^2\delta^{\gamma\rho}
    \partial_\rho z^{1-d}
    \partial_\gamma(z^2 h_{kl})
    \\
    \left.
    -
    2 z^{5-d}
    \left(
        k_i k^l h_{lj}
        -
        k_i k_j \delta^{kl}h_{kl}
    \right)
    \right].
\end{multline}
The corresponding free equation of motion for $h_{ij}$ in Fourier space reads
\begin{multline}
  \left\{
  \left(
      \delta^k_i\delta^l_j
      -
      \delta_{ij}\delta^{kl}
  \right)
  \left[
      z^2\partial_z^2
      +
      (5-d)z\partial_z
      +
      2(2-d)
  \right]
  \right.
  \\
  \left.
  +
  k^2z^2
  \left[
      \pi^k_i\pi^l_j
      -
      \pi_{ij}\pi^{kl}
      -
      2
      \left(
          \delta^k_i\delta^l_j
          -
          \delta_{ij}\delta^{kl}
      \right)
  \right]
  \right\}
  h_{kl}
  =
  0 .
\end{multline}

\vskip 4pt
As for the scalar and spin-one cases we solve for the bulk-to-bulk propagator in Mellin
space,
\begin{equation}\label{EAdS GB prop 2}
   G^{\text{AdS}}_{i_1i_2;j_1j_2}(u,\bar{u};{\bf k})
   =
   \int^\infty_0
   \frac{{\rm d}z}{z}
   \frac{{\rm d}{\bar z}}{\bar z}\,
   G^{\text{AdS}}_{i_1i_2;j_1j_2}(z,\bar{z};{\bf k})\,
   z^{2 u-\frac{d}{2}+2}
   \bar{z}^{2 \bar{u}-\frac{d}{2}+2}.
\end{equation}
The propagator equation reads
\begin{multline}
 \left(
     \delta^{j^\prime_1}_{i_1}\delta^{j^\prime_2}_{i_2}
     -
     \delta_{i_1 i_2} \delta^{j^\prime_1 j^\prime_2}
 \right)
 \left[
     \left(\tfrac{d}{2}-2u-2\right)
     \left(\tfrac{d}{2}-2u-3\right)
     +
     (5-d)
     \left(\tfrac{d}{2}-2u-2\right)
     +
     2(2-d)
 \right]
 G^{\text{AdS}}_{i_1i_2;j^\prime_1j^\prime_2}(u,\bar{u};{\bf k})
 \\
 +
 k^2
 \left[
     \pi^{j^\prime_1}_{i_1}\pi^{j^\prime_2}_{i_2}
     -
     \pi_{i_1 i_2}\pi^{j^\prime_1 j^\prime_2}
     -
     2
     \left(
         \delta^{j^\prime_1}_{i_1}\delta^{j^\prime_2}_{i_2}
         -
         \delta_{i_1 i_2}\delta^{j^\prime_1 j^\prime_2}
     \right)
 \right]
 G^{\text{AdS}}_{j^\prime_1j^\prime_2;j_1j_2}(u+1,\bar{u};{\bf k})
 \\
 =
 -i
 \left(
     \delta_{i_1j_1}\delta_{i_2j_2}
     +
     \delta_{i_1j_2}\delta_{i_2j_1}
 \right)
 \pi\delta(u+\bar u).
\end{multline}
The solution can be written as\footnote{Note that the three terms in the square bracket
in \eqref{Spin 2 bubu Mellin} do not share the same Mellin integration contour. This is
discussed in more detail in appendix \ref{app::mellincontour}.}
\begin{multline}\label{Spin 2 bubu Mellin}
   G^{\text{AdS}\,\Delta_h}_{i_1i_2;j_1j_2}(u,\bar{u};{\bf k})
   =
   \frac1{16\pi}
   \Big[
        P^{(0)}_{i_1i_2;j_1j_2}\csc(\pi(u+\bar{u}))
        +
        P^{(1)}_{i_1i_2;j_1j_2}
        \csc(\pi(u+\bar{u}+1))k^{-2}
        \\
        +
        P^{(2)}_{i_1i_2;j_1j_2}
        \csc(\pi(u+\bar{u}+2))k^{-4}
   \Big]
   \\
   \times
   \omega_{\Delta_h}(u,\bar{u})
   \Gamma\left(u\pm\frac{1}{2}\left(\Delta_h-\frac d2\right)\right)
   \Gamma\left(\bar{u}\pm\frac{1}{2}\left(\Delta_h-\frac d2\right)\right)
   \left(\frac{k}{2}\right)^{-2u-2\bar{u}} .
\end{multline}
Here $\Delta_h$ is to be set equal to either $\Delta_h^+=d$ or $\Delta_h^-=0$. The
factor $\omega_{\Delta_h}(u,\bar u)$ is the same spectral weight \eqref{projector bc} as in the scalar case,
written as a function of the scaling dimension. The tensor structures are most
conveniently expressed by contracting with constant auxiliary vectors $w^i$ and
$\bar w^i$, see the conventions \eqref{indexfree}:
\begin{subequations}
\begin{align}
    P^{(0)}(w,\bar w)
    &=
    (w\cdot \bar w)^2
    -
    \frac{w\cdot w\,\bar w\cdot \bar w}{d-1},
    \\
    P^{(1)}(w,\bar w)
    &=
    \frac{\bar w\cdot \bar w\,(k\cdot w)^2}{d-1}
    +
    \frac{w\cdot w\,(k\cdot \bar w)^2}{d-1}
    -
    2 k\cdot w\, k\cdot \bar w\, w\cdot \bar w,
    \\
    P^{(2)}(w,\bar w)
    &=
    \frac{d-2}{d-1}
    (k\cdot w)^2
    (k\cdot \bar w)^2 .
\end{align}
\end{subequations}
As a further check, we compare with other expressions available in the literature in
appendix \ref{app::raju}.

\vskip 4pt
As before, considering an asymptotic expansion in $\bar z/z\ll 1$, one obtains the
bulk-to-boundary propagator:
\begin{align}
    K_{i_1i_2;j_1j_2}^{\text{AdS}\,\Delta_h}(z;k)
    &=
    \frac{
    \pi_{i_1j_1}\pi_{i_2j_2}
    +
    \pi_{i_2j_1}\pi_{i_1j_2}
    -
    \frac2{d-1}\pi_{i_1i_2}\pi_{j_1j_2}
    }
    {
    \Gamma\left(\Delta_h-\frac d2+1\right)
    }
    z^{\frac{d}{2}-2}
    \left(\frac{k}{2}\right)^{\Delta_h-\frac d2}
    K_{\Delta_h-\frac d2}(kz)
    \nonumber
    \\
    &\quad
    -
    \frac{
    (d-2)z^{d-\Delta_h-2}
    \left[
        k^2z^2
        +
        4\left(1-\Delta_h+\frac d2\right)
    \right]
    }
    {
    4(d-1)k^4
    \left(\Delta_h-\frac d2\right)
    \left(1-\Delta_h+\frac d2\right)
    }
    k_{i_1}k_{i_2}k_{j_1}k_{j_2}
    \nonumber
    \\
    &\quad
    -
    \frac{
    z^{d-\Delta_h-2}
    }
    {
    \left(\Delta_h-\frac d2\right)(d-1)k^2
    }
    T_{i_1i_2;j_1j_2}.
\end{align}
The first line is the transverse component and the remaining terms are longitudinal
components. The tensor structure $T_{i_1i_2;j_1j_2}$ is most conveniently expressed by
contracting with auxiliary vectors $w^i$ and $\bar w^i$:
\begin{equation}
    T(w,\bar w)
    =
    -2(d-1) k\cdot w\, k\cdot \bar w\, w\cdot \bar w
    +
    \bar w\cdot \bar w\,(k\cdot w)^2
    +
    w\cdot w\,(k\cdot \bar w)^2 .
\end{equation}
As far as we are aware this expression, which includes all longitudinal components, is
new.

\subsection*{From dS to EAdS.}

Drawing lessons from the scalar and gauge-boson examples, we can immediately write down
the relation between Schwinger-Keldysh propagators for gravitons in dS and the graviton
bulk-to-bulk propagator \eqref{EAdS GB prop 2} in EAdS under Wick rotation:
\begin{shaded}
\begin{multline}\label{dS graviton prop to EAdS}
    G^{\pm\hat{\pm}}_{i_1i_2;j_1j_2}(\eta,\bar{\eta};{\bf k})
    \to
    c^{\text{dS-AdS}}_{\Delta_h^+}
    e^{\mp \left(\Delta_h^+-2\right)\frac{\pi i}{2}}
    e^{\hat{\mp} \left(\Delta_h^+-2\right)\frac{\pi i}{2}}
    G^{\text{AdS}\,\Delta_h^+}_{i_1i_2;j_1j_2}(z,\bar{z};{\bf k})
    \\
    +
    c^{\text{dS-AdS}}_{\Delta_h^-}
    e^{\mp \left(\Delta_h^--2\right)\frac{\pi i}{2}}
    e^{\hat{\mp} \left(\Delta_h^--2\right)\frac{\pi i}{2}}
    G^{\text{AdS}\,\Delta_h^-}_{i_1i_2;j_1j_2}(z,\bar{z};{\bf k}) .
\end{multline}
\end{shaded}
\noindent Here
\begin{equation}
    \Delta_h^+=d,
    \qquad
    \Delta_h^-=0 .
\end{equation}
The $\Delta_h^+=d$ contribution corresponds to the boundary stress tensor, while
$\Delta_h^-=0$ corresponds to the shadow falloff.

\vskip 4pt
For the bulk-to-boundary propagators we have
\begin{shaded}
\begin{equation}
   K^{\pm\,\Delta_h}_{i_1i_2;j_1j_2}(\eta;{\bf k})
   \to
   e^{\mp \left(\Delta_h-2\right)\frac{\pi i}{2}}
   c^{\text{dS-AdS}}_{\Delta_h}
   K^{\text{AdS}\,\Delta_h}_{i_1i_2;j_1j_2}(z;{\bf k}),
   \qquad
   \Delta_h=\Delta_h^+,\Delta_h^- .
\end{equation}
\end{shaded}

\vskip 4pt
The results presented in this section show that the general relations \eqref{bubuwick}
and \eqref{bubowick} between spin-$J$ propagators for late-time correlators and
propagators for EAdS Witten diagrams presented in
\cite{Sleight:2020obc,Sleight:2021plv} extend to theories of gravitons as well.

\subsection{The case $\nu \in -i \mathbb{N}$.}
\label{subsec::nuimint}

Recall that the parameter $\nu$ labels the irreducible representation of the dS isometry group, where $\Delta_\pm = \frac{d}{2}\pm i\nu$. For scalar fields, the unitary values fall into three categories \cite{Dobrev:1977qv,Basile:2016aen}:
\begin{align}\nonumber
  &\hspace*{-0.5cm}\bullet \:\: \text{Principal Series:} \:\:\: \text{Massive Particles},\:\:\: \nu \in \mathbb{R}, \:\:\: m^2 \geq \left(\tfrac{d}{2}\right)^2.  \\ \nonumber
  &\hspace*{-0.5cm}\bullet \:\: \text{Complementary Series:} \:\:\: \text{Light Particles}, \:\:\: \nu \in i\mathbb{R}, \:\:\: |\nu| \in \left(0,\tfrac{d}{2}\right), \:\:\: 0 < m^2 < \left(\tfrac{d}{2}\right)^2. \\ \nonumber 
  &\hspace*{-0.5cm}\bullet \:\: \text{Exceptional Series:} \:\:\: \nu = \pm i\left(\tfrac{d}{2}+n\right), \:\:\: n \in \mathbb{N}_0. 
\end{align}
Massless scalar particles correspond to the first exceptional value, $n=0$, i.e. $\nu=\pm i\frac{d}{2}$. As we have seen, gauge bosons correspond to $\nu=\pm i\left(\tfrac{d}{2}-1\right)$ and gravitons $\nu=\pm i\frac{d}{2}$.

\vskip 4pt
In this section we discuss the case $\nu=-i n, \, n \in \mathbb{N}$, which therefore applies to some points in the complementary series for scalar fields, and in even $d$ to scalar fields carrying exceptional series representations (including massless scalars), gauge bosons and gravitons. For such values of $\nu$ the coefficient \eqref{cseadstods} is divergent:
\begin{equation}\label{cscdiv}
    c^{\text{dS-AdS}}_{\frac{d}{2}+n} =\frac{1}{2}\text{csc}\left(n \pi \right).
\end{equation}
This is a feature of the decomposition in terms of EAdS propagators $G^{\text{AdS}}_{\frac{d}{2}\pm i\nu}(z,\bar{z})$, since the dS propagators themselves are finite for such values.

\begin{figure}[t]
    \centering
    \includegraphics[width=0.9\textwidth]{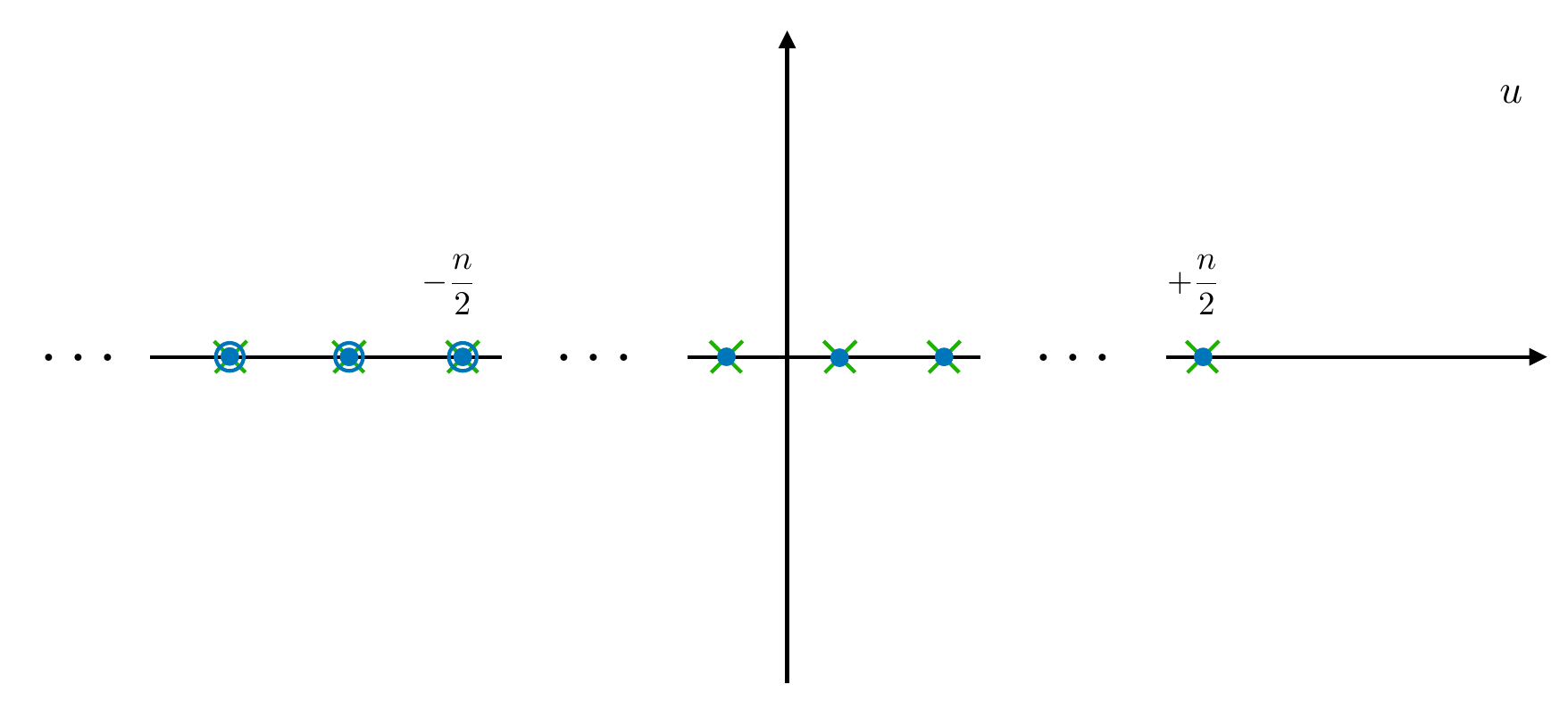}
    \caption{For $\nu = -in$ the two families of poles in the Mellin variable $u$, illustrated in figure \ref{fig::upoles}, both collapse along the real axis and coincide for all but a finite number of poles. This gives an infinite number of double poles and a finite number of single poles.}
    \label{fig::npoles}
\end{figure}

\vskip 4pt
 The divergence \eqref{cscdiv} arises because the two solutions $\Delta_\pm$ of the EAdS propagator equation, in terms of which we have expanded the dS propagators \eqref{bubuwick}, coincide for $\nu \in -i \mathbb{N}$. This is straightforward to see at the level of the Mellin representation, where the projectors \eqref{projector bc} for the two falloffs are indistinguishable in this case:
\begin{align}
    \omega_{+in}(u,\bar{u})= \omega_{-in}(u,\bar{u}), \qquad n \in \mathbb{N}.
\end{align}
The poles $\Gamma(u\pm \tfrac{i\nu}2)$ overlap to give double poles, which is illustrated in figure \ref{fig::npoles}. The solution can be naturally interpreted as the one corresponding to the $\Delta_+$ falloff, since all the poles in $u$ generating this boundary behaviour are double poles and hence survive the action of the projector \eqref{projector bc}. To obtain the $\Delta_-$ falloff one adds a homogeneous solution to the propagator equation given by the EAdS harmonic function \eqref{splitrep}, which we denote by $\Omega^{\text{AdS}}_{\nu}(z,\bar{z})$. A well defined EAdS decomposition of the dS propagators \eqref{bubuwick} in the case $\nu \in -i\mathbb{N}$ is then obtained by expressing them as a linear combination of $G^{\text{AdS}}_{\frac{d}{2}+i\nu}(z,\bar{z})$ and the EAdS harmonic function $\Omega^{\text{AdS}}_{\nu}(z,\bar{z})$.

\vskip 4pt
Since the dS propagators themselves are non-singular for $\nu \in i \mathbb{N}$ the divergence \eqref{cscdiv} cancels when summing the two terms in \eqref{bubuwick}. Indeed, setting $\nu=-in+\epsilon$ and expanding the divergences of the two terms cancel. This is manifest using the identity 
\begin{equation}\label{harmsubtr}
    G^{\text{AdS}}_{\frac{d}{2}-i\nu}(z,\bar{z}) = G^{\text{AdS}}_{\frac{d}{2}+i\nu}(z,\bar{z}) +\frac{2\pi i}{\nu} \Omega^{\text{AdS}}_{\nu}(z,\bar{z}),
\end{equation}
which gives, for the $\pm \pm$ propagators:
\begin{shaded}
 \begin{equation}\nonumber
   G^{\pm\pm}(\eta,\bar{\eta};{\bf k})\to \pm i e^{\mp \left( d-2J \right)\frac{\pi i}{2}}\, G^{\text{AdS}}_{\frac{d}{2}+i\nu}(z,\bar{z};{\bf k})
   + e^{\mp\left(\frac{d}{2}-i\nu-J\right) \pi i}\, \Gamma\left(+i\nu\right)\Gamma\left(-i\nu\right)
   \Omega^{\text{AdS}}_\nu(z,\bar{z};{\bf k}),
\end{equation}   
\end{shaded}
\noindent and for the $\pm \mp$ propagators \cite{Sleight:2019mgd,Sleight:2019hfp}:
\begin{shaded}
 \begin{equation}\nonumber
   G^{\pm\mp}(\eta,\bar{\eta};{\bf k})\to  \Gamma\left(+i\nu\right)\Gamma\left(-i\nu\right)
   \Omega^{\text{AdS}}_\nu(z,\bar{z};{\bf k}).
\end{equation}   
\end{shaded}
\noindent Notice that the harmonic function $\Omega^{\text{AdS}}_\nu$ is vanishing for $\nu = -i \mathbb{N}$, which can be seen from their representation \eqref{splitrep}.\footnote{It can also be seen from the identity \eqref{harmsubtr}, since the propagators $G^{\text{AdS}}_{\frac{d}{2}\pm i\nu}(z,\bar{z})$ are equal for $\nu = -i \mathbb{N}$.} The combination  $\Gamma\left(-i\nu\right)\Omega^{\text{AdS}}_\nu$ ensures a non-zero and finite result for $\nu = -i \mathbb{N}$.

\vskip 4pt
For the bulk-to-boundary propagators, taking the boundary limit of the above and setting $\nu = -in$ with $n \in \mathbb{N}$, for the $\Delta_+ = \frac{d}{2}+n$ falloff we have 
\begin{shaded}
\begin{align}
     K^{\pm}_{\frac{d}{2}+n}(\eta;{\bf k}) &\to \pm i e^{\mp \left(n -J\right)\frac{\pi i}{2}} K^{\text{AdS}}_{\frac{d}{2}+n}(z;{\bf k}).
\end{align}    
\end{shaded}
\noindent For the $\Delta_- = \frac{d}{2}-n$ falloff instead the original relation is unchanged: 
\begin{shaded}
  \begin{equation}
    K^{\pm}_{\frac{d}{2}-n}(\eta;{\bf k}) \to e^{\mp \left(\frac{d}{2}-n-J\right) \frac{\pi i}{2}} c^{\text{dS-AdS}}_{\frac{d}{2}-n-i\epsilon}K^{\text{AdS}}_{\frac{d}{2}-n-i\epsilon}(z;{\bf k}).
\end{equation}   
\end{shaded}
\noindent Like for the EAdS harmonic function above, the divergence \eqref{cscdiv} in this case is canceled by the vanishing of the EAdS bulk-to-boundary propagator for these values (see equation \eqref{EAdS bubo scalar}), giving a finite expression.

\section{Scalar QED}
\label{sec::scalarqed}

In the following sections we apply the prescription outlined in the previous sections to
recast late-time correlators in specific theories of gauge bosons and gravitons in terms of
Witten diagrams in EAdS.

\vskip 4pt
We begin with scalar QED in dS$_{d+1}$, whose Lagrangian is
\begin{equation}
    {\cal L}
    =
    -\frac{1}{4}g^{\mu\rho}g^{\nu\sigma}F_{\rho\sigma}F_{\mu\nu}
    -g^{\mu\nu}\left(D_\mu\varphi\right)^*
    \left(D_\nu\varphi\right)
    -m^2\varphi^*\varphi ,
\end{equation}
where $D_\mu=\nabla_\mu+ieA_\mu$. Writing
\begin{equation}
    \varphi=\frac{1}{\sqrt{2}}\left(\phi_1+i\phi_2\right),
\end{equation}
with real scalar fields $\phi_{1,2}$ of equal mass $m$, the interaction vertices are
\begin{subequations}
\begin{align}
    V_{A\phi_1\phi_2}
    &=
    e\,g^{\mu\nu}A_\mu
    \left[
        \left(\partial_\nu\phi_1\right)\phi_2
        -
        \phi_1\left(\partial_\nu\phi_2\right)
    \right],
    \label{qed 3pt vertex}
    \\
    V_{AA\phi_1\phi_2}
    &=
    -\frac{e^2}{2}g^{\mu\nu}A_\mu A_\nu
    \left(\phi_1^2+\phi_2^2\right).
    \label{qed 4pt vertex}
\end{align}
\end{subequations}
To determine their rotation under \eqref{wickzeta}, we go to Poincar\'e coordinates
\eqref{poincare}. In temporal gauge $A_\eta=0$, these become
\begin{align}
    V_{A\phi_1\phi_2}
    &=
    e\left(-\eta\right)^2\delta^{ij}A_i
    \left[
        \left(\partial_j\phi_1\right)\phi_2
        -
        \phi_1\left(\partial_j\phi_2\right)
    \right],
    \nonumber
    \\
    V_{AA\phi_1\phi_2}
    &=
    -\frac{e^2}{2}
    \left(-\eta\right)^2\delta^{ij}A_i A_j
    \left(\phi_1^2+\phi_2^2\right).
    \label{QED vertices}
\end{align}

\vskip 4pt
We denote the two gauge-field falloffs by
\begin{equation}
    \Delta_A^+=d-1,
    \qquad
    \Delta_A^-=1,
\end{equation}
and the common scalar falloffs by
\begin{equation}
    \Delta_\phi^+=\frac d2+i\nu,
    \qquad
    \Delta_\phi^-=\frac d2-i\nu,
    \qquad
    \Delta_\phi^+\Delta_\phi^-=m^2 .
\end{equation}
The equality of the scalar masses is reflected in the fact that both $\phi_1$ and $\phi_2$
are associated with the same pair $\Delta_\phi^\pm$. 

\vskip 4pt
Combined with the propagators in the previous section, the rules to recast perturbative
late-time correlators in scalar QED in terms of Witten diagrams in EAdS under the Wick
rotations \eqref{wickzeta} are as follows:
\begin{itemize}

    \item {\bf Gauge-boson propagators}:
    \begin{multline}
    G^{\pm\hat{\pm}}_{i;j}(\eta,\bar{\eta};{\bf k})
    \to
    c^{\text{dS-AdS}}_{\Delta_A^+}
    e^{\mp\left(\Delta_A^+-1\right)\frac{\pi i}{2}}
    e^{\hat{\mp}\left(\Delta_A^+-1\right)\frac{\pi i}{2}}
    G^{\text{AdS}\,\Delta_A^+}_{i;j}(z,\bar{z};{\bf k})
    \\
    +
    c^{\text{dS-AdS}}_{\Delta_A^-}
    e^{\mp\left(\Delta_A^--1\right)\frac{\pi i}{2}}
    e^{\hat{\mp}\left(\Delta_A^--1\right)\frac{\pi i}{2}}
    G^{\text{AdS}\,\Delta_A^-}_{i;j}(z,\bar{z};{\bf k}) .
    \end{multline}
    Similarly, for the bulk-to-boundary propagators,
    \begin{equation}
    K^{\pm\,\Delta_A}_{i;j}(\eta;{\bf k})
    \to
    e^{\mp\left(\Delta_A-1\right)\frac{\pi i}{2}}
    c^{\text{dS-AdS}}_{\Delta_A}
    K^{\text{AdS}\,\Delta_A}_{i;j}(z;{\bf k}),
    \qquad
    \Delta_A=\Delta_A^+,\Delta_A^- .
    \end{equation}
    For even boundary dimension, the coefficients
    $c^{\text{dS-AdS}}_{\Delta_A^\pm}$ are understood with the limiting prescription of
    section \ref{subsec::nuimint}.

    \item {\bf Scalar propagators}:
    \begin{multline}\label{bubu scalar qed sk}
    G^{\pm\hat{\pm}}(\eta,\bar{\eta};{\bf k})
    \to
    c^{\text{dS-AdS}}_{\Delta_\phi^+}
    e^{\mp\Delta_\phi^+\frac{\pi i}{2}}
    e^{\hat{\mp}\Delta_\phi^+\frac{\pi i}{2}}
    G^{\text{AdS}\,\Delta_\phi^+}(z,\bar{z};{\bf k})
    \\
    +
    c^{\text{dS-AdS}}_{\Delta_\phi^-}
    e^{\mp\Delta_\phi^-\frac{\pi i}{2}}
    e^{\hat{\mp}\Delta_\phi^-\frac{\pi i}{2}}
    G^{\text{AdS}\,\Delta_\phi^-}(z,\bar{z};{\bf k}) .
    \end{multline}
    Similarly,
    \begin{equation}
    K^{\pm}_{\Delta_\phi}(\eta;{\bf k})
    \to
    e^{\mp\Delta_\phi\frac{\pi i}{2}}
    c^{\text{dS-AdS}}_{\Delta_\phi}
    K^{\text{AdS}}_{\Delta_\phi}(z;{\bf k}),
    \qquad
    \Delta_\phi=\Delta_\phi^+,\Delta_\phi^- .
    \end{equation}
    For definiteness, in the following we take $\nu>0$, so that
\begin{equation}
    c^{\text{dS-AdS}}_{\Delta_\phi^+}
    =
    \frac{i}{2\sinh(\pi\nu)},
    \qquad
    c^{\text{dS-AdS}}_{\Delta_\phi^-}
    =
    -\frac{i}{2\sinh(\pi\nu)}.
\end{equation}

    \item {\bf Vertices}:
    \begin{equation}
      V_{A\phi_1\phi_2}(\eta)
      \to
      e^{\mp\pi i}V_{A\phi_1\phi_2}(z),
      \qquad
      V_{AA\phi_1\phi_2}(\eta)
      \to
      e^{\mp\pi i}V_{AA\phi_1\phi_2}(z).
    \end{equation}
    
\end{itemize}

\vskip 4pt
The above rules are equivalently generated by the EAdS Lagrangian (see section
\ref{sec::SKPI}),
\begin{equation}\label{scalarQEDL}
    {\cal L}_{\text{EAdS}}
    =
    {\cal L}^{(2)}_{\text{EAdS}}
    +
    {\cal L}^{(3)}_{\text{EAdS}}
    +
    {\cal L}^{(4)}_{\text{EAdS}} .
\end{equation}
In the following we write $A_{d-1}$ and $A_1$ for the gauge fields with falloffs
$\Delta_A^+=d-1$ and $\Delta_A^-=1$, respectively. The quadratic terms are
\begin{multline}\label{scalarQEDL2}
{\cal L}^{(2)}_{\text{EAdS}}
=
\frac14
\operatorname{sgn}\!\left[c^{\text{dS-AdS}}_{d-1}\right]
\left[
F^{\mu\nu}\!\left[A_{d-1}\right]
F_{\mu\nu}\!\left[A_{d-1}\right]
-
F^{\mu\nu}\!\left[A_{1}\right]
F_{\mu\nu}\!\left[A_{1}\right]
\right]
\\
+
i\sum_{a=1}^{2}
\left(
    \frac{1}{2}
    \nabla_\mu\phi_{a,\Delta_\phi^+}
    \nabla^\mu\phi_{a,\Delta_\phi^+}
    -
    \frac{1}{2}m^2
    \phi_{a,\Delta_\phi^+}^{\,2}
\right)
\\
-
i\sum_{a=1}^{2}
\left(
    \frac{1}{2}
    \nabla_\mu\phi_{a,\Delta_\phi^-}
    \nabla^\mu\phi_{a,\Delta_\phi^-}
    -
    \frac{1}{2}m^2
    \phi_{a,\Delta_\phi^-}^{\,2}
\right).
\end{multline}

It is convenient to write the cubic interaction in terms of the current
\begin{equation}\label{scalarQEDcurrent}
    J^\mu_{\alpha\beta}
    =
    \phi_{2,\Delta_\phi^\beta}
    \nabla^\mu\phi_{1,\Delta_\phi^\alpha}
    -
    \phi_{1,\Delta_\phi^\alpha}
    \nabla^\mu\phi_{2,\Delta_\phi^\beta},
    \qquad
    \alpha,\beta=\pm .
\end{equation}
With this convention, the cubic interaction is\footnote{Notice in particular that the mixed scalar-shadow couplings to $A_1$, i.e. $A_1\,J^\mu_{+-},\,
    A_1\,J^\mu_{-+}$, are absent.}
\begin{multline}\label{scalarQEDL3}
{\cal L}^{(3)}_{\text{EAdS}}
=
e\sqrt{\left|c^{\text{dS-AdS}}_{d-1}\right|}
\operatorname{csch}(\pi\nu)
\left(A_{d-1}\right)_\mu
\Bigg[
\sin\left(\Delta_\phi^+\pi\right)
J^\mu_{++}
\\
+
\sin\left(\frac{d\pi}{2}\right)
\left(
    J^\mu_{+-}
    +
    J^\mu_{-+}
\right)
+
\sin\left(\Delta_\phi^-\pi\right)
J^\mu_{--}
\Bigg]
\\
+
i e\sqrt{\left|c^{\text{dS-AdS}}_{1}\right|}
\left(A_1\right)_\mu
\left(
    J^\mu_{--}
    -
    J^\mu_{++}
\right).
\end{multline}

Finally, the quartic interaction is\footnote{For integer $d$, the quartic couplings
\begin{equation}
    A_{d-1}A_{d-1}\,\phi_{\Delta_\phi^+}\phi_{\Delta_\phi^-},
    \qquad
    A_1A_1\,\phi_{\Delta_\phi^+}\phi_{\Delta_\phi^-}
\end{equation}
are absent. In dimensional continuation, the first is proportional to $\sin(d\pi)$ and
should be kept until the integer-$d$ limit is taken, while the second vanishes identically.}
\begin{multline}\label{scalarQEDL4}
{\cal L}^{(4)}_{\text{EAdS}}
=
\frac{e^2}{2}
\left|c^{\text{dS-AdS}}_{d-1}\right|
\operatorname{csch}(\pi\nu)
\left(A_{d-1}\right)_\mu
\left(A_{d-1}\right)^\mu
\Bigg[
\sin\left[
    \pi
    \left(
        \Delta_\phi^+ + \frac{d}{2}
    \right)
\right]
\left(
    \phi_{1,\Delta_\phi^+}^{\,2}
    +
    \phi_{2,\Delta_\phi^+}^{\,2}
\right)
\\
+
\sin\left[
    \pi
    \left(
        \Delta_\phi^- + \frac{d}{2}
    \right)
\right]
\left(
    \phi_{1,\Delta_\phi^-}^{\,2}
    +
    \phi_{2,\Delta_\phi^-}^{\,2}
\right)
\\
+
2\sin\left(
    d \pi
\right)
\left(
    \phi_{1,\Delta_\phi^+}\phi_{1,\Delta_\phi^-}
    +
    \phi_{2,\Delta_\phi^+}\phi_{2,\Delta_\phi^-}
\right)
\Bigg]
\\
+
\frac{i e^2}{2}
\left|c^{\text{dS-AdS}}_{1}\right|
\left(A_1\right)_\mu
\left(A_1\right)^\mu
\left[
    \phi_{1,\Delta_\phi^+}^{\,2}
    +
    \phi_{2,\Delta_\phi^+}^{\,2}
    -
    \phi_{1,\Delta_\phi^-}^{\,2}
    -
    \phi_{2,\Delta_\phi^-}^{\,2}
\right]
\\
-
e^2
\sqrt{
\left|c^{\text{dS-AdS}}_{d-1}\right|
\left|c^{\text{dS-AdS}}_{1}\right|
}
\operatorname{csch}(\pi\nu)
\left(A_{d-1}\right)_\mu
\left(A_1\right)^\mu
\Bigg[
\sin\left(
    \pi\Delta_\phi^+
\right)
\left(
    \phi_{1,\Delta_\phi^+}^{\,2}
    +
    \phi_{2,\Delta_\phi^+}^{\,2}
\right)
\\
+
\sin\left(
    \pi\Delta_\phi^-
\right)
\left(
    \phi_{1,\Delta_\phi^-}^{\,2}
    +
    \phi_{2,\Delta_\phi^-}^{\,2}
\right)
\\
+
2\sin\left(
    \frac{d\pi}{2}
\right)
\left(
    \phi_{1,\Delta_\phi^+}\phi_{1,\Delta_\phi^-}
    +
    \phi_{2,\Delta_\phi^+}\phi_{2,\Delta_\phi^-}
\right)
\Bigg].
\end{multline}
For odd boundary dimension one may further use
$\sqrt{|c^{\text{dS-AdS}}_{1}|}
=
\sqrt{|c^{\text{dS-AdS}}_{d-1}|}
=1/\sqrt2$ and
$\operatorname{sgn}(c^{\text{dS-AdS}}_{d-1})=(-1)^{(d-1)/2}$.
For even $d$, the gauge-field coefficients are understood with the limiting prescription
described in section \ref{subsec::nuimint}.

\vskip 4pt
In the following we give some simple examples.

\begin{figure}[t]
    \centering
    \includegraphics[width=0.65\textwidth]{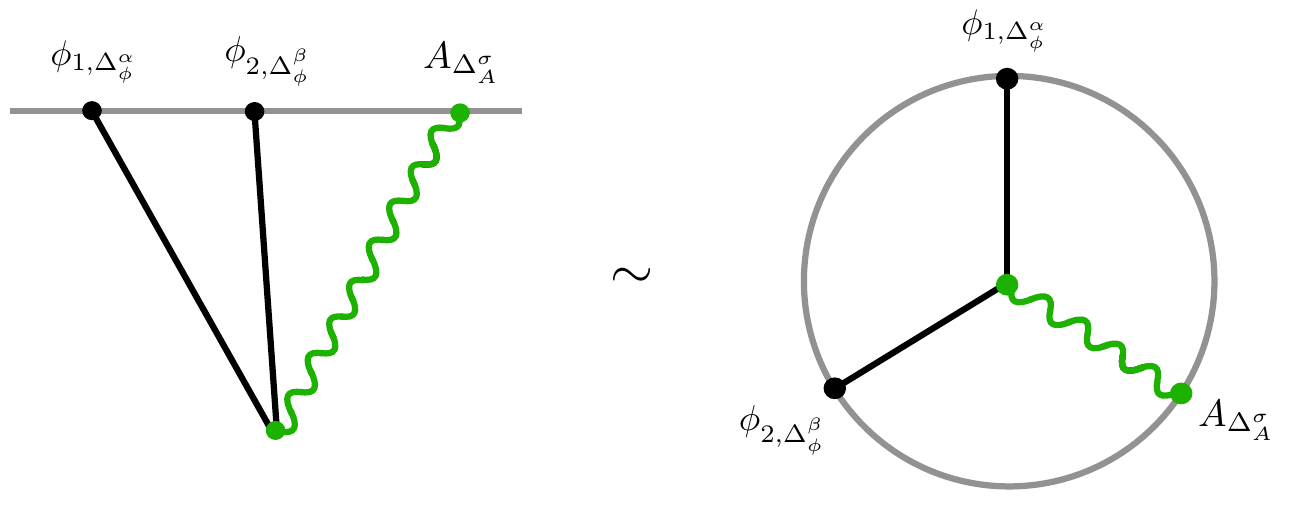}
    \caption{The three-point contact diagram in dS scalar QED is proportional to the corresponding three-point contact Witten diagram in EAdS. The proportionality constant is given in \eqref{3pt qed ds to ads}.}
    \label{fig::3ptqed}
\end{figure}

\paragraph{Contact diagrams.}

The simplest example is the three-point contact diagram of the two real scalar components
$\phi_1$ and $\phi_2$ coupled to a photon through the cubic vertex
\eqref{qed 3pt vertex}. This is illustrated in figure \ref{fig::3ptqed}. For external
falloffs $\Delta_\phi^\alpha$, $\Delta_\phi^\beta$ and $\Delta_A^\sigma$, with
$\alpha,\beta,\sigma=\pm$, the two Schwinger-Keldysh branch contributions combine to give
\begin{multline}\label{3pt qed ds to ads}
    \left\langle
        \phi_{1,\Delta_\phi^\alpha}({\bf k}_1)
        \phi_{2,\Delta_\phi^\beta}({\bf k}_2)
        A_{\Delta_A^\sigma}({\bf k}_3)
    \right\rangle_{\text{dS},\,\text{contact}}
    \\
    =
    2\sin\left[
    \left(
        -d+1+\Delta_\phi^\alpha+\Delta_\phi^\beta+\Delta_A^\sigma
    \right)
    \frac{\pi}{2}
    \right]
    c^{\text{dS-AdS}}_{\Delta_\phi^\alpha}
    c^{\text{dS-AdS}}_{\Delta_\phi^\beta}
    c^{\text{dS-AdS}}_{\Delta_A^\sigma}
    \\
    \times
    \left\langle
        \phi_{1,\Delta_\phi^\alpha}({\bf k}_1)
        \phi_{2,\Delta_\phi^\beta}({\bf k}_2)
        A_{\Delta_A^\sigma}({\bf k}_3)
    \right\rangle_{\text{EAdS},\,\text{contact}} .
\end{multline}
Using
\begin{equation}
    \Delta_A^+=d-1,
    \qquad
    \Delta_A^-=1,
    \qquad
    \Delta_\phi^++\Delta_\phi^-=d,
    \qquad
    \Delta_\phi^\pm=\frac d2\pm i\nu,
\end{equation}
the sine factors for the $\Delta_A^+=d-1$ photon falloff are
\begin{align}
    &2\sin\left[
    \left(
        -d+1+\Delta_\phi^+ +\Delta_\phi^+ +\Delta_A^+
    \right)
    \frac{\pi}{2}
    \right]
    =
    2\sin\left(\pi\Delta_\phi^+\right),
    \\
    &2\sin\left[
    \left(
        -d+1+\Delta_\phi^- +\Delta_\phi^- +\Delta_A^+
    \right)
    \frac{\pi}{2}
    \right]
    =
    2\sin\left(\pi\Delta_\phi^-\right),
    \\
    &2\sin\left[
    \left(
        -d+1+\Delta_\phi^+ +\Delta_\phi^- +\Delta_A^+
    \right)
    \frac{\pi}{2}
    \right]
    =
    2\sin\left(
        \frac{d\pi}{2}
    \right).
\end{align}
For the $\Delta_A^-=1$ photon falloff, one obtains
\begin{align}
    &2\sin\left[
    \left(
        -d+1+\Delta_\phi^+ +\Delta_\phi^+ +\Delta_A^-
    \right)
    \frac{\pi}{2}
    \right]
    =
    -2i\sinh(\pi\nu),
    \\
    &2\sin\left[
    \left(
        -d+1+\Delta_\phi^- +\Delta_\phi^- +\Delta_A^-
    \right)
    \frac{\pi}{2}
    \right]
    =
    +2i\sinh(\pi\nu),
    \\
    &2\sin\left[
    \left(
        -d+1+\Delta_\phi^+ +\Delta_\phi^- +\Delta_A^-
    \right)
    \frac{\pi}{2}
    \right]
    =
    0 .
\end{align}
Thus the cubic contact diagram with the $\Delta_A^-=1$ photon falloff has no mixed
scalar-shadow contribution. This is precisely the absence of the $A_1J^\mu_{+-}$ and
$A_1J^\mu_{-+}$ couplings in \eqref{scalarQEDL3}. 

\vskip 4pt
Similarly, the four-point contact diagram generated by the quartic vertex
\eqref{qed 4pt vertex} involves two photons and two scalars of the same real component.
For example, for external $\phi_1$ fields one obtains
\begin{multline}\label{4pt qed ds to ads}
    \left\langle
        \phi_{1,\Delta_\phi^\alpha}({\bf k}_1)
        A_{\Delta_A^\sigma}({\bf k}_2)
        \phi_{1,\Delta_\phi^\beta}({\bf k}_3)
        A_{\Delta_A^\rho}({\bf k}_4)
    \right\rangle_{\text{dS},\,\text{contact}}
    \\
    =
    2\sin\left[
    \left(
        -d
        +
        \Delta_\phi^\alpha
        +
        \Delta_A^\sigma
        +
        \Delta_\phi^\beta
        +
        \Delta_A^\rho
    \right)
    \frac{\pi}{2}
    \right]
    c^{\text{dS-AdS}}_{\Delta_\phi^\alpha}
    c^{\text{dS-AdS}}_{\Delta_A^\sigma}
    c^{\text{dS-AdS}}_{\Delta_\phi^\beta}
    c^{\text{dS-AdS}}_{\Delta_A^\rho}
    \\
    \times
    \left\langle
        \phi_{1,\Delta_\phi^\alpha}({\bf k}_1)
        A_{\Delta_A^\sigma}({\bf k}_2)
        \phi_{1,\Delta_\phi^\beta}({\bf k}_3)
        A_{\Delta_A^\rho}({\bf k}_4)
    \right\rangle_{\text{EAdS},\,\text{contact}} .
\end{multline}
The same formula holds with $\phi_1$ replaced everywhere by $\phi_2$. For mixed scalar falloffs, $\Delta_\phi^+$ and $\Delta_\phi^-$, the sine factor reduces to
\begin{equation}
    2\sin\left[
    \left(
        \Delta_A^\sigma+\Delta_A^\rho
    \right)
    \frac{\pi}{2}
    \right].
\end{equation}
Using $\Delta_A^+=d-1$ and $\Delta_A^-=1$, the three possible gauge-field assignments are
\begin{align}
    &2\sin\left[
    \left(
        \Delta_A^+ + \Delta_A^+
    \right)
    \frac{\pi}{2}
    \right]
    =
    2\sin\left((d-1)\pi\right)
    =
    -2\sin\left[
        \pi
        \left(
            \Delta_\phi^+ + \Delta_\phi^-
        \right)
    \right],
    \\
    &2\sin\left[
    \left(
        \Delta_A^+ + \Delta_A^-
    \right)
    \frac{\pi}{2}
    \right]
    =
    2\sin\left(\frac{d\pi}{2}\right)
    =
    2\sin\left[
        \frac{\pi}{2}
        \left(
            \Delta_\phi^+ + \Delta_\phi^-
        \right)
    \right],
    \\
    &2\sin\left[
    \left(
        \Delta_A^- + \Delta_A^-
    \right)
    \frac{\pi}{2}
    \right]
    =
    2\sin(\pi)
    =
    0 .
\end{align}
This agrees with the structure of \eqref{scalarQEDL4}.

\begin{figure}[t]
    \centering
    \includegraphics[width=\textwidth]{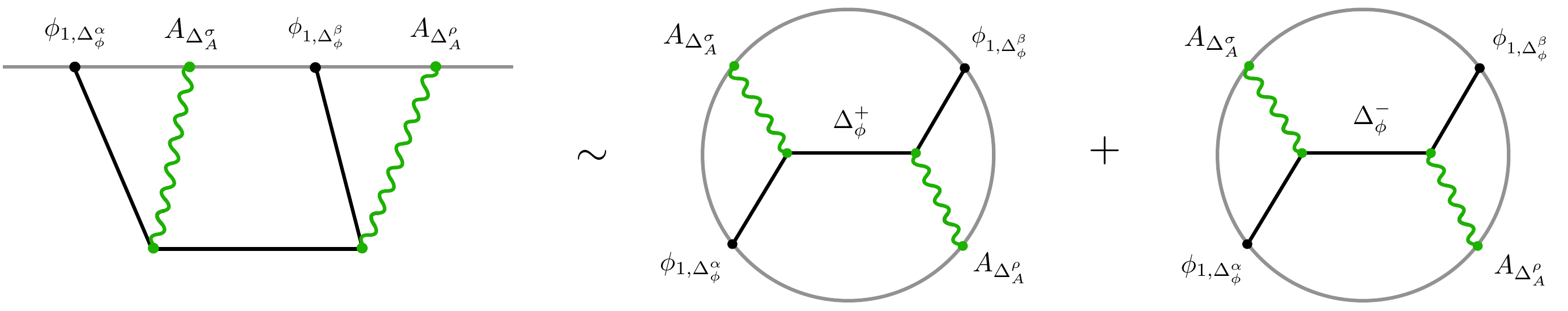}
    \caption{The four-point exchange in dS scalar QED can be recast as a sum of two four-point exchange Witten diagrams in EAdS for the two possible $\Delta_\pm$ boundary conditions on the exchanged scalar. The coefficient of each exchange Witten diagram is given in \eqref{comptonexch}  }
    \label{fig::4ptexch}
\end{figure}

\paragraph{Compton scattering.}

One proceeds in a similar fashion for exchange diagrams. As an example, consider the
scalar exchange contribution to Compton scattering generated by two cubic vertices
\eqref{qed 3pt vertex}, see figure \ref{fig::4ptexch}. Since the cubic vertex couples
$A_\mu$ to one $\phi_1$ and one $\phi_2$, a scalar exchange diagram with two cubic
vertices connects external scalars of the same real component. For definiteness we take
the external scalars to be $\phi_1$; the case with external $\phi_2$ is identical, with
the exchanged scalar replaced by $\phi_1$.

\vskip 4pt
For external falloffs $\Delta_\phi^\alpha$, $\Delta_\phi^\beta$ and photon falloffs
$\Delta_A^\sigma$, $\Delta_A^\rho$, with $\alpha,\beta,\sigma,\rho=\pm$, the exchange
diagram is
\begin{multline}\label{comptonexch}
    \left\langle
        \phi_{1,\Delta_\phi^\alpha}({\bf k}_1)
        A_{\Delta_A^\sigma}({\bf k}_2)
        \phi_{1,\Delta_\phi^\beta}({\bf k}_3)
        A_{\Delta_A^\rho}({\bf k}_4)
    \right\rangle_{\text{dS},\,\text{exch}}
    \\
    =
    c^{\text{dS-AdS}}_{\Delta_\phi^\alpha}
    c^{\text{dS-AdS}}_{\Delta_A^\sigma}
    c^{\text{dS-AdS}}_{\Delta_\phi^\beta}
    c^{\text{dS-AdS}}_{\Delta_A^\rho}
    \sum_{\gamma=\pm}
    c^{\text{dS-AdS}}_{\Delta_\phi^\gamma}
    \\
    \times
    2\sin\left[
        \left(
            -d+1
            +
            \Delta_\phi^\alpha
            +
            \Delta_A^\sigma
            +
            \Delta_\phi^\gamma
        \right)
        \frac{\pi}{2}
    \right]
    2\sin\left[
        \left(
            -d+1
            +
            \Delta_\phi^\beta
            +
            \Delta_A^\rho
            +
            \Delta_\phi^\gamma
        \right)
        \frac{\pi}{2}
    \right]
    \\
    \times
    \left\langle
        \phi_{1,\Delta_\phi^\alpha}({\bf k}_1)
        A_{\Delta_A^\sigma}({\bf k}_2)
        \phi_{1,\Delta_\phi^\beta}({\bf k}_3)
        A_{\Delta_A^\rho}({\bf k}_4)
    \right\rangle_{\text{EAdS},\,\text{exch};\,\Delta_\phi^\gamma}.
\end{multline}
Here $\Delta_\phi^\gamma$ is the falloff of the exchanged scalar. The two sine factors
are precisely the factors appearing in the two cubic subdiagrams. This is the expected
factorisation of the dS-to-EAdS map on the scalar exchange channel. The vertex factors in \eqref{comptonexch} are exactly those of the three-point contact
diagram. In particular, a vertex containing the $\Delta_A^-=1$ photon falloff has no
mixed scalar-shadow contribution:
\begin{equation}
    2\sin\left[
        \left(
            -d+1+\Delta_\phi^+ + \Delta_A^- + \Delta_\phi^-
        \right)
        \frac{\pi}{2}
    \right]
    =
    0 .
\end{equation}
Thus an $A_1$ vertex forces the scalar falloffs on the two scalar legs of that vertex to
be the same. Equivalently, this is the exchange-diagram counterpart of the absence of the
$A_1J^\mu_{+-}$ and $A_1J^\mu_{-+}$ couplings in \eqref{scalarQEDL3}.

\vskip 4pt
For example, if both photon legs have the $\Delta_A^-=1$ falloff, the only non-vanishing
terms in \eqref{comptonexch} are those for which the exchanged scalar has the same
falloff as both external scalars:
\begin{equation}
    \alpha=\beta=\gamma .
\end{equation}
If one of the photon legs has falloff $\Delta_A^+=d-1$, mixed scalar falloffs are allowed
at that vertex, with coefficient proportional to
\begin{equation}
    \sin\left[
        \frac{\pi}{2}
        \left(
            \Delta_\phi^+ + \Delta_\phi^-
        \right)
    \right]
    =
    \sin\left(\frac{d\pi}{2}\right).
\end{equation}
Thus the selection rules of the exchange diagram are just the factorised selection rules
of the cubic EAdS interaction \eqref{scalarQEDL3}.

\section{Pure Yang-Mills}
\label{sec::pureYM}

In this section we consider pure Yang-Mills theory with $SU(N)$ gauge group. The Lagrangian is,\footnote{We are following the conventions of the QFT book by Srednicki \cite{Srednicki:2007qs}.}
\begin{align}
    {\cal L}=&-\frac{1}{2}\text{tr}(F^{\mu\nu}F_{\mu\nu}),
\end{align}
where, 
\begin{subequations}
 \begin{align}
    D_\mu&:=\partial_\mu-i\mathrm{g}A_\mu,\\
    F_{\mu\nu}&=t^{\mathrm{a}}F^\mathrm{a}_{\mu\nu}:=\frac{i}{\mathrm{g}}[D_\mu,D_\nu]=\partial_\mu A_\nu-\partial_\nu A_\mu -i\mathrm{g}[A_\mu,A_\nu],
\end{align}   
\end{subequations}
with generators $t^\mathrm{a}\in SU(N)$ where  $\text{tr}(t^\mathrm{a}t^\mathrm{b})=\frac{\delta^{\mathrm{{ab}}}}{2}$ and $[t^\mathrm{a},t^\mathrm{b}]=if^{\mathrm{{abc}}}t^\mathrm{c}$. In the temporal gauge $A_\eta=0$ the theory has the following 3-gluon and 4-gluon interaction vertices,
\begin{subequations}\label{YM vertices}
 \begin{align}\label{YM cubic}
    V_{AAA}\left(\eta\right)=& -\mathrm{g}f^{\mathrm{{abc}}}\left(-\eta\right)^4\delta^{ik}\delta^{jl}A^{\mathrm{a}}_k A^{\mathrm{b}}_l(\partial_i A^{\mathrm{c}}_j),\\ \label{YM quartic}
    V_{AAAA}\left(\eta\right)=& -\frac{\mathrm{g}^2}{4}f^{\mathrm{{abe}}}f^{\mathrm{cde}}\left(-\eta\right)^4\delta^{ik}\delta^{jl}A^{\mathrm{a}}_i A^{\mathrm{b}}_j A^{\mathrm{c}}_k A^{\mathrm{d}}_l.
\end{align}   
\end{subequations}

\vskip 4pt
We denote the two gauge-field falloffs by
\begin{equation}
    \Delta_A^+=d-1,
    \qquad
    \Delta_A^-=1.
\end{equation}
The rules to recast perturbative late-time correlators of $A_\mu$ in terms of Witten
diagrams in EAdS under the Wick rotations \eqref{wickzeta} are then as follows:
\begin{itemize}
    \item {\bf Gauge-boson propagators}:
    \begin{multline}\label{gbbubu1}
    G^{\pm\hat{\pm}}_{i \mathrm{a}; j \mathrm{b}}(\eta,\bar{\eta};{\bf k})
    \to
    c^{\text{dS-AdS}}_{\Delta_A^+}
    e^{\mp \left(\Delta_A^+-1\right)\frac{\pi i}{2}}
    e^{\hat{\mp} \left(\Delta_A^+-1\right)\frac{\pi i}{2}}
    G^{\text{AdS}\,\Delta_A^+}_{i \mathrm{a}; j \mathrm{b}}(z,\bar{z};{\bf k})
    \\
    +
    c^{\text{dS-AdS}}_{\Delta_A^-}
    e^{\mp \left(\Delta_A^--1\right)\frac{\pi i}{2}}
    e^{\hat{\mp} \left(\Delta_A^--1\right)\frac{\pi i}{2}}
    G^{\text{AdS}\,\Delta_A^-}_{i \mathrm{a}; j \mathrm{b}}(z,\bar{z};{\bf k}) .
    \end{multline}
    Similarly,
    \begin{equation}\label{gbbubo1}
    K^{\pm\,\Delta_A}_{i \mathrm{a}; j \mathrm{b}}(\eta;{\bf k})
    \to
    e^{\mp \left(\Delta_A-1\right)\frac{\pi i}{2}}
    c^{\text{dS-AdS}}_{\Delta_A}
    K^{\text{AdS}\,\Delta_A}_{i \mathrm{a}; j \mathrm{b}}(z;{\bf k}),
    \qquad
    \Delta_A=\Delta_A^+,\Delta_A^- .
    \end{equation}
For even boundary dimension, the coefficients
    $c^{\text{dS-AdS}}_{\Delta_A^\pm}$ are understood with the limiting prescription of
    section \ref{subsec::nuimint}
    \item {\bf Vertices}:
    \begin{equation}\label{YMvert1}
      {\cal V}_{AAA}\left(\eta\right) \to {\cal V}_{AAA}\left(z\right),
      \qquad
      {\cal V}_{AAAA}\left(\eta\right) \to {\cal V}_{AAAA}\left(z\right).
    \end{equation}
\end{itemize}

These rules are equivalently generated by the EAdS Lagrangian, see section
\ref{sec::SKPI},
\begin{equation}\label{YML}
    {\cal L}_{\text{EAdS}}
    =
    {\cal L}^{(2)}_{\text{EAdS}}
    +
    {\cal L}^{(3)}_{\text{EAdS}}
    +
    {\cal L}^{(4)}_{\text{EAdS}} .
\end{equation}
Writing $A_{d-1}$ and $A_1$ for the fields with falloffs $\Delta_A^+=d-1$ and
$\Delta_A^-=1$, respectively, the quadratic terms are
\begin{equation}\label{YML2}
    {\cal L}^{(2)}_{\text{EAdS}}
    =
    \frac14
    \operatorname{sgn}\!\left[c^{\text{dS-AdS}}_{d-1}\right]
    \left[
        F^{\mu\nu}\!\left[A_{d-1}\right]
        F_{\mu\nu}\!\left[A_{d-1}\right]
        -
        F^{\mu\nu}\!\left[A_1\right]
        F_{\mu\nu}\!\left[A_1\right]
    \right],
\end{equation}
where here $F_{\mu\nu}[A]=\partial_\mu A_\nu-\partial_\nu A_\mu$ denotes the linearised
field strength.

The cubic interaction is
\begin{multline}\label{YML3}
{\cal L}^{(3)}_{\text{EAdS}}
=
2\mathrm{g} f^{\mathrm{a}\mathrm{b}\mathrm{c}}
\sum_{\sigma_1,\sigma_2,\sigma_3=\pm}
\sqrt{
\left|c^{\text{dS-AdS}}_{\Delta_A^{\sigma_1}}\right|
\left|c^{\text{dS-AdS}}_{\Delta_A^{\sigma_2}}\right|
\left|c^{\text{dS-AdS}}_{\Delta_A^{\sigma_3}}\right|
}
\\
\times
\sin\left[
\left(
    \Delta_A^{\sigma_1}
    +
    \Delta_A^{\sigma_2}
    +
    \Delta_A^{\sigma_3}
    -
    d
    -
    3
\right)
\frac{\pi}{2}
\right]
\\
\times
\partial^\mu\left(A_{\Delta_A^{\sigma_1}}\right)^{\mathrm{a}\nu}
\left(A_{\Delta_A^{\sigma_2}}\right)^{\mathrm{b}}_{\mu}
\left(A_{\Delta_A^{\sigma_3}}\right)^{\mathrm{c}}_{\nu}.
\end{multline}
The quartic interaction is
\begin{multline}\label{YML4}
{\cal L}^{(4)}_{\text{EAdS}}
=
\frac12 \mathrm{g}^2
f^{\mathrm{e}\mathrm{a}\mathrm{b}}
f^{\mathrm{e}\mathrm{c}\mathrm{d}}
\sum_{\sigma_1,\sigma_2,\sigma_3,\sigma_4=\pm}
\sqrt{
\left|c^{\text{dS-AdS}}_{\Delta_A^{\sigma_1}}\right|
\left|c^{\text{dS-AdS}}_{\Delta_A^{\sigma_2}}\right|
\left|c^{\text{dS-AdS}}_{\Delta_A^{\sigma_3}}\right|
\left|c^{\text{dS-AdS}}_{\Delta_A^{\sigma_4}}\right|
}
\\
\times
\sin\left[
\left(
    \Delta_A^{\sigma_1}
    +
    \Delta_A^{\sigma_2}
    +
    \Delta_A^{\sigma_3}
    +
    \Delta_A^{\sigma_4}
    -
    d
    -
    4
\right)
\frac{\pi}{2}
\right]
\\
\times
\left(A_{\Delta_A^{\sigma_1}}\right)^{\mathrm{a}\mu}
\left(A_{\Delta_A^{\sigma_2}}\right)^{\mathrm{b}\nu}
\left(A_{\Delta_A^{\sigma_3}}\right)^{\mathrm{c}}_{\mu}
\left(A_{\Delta_A^{\sigma_4}}\right)^{\mathrm{d}}_{\nu}.
\end{multline}
For odd boundary dimension one may further use
$\sqrt{|c^{\text{dS-AdS}}_{1}|}
=
\sqrt{|c^{\text{dS-AdS}}_{d-1}|}
=1/\sqrt2$ and
$\operatorname{sgn}(c^{\text{dS-AdS}}_{d-1})=(-1)^{(d-1)/2}$.
For even $d$, the gauge-field coefficients are understood with the limiting prescription
described in section \ref{subsec::nuimint}.

\paragraph{Contact diagrams.}

For the three-point contact diagram generated by the cubic vertex \eqref{YM cubic}, the
two Schwinger-Keldysh branch contributions combine to give
\begin{multline}\label{YM 3pt contact}
    \left\langle
        A_{\Delta_A^{\sigma_1}}({\bf k}_1)
        A_{\Delta_A^{\sigma_2}}({\bf k}_2)
        A_{\Delta_A^{\sigma_3}}({\bf k}_3)
    \right\rangle_{\text{dS},\,\text{contact}}
    \\
    =
    2
    \sin\left[
    \left(
        \Delta_A^{\sigma_1}
        +
        \Delta_A^{\sigma_2}
        +
        \Delta_A^{\sigma_3}
        -
        d
        -
        3
    \right)
    \frac{\pi}{2}
    \right]
    \prod_{i=1}^{3}
    c^{\text{dS-AdS}}_{\Delta_A^{\sigma_i}}
    \\
    \times
    \left\langle
        A_{\Delta_A^{\sigma_1}}({\bf k}_1)
        A_{\Delta_A^{\sigma_2}}({\bf k}_2)
        A_{\Delta_A^{\sigma_3}}({\bf k}_3)
    \right\rangle_{\text{EAdS},\,\text{contact}} .
\end{multline}
Using
\begin{equation}
    \Delta_A^+=d-1,
    \qquad
    \Delta_A^-=1,
    \qquad
    \Delta_A^+ + \Delta_A^- = d,
\end{equation}
the various cases are as follows:
\begin{equation}
\sin\left[
    \left(
        \sum_{i=1}^{3}\Delta_A^{\sigma_i}
        -
        d
        -
        3
    \right)
    \frac{\pi}{2}
\right]
=
\begin{cases}
\sin(\pi\Delta_A^+)
=
-\sin(d\pi),
&
\Delta_A^+\Delta_A^+\Delta_A^+,
\\[4pt]
\sin\left[
    \frac{\pi}{2}
    \left(
        \Delta_A^+ + \Delta_A^-
    \right)
\right]
=
\sin\left(\frac{d\pi}{2}\right),
&
\Delta_A^+\Delta_A^+\Delta_A^-,
\\[4pt]
0,
&
\Delta_A^+\Delta_A^-\Delta_A^-,
\\[4pt]
-\sin\left[
    \frac{\pi}{2}
    \left(
        \Delta_A^+ + \Delta_A^-
    \right)
\right]
=
-\sin\left(\frac{d\pi}{2}\right),
&
\Delta_A^-\Delta_A^-\Delta_A^- .
\end{cases}
\end{equation}
Thus the $\Delta_A^+\Delta_A^-\Delta_A^-$ coupling vanishes identically. The
$\Delta_A^+\Delta_A^+\Delta_A^+$ coupling is proportional to
$\sin(\pi\Delta_A^+)=-\sin(d\pi)$, and therefore vanishes after specialising $d$ to an
integer. The remaining two cases are proportional to $\sin(d\pi/2)$, and vanish when $d$
is an even integer.

\vskip 4pt
The sine factors above should be understood before specialising $d$ to an integer. In
particular, a zero of the sine factor can multiply a divergence of the corresponding EAdS
contact integral. Thus a vanishing sine factor implies that the non-local part of the
late-time correlator vanishes, but it does not by itself exclude local terms generated by
renormalisation. These possible local terms can be analysed using the expression \eqref{EAdS bubo GB} for
the EAdS bulk-to-boundary propagator in terms of Bessel-$K$ functions. The relevant EAdS
three-point contact diagrams reduce to integrated products of Bessel-$K$ functions,
namely triple-$K$ integrals, whose convergence conditions are well understood
\cite{Bzowski:2015pba,Bzowski:2017poo,Bzowski:2018fql}. Applying these conditions to
the gauge-field falloffs $\Delta_A^+=d-1$ and $\Delta_A^-=1$, one finds that IR
divergences are present only for the falloff assignments
\begin{equation}
    \Delta_A^+\Delta_A^+\Delta_A^+
    \qquad \text{for even } d,
\end{equation}
and
\begin{equation}
    \Delta_A^-\Delta_A^-\Delta_A^-
    \qquad \text{for even } d\geq 4 .
\end{equation}
In these cases one should check whether the zero of the sine factor combines with the
divergence of the EAdS contact diagram to produce a finite local contribution after
renormalisation.

\vskip 4pt
Similarly, for the four-point contact diagram generated by the quartic vertex
\eqref{YM quartic}, one obtains
\begin{multline}\label{YM 4pt contact}
    \left\langle
        A_{\Delta_A^{\sigma_1}}({\bf k}_1)
        A_{\Delta_A^{\sigma_2}}({\bf k}_2)
        A_{\Delta_A^{\sigma_3}}({\bf k}_3)
        A_{\Delta_A^{\sigma_4}}({\bf k}_4)
    \right\rangle_{\text{dS},\,\text{contact}}
    \\
    =
    2
    \sin\left[
    \left(
        \Delta_A^{\sigma_1}
        +
        \Delta_A^{\sigma_2}
        +
        \Delta_A^{\sigma_3}
        +
        \Delta_A^{\sigma_4}
        -
        d
        -
        4
    \right)
    \frac{\pi}{2}
    \right]
    \prod_{i=1}^{4}
    c^{\text{dS-AdS}}_{\Delta_A^{\sigma_i}}
    \\
    \times
    \left\langle
        A_{\Delta_A^{\sigma_1}}({\bf k}_1)
        A_{\Delta_A^{\sigma_2}}({\bf k}_2)
        A_{\Delta_A^{\sigma_3}}({\bf k}_3)
        A_{\Delta_A^{\sigma_4}}({\bf k}_4)
    \right\rangle_{\text{EAdS},\,\text{contact}} .
\end{multline}
The corresponding sine factor is
\begin{equation}
\sin\left[
    \left(
        \sum_{i=1}^{4}\Delta_A^{\sigma_i}
        -
        d
        -
        4
    \right)
    \frac{\pi}{2}
\right]
=
\begin{cases}
\sin\left[
    \frac{3\pi}{2}
    \left(
        \Delta_A^+ + \Delta_A^-
    \right)
\right]
=
\sin\left(\frac{3d\pi}{2}\right),
&
\Delta_A^+\Delta_A^+\Delta_A^+\Delta_A^+,
\\[4pt]
\sin(\pi\Delta_A^+)
=
-\sin(d\pi),
&
\Delta_A^+\Delta_A^+\Delta_A^+\Delta_A^-,
\\[4pt]
\sin\left[
    \frac{\pi}{2}
    \left(
        \Delta_A^+ + \Delta_A^-
    \right)
\right]
=
\sin\left(\frac{d\pi}{2}\right),
&
\Delta_A^+\Delta_A^+\Delta_A^-\Delta_A^-,
\\[4pt]
0,
&
\Delta_A^+\Delta_A^-\Delta_A^-\Delta_A^-,
\\[4pt]
-\sin\left[
    \frac{\pi}{2}
    \left(
        \Delta_A^+ + \Delta_A^-
    \right)
\right]
=
-\sin\left(\frac{d\pi}{2}\right),
&
\Delta_A^-\Delta_A^-\Delta_A^-\Delta_A^- .
\end{cases}
\end{equation}
As in the three-point case, the zeros of these sine factors should be interpreted in
dimensional continuation. A zero removes the non-local part of the correlator, but local
terms may still arise if the corresponding EAdS contact integral develops a divergence
and requires renormalisation.

\section{Gravity}
\label{sec::EHGr}

In this section we consider Einstein-Hilbert gravity. The expansion of the latter around a given background is an infinite series in weak field fluctuations $h_{\mu \nu}$. In the following we consider the expansion up to cubic order, where the cubic Lagrangian reads: 
\begin{align}\label{cubic EH}
    {\cal V}_{hhh}= & \kappa \left(\tfrac{1}{2} h^{\mu \nu} \nabla_{\mu}h^{\rho \sigma} \nabla_{\nu}h_{\rho \sigma} -  \tfrac{1}{2} h^{\mu \nu} \nabla_{\mu}h^{\rho}{}_{\rho} \nabla_{\nu}h^{\sigma}{}_{\sigma} + \tfrac{3}{2} h^{\mu \nu} \nabla_{\nu}h^{\sigma}{}_{\sigma} \nabla_{\rho}h_{\mu}{}^{\rho} + \tfrac{1}{2} h^{\mu \nu} \nabla_{\nu}h_{\mu}{}^{\rho} \nabla_{\rho}h^{\sigma}{}_{\sigma}\nonumber \right. \\ - & h^{\mu \nu} \nabla_{\rho}h^{\sigma}{}_{\sigma} \nabla^{\rho}h_{\mu \nu} + \tfrac{1}{4} h^{\mu}{}_{\mu} \nabla_{\rho}h^{\sigma}{}_{\sigma} \nabla^{\rho}h^{\nu}{}_{\nu} -  h^{\mu \nu} \nabla_{\rho}h_{\mu}{}^{\rho} \nabla_{\sigma}h_{\nu}{}^{\sigma} -  h^{\mu \nu} \nabla_{\nu}h_{\mu}{}^{\rho} \nabla_{\sigma}h_{\rho}{}^{\sigma} \nonumber\\+& \tfrac{1}{2} h^{\mu}{}_{\mu} \nabla_{\nu}h^{\nu \rho} \nabla_{\sigma}h_{\rho}{}^{\sigma} + h^{\mu \nu} \nabla^{\rho}h_{\mu \nu} \nabla_{\sigma}h_{\rho}{}^{\sigma} -  \tfrac{1}{2} h^{\mu}{}_{\mu} \nabla^{\rho}h^{\nu}{}_{\nu} \nabla_{\sigma}h_{\rho}{}^{\sigma} -  h^{\mu \nu} \nabla_{\nu}h_{\rho \sigma} \nabla^{\sigma}h_{\mu}{}^{\rho} \nonumber\\+& \left. h^{\mu \nu} \nabla_{\sigma}h_{\nu \rho} \nabla^{\sigma}h_{\mu}{}^{\rho} -  \tfrac{1}{4} h^{\mu}{}_{\mu} \nabla_{\sigma}h_{\nu \rho} \nabla^{\sigma}h^{\nu \rho}\right),
\end{align}
which can be obtained either by expanding the Einstein-Hilbert Lagrangian to cubic order or by applying the Noether procedure. This can be written in Poincar\'e coordinates using the identity: 
\begin{equation}
    \nabla_{\sigma}h_{\mu \nu}
    =
    \frac1{\eta^2}
    \left(
        \partial_{\sigma}(\eta^2h_{\mu \nu})
        +
        \eta\delta_\mu^{\eta}h_{\sigma\nu}
        +
        \eta\delta_\nu^{\eta}h_{\sigma\mu}
    \right).
\end{equation}
To extract the phase \eqref{vertexwick} in the rotation of the vertex to EAdS, it is sufficient to consider the on-shell vertex, which in the temporal gauge is simply: 
\begin{equation}
     {\cal V}_{hhh} \approx  \kappa (-\eta)^8\left[ -\frac{1}{2}\delta^{i i_1} \delta^{k k_1} \delta^{l l_1} \delta^{j j_1} + \delta^{i i_1} \delta^{j k_1} \delta^{j_1 k} \delta^{l_1 l}\right]h_{i_1 j_1} \partial_i h_{k_1 l_1} \partial_{j} h_{k l}.
\end{equation}

\vskip 4pt
We denote the two graviton falloffs by
\begin{equation}
    \Delta_h^+=d,
    \qquad
    \Delta_h^-=0 .
\end{equation}
The rules to recast perturbative late-time correlators of $h_{\mu\nu}$ in terms of
Witten diagrams in EAdS under the Wick rotations \eqref{wickzeta} then read as follows
up to cubic order:
\begin{itemize}
    \item {\bf Graviton propagators}:
    \begin{multline}\label{bubugrav1}
    G^{\pm\hat{\pm}}_{i_1 i_2 ; j_1 j_2}(\eta,\bar{\eta};{\bf k})
    \to
    c^{\text{dS-AdS}}_{\Delta_h^+}
    e^{\mp \left(\Delta_h^+-2\right)\frac{\pi i}{2}}
    e^{\hat{\mp} \left(\Delta_h^+-2\right)\frac{\pi i}{2}}
    G^{\text{AdS}\,\Delta_h^+}_{i_1 i_2 ; j_1 j_2}(z,\bar{z};{\bf k})
    \\
    +
    c^{\text{dS-AdS}}_{\Delta_h^-}
    e^{\mp \left(\Delta_h^--2\right)\frac{\pi i}{2}}
    e^{\hat{\mp} \left(\Delta_h^--2\right)\frac{\pi i}{2}}
    G^{\text{AdS}\,\Delta_h^-}_{i_1 i_2 ; j_1 j_2}(z,\bar{z};{\bf k}) .
    \end{multline}
    Similarly,
    \begin{equation}\label{bubograv1}
    K^{\pm\,\Delta_h}_{i_1 i_2 ; j_1 j_2}(\eta;{\bf k})
    \to
    e^{\mp \left(\Delta_h-2\right)\frac{\pi i}{2}}
    c^{\text{dS-AdS}}_{\Delta_h}
    K^{\text{AdS}\,\Delta_h}_{i_1 i_2 ; j_1 j_2}(z;{\bf k}),
    \qquad
    \Delta_h=\Delta_h^+,\Delta_h^- .
    \end{equation}

    \item {\bf Vertices}:
    \begin{equation}\label{gravvert1}
      {\cal V}_{hhh}\left(\eta\right) \to {\cal V}_{hhh}\left(z\right).
    \end{equation}
    One proceeds similarly for higher-order vertices in the fluctuations $h_{\mu\nu}$,
    expanding the Einstein-Hilbert Lagrangian to the desired order and applying the
    rotations \eqref{wickzeta}.
\end{itemize}

These rules are equivalently generated by the following EAdS Lagrangian up to cubic
order, see section \ref{sec::SKPI},
\begin{equation}\label{EHL}
    {\cal L}_{\text{EAdS}}
    =
    {\cal L}^{(2)}_{\text{EAdS}}
    +
    {\cal L}^{(3)}_{\text{EAdS}}
    +
    \ldots .
\end{equation}
Writing $h_d$ and $h_0$ for the fields with falloffs $\Delta_h^+=d$ and
$\Delta_h^-=0$, respectively, the quadratic terms are
\begin{equation}\label{EHL2}
    {\cal L}^{(2)}_{\text{EAdS}}
    =
    \operatorname{sgn}\!\left[c^{\text{dS-AdS}}_{d}\right]
    \left(
        {\cal L}_{\text{lin EH}}\!\left[h_d\right]
        -
        {\cal L}_{\text{lin EH}}\!\left[h_0\right]
    \right),
\end{equation}
where ${\cal L}_{\text{lin EH}}$ is the linearised Einstein-Hilbert Lagrangian
\eqref{linEH}. 

The cubic interaction is
\begin{multline}\label{EHL3}
    {\cal L}^{(3)}_{\text{EAdS}}
    =
    2
    \sum_{\sigma_1,\sigma_2,\sigma_3=\pm}
    \sqrt{
    \left|c^{\text{dS-AdS}}_{\Delta_h^{\sigma_1}}\right|
    \left|c^{\text{dS-AdS}}_{\Delta_h^{\sigma_2}}\right|
    \left|c^{\text{dS-AdS}}_{\Delta_h^{\sigma_3}}\right|
    }
    \\
    \times
    \sin\left[
    \left(
        \Delta_h^{\sigma_1}
        +
        \Delta_h^{\sigma_2}
        +
        \Delta_h^{\sigma_3}
        -
        d
        -
        6
    \right)
    \frac{\pi}{2}
    \right]
    {\cal V}_{hhh}
    \left[
        h_{\Delta_h^{\sigma_1}},
        h_{\Delta_h^{\sigma_2}},
        h_{\Delta_h^{\sigma_3}}
    \right].
\end{multline}
For odd boundary dimension one may further use
$\sqrt{|c^{\text{dS-AdS}}_{0}|}
=
\sqrt{|c^{\text{dS-AdS}}_{d}|}
=1/\sqrt{2}$ and
$\operatorname{sgn}(c^{\text{dS-AdS}}_{d})=(-1)^{(d+1)/2}$.
For even $d$, the coefficients associated with the graviton falloffs are understood with
the limiting prescription described in section \ref{subsec::nuimint}.

\paragraph{Three-point contact diagram.}

The three-point contact diagram generated by the cubic vertex \eqref{cubic EH} is
related to its EAdS counterpart by
\begin{multline}\label{dS graviton 3pt}
    \left\langle
        h_{\Delta_h^{\sigma_1}}({\bf k}_1)
        h_{\Delta_h^{\sigma_2}}({\bf k}_2)
        h_{\Delta_h^{\sigma_3}}({\bf k}_3)
    \right\rangle_{\text{dS},\,\text{contact}}
    \\
    =
    2
    \sin\left[
    \left(
        \Delta_h^{\sigma_1}
        +
        \Delta_h^{\sigma_2}
        +
        \Delta_h^{\sigma_3}
        -
        d
        -
        6
    \right)
    \frac{\pi}{2}
    \right]
    \prod_{i=1}^{3}
    c^{\text{dS-AdS}}_{\Delta_h^{\sigma_i}}
    \\
    \times
    \left\langle
        h_{\Delta_h^{\sigma_1}}({\bf k}_1)
        h_{\Delta_h^{\sigma_2}}({\bf k}_2)
        h_{\Delta_h^{\sigma_3}}({\bf k}_3)
    \right\rangle_{\text{EAdS},\,\text{contact}} .
\end{multline}
Using
\begin{equation}
    \Delta_h^+=d,
    \qquad
    \Delta_h^-=0,
    \qquad
    \Delta_h^++\Delta_h^-=d,
\end{equation}
the sine factor depends only on the number of $\Delta_h^+$ falloffs:
\begin{equation}
\sin\left[
    \left(
        \sum_{i=1}^{3}\Delta_h^{\sigma_i}
        -
        d
        -
        6
    \right)
    \frac{\pi}{2}
\right]
=
\begin{cases}
-\sin(\pi\Delta_h^+)
=
-\sin(d\pi),
&
\Delta_h^+\Delta_h^+\Delta_h^+,
\\[4pt]
-\sin\left[
    \frac{\pi}{2}
    \left(
        \Delta_h^+ + \Delta_h^-
    \right)
\right]
=
-\sin\left(\frac{d\pi}{2}\right),
&
\Delta_h^+\Delta_h^+\Delta_h^-,
\\[4pt]
0,
&
\Delta_h^+\Delta_h^-\Delta_h^-,
\\[4pt]
\sin\left[
    \frac{\pi}{2}
    \left(
        \Delta_h^+ + \Delta_h^-
    \right)
\right]
=
\sin\left(\frac{d\pi}{2}\right),
&
\Delta_h^-\Delta_h^-\Delta_h^- .
\end{cases}
\end{equation}
Thus the $\Delta_h^+\Delta_h^-\Delta_h^-$ coupling vanishes identically. The
$\Delta_h^+\Delta_h^+\Delta_h^+$ coupling is proportional to $-\sin(d\pi)$, and
therefore vanishes after specialising $d$ to an integer. The remaining two cases are
proportional to $\sin(d\pi/2)$, and vanish when $d$ is an even integer.

\vskip 4pt
As in Yang-Mills theory, the sine factors should be understood before specialising $d$
to an integer. A vanishing sine factor implies that the non-local part of the late-time
correlator vanishes, but it does not by itself exclude local terms generated by
renormalisation. Applying the triple-$K$ convergence criteria
\cite{Bzowski:2015pba,Bzowski:2017poo,Bzowski:2018fql}, one finds an IR divergence in
the $\Delta_h^+\Delta_h^+\Delta_h^+$ three-point function for even $d\geq 2$. In this
case one should check whether the zero of the sine factor combines with the divergence
of the EAdS contact diagram to produce a finite local contribution after renormalisation.

\paragraph{Four-point graviton exchange.}

As before, the tree-level four-graviton exchange factorises into corresponding EAdS
exchange diagrams. For a fixed exchange channel,
\begin{multline}\label{graviton exchange}
    \left\langle
        h_{\Delta_h^{\sigma_1}}({\bf k}_1)
        h_{\Delta_h^{\sigma_2}}({\bf k}_2)
        h_{\Delta_h^{\sigma_3}}({\bf k}_3)
        h_{\Delta_h^{\sigma_4}}({\bf k}_4)
    \right\rangle_{\text{dS},\,\text{exch}}
    \\
    =
    \prod_{i=1}^{4}
    c^{\text{dS-AdS}}_{\Delta_h^{\sigma_i}}
    \sum_{\gamma=\pm}
    c^{\text{dS-AdS}}_{\Delta_h^\gamma}
    \\
    \times
    2\sin\left[
    \left(
        \Delta_h^{\sigma_1}
        +
        \Delta_h^{\sigma_2}
        +
        \Delta_h^\gamma
        -
        d
        -
        6
    \right)
    \frac{\pi}{2}
    \right]
    2\sin\left[
    \left(
        \Delta_h^{\sigma_3}
        +
        \Delta_h^{\sigma_4}
        +
        \Delta_h^\gamma
        -
        d
        -
        6
    \right)
    \frac{\pi}{2}
    \right]
    \\
    \times
    \left\langle
        h_{\Delta_h^{\sigma_1}}({\bf k}_1)
        h_{\Delta_h^{\sigma_2}}({\bf k}_2)
        h_{\Delta_h^{\sigma_3}}({\bf k}_3)
        h_{\Delta_h^{\sigma_4}}({\bf k}_4)
    \right\rangle_{\text{EAdS},\,\text{exch};\,\Delta_h^\gamma}.
\end{multline}
The other exchange channels are obtained by permuting the external legs. The coefficient
multiplying each EAdS exchange is the product of the two corresponding three-point
contact coefficients, as expected from factorisation.

\vskip 4pt
Using the rules \eqref{bubugrav1}, \eqref{bubograv1} and \eqref{gravvert1}, and their
higher-order analogues, one can proceed similarly to recast any late-time correlator of
gravitons in terms of corresponding EAdS Witten diagrams.

\section{Conclusions}

In this work we revisited the map \cite{Sleight:2020obc,Sleight:2021plv} between late-time correlators in dS and boundary correlators in Euclidean AdS for the cases of gauge bosons and gravitons. Particular attention was given to the subtleties associated with massless representations in even boundary dimensions, clarifying how these cases can be consistently accommodated within the framework, providing a streamlined reformulation of the in–in Feynman rules for scalar QED, pure Yang–Mills theory, and Einstein gravity in terms of Witten diagrams in EAdS.

\vskip 4pt
In this framework, each gauge field in dS corresponds to a pair of gauge fields in EAdS, one obeying Dirichlet boundary conditions and the other obeying Neumann boundary conditions. The latter Neumann mode induces a dynamical boundary gauge field, and correlators involving this boundary gauge boson (and, more generally, correlators of charged local operators) are gauge dependent unless appropriate non-local dressings are included. This is not a pathology of our construction but a standard feature of gauge theories (and gravity): Gauss' law obstructs strictly local charged operators, so gauge-invariant charged observables necessarily involve non-local dressings/intertwiners. See e.g.\ \cite{strocchiBook} for a classic discussion and \cite{Grassi:2024vkb} for recent progress. One may either work in a manifestly non-local gauge (where non-locality appears in commutators), or retain locality in an enlarged Hilbert space and impose Gauss’ law on physical states; in the latter viewpoint, strictly local fields generate a local field algebra, while charged sectors require non-local intertwiners.

\vskip 4pt
The focus of the present paper is on gauge-fixed local correlators and their reformulation in EAdS. These quantities should be regarded as local building blocks, rather than as already-dressed gauge-invariant observables. In particular, correlators involving the boundary gauge boson or graviton, and more generally correlators of charged local operators, require appropriate non-local dressings/intertwiners in order to define gauge-invariant observables. Our results provide the local ingredients needed for such constructions, while the explicit incorporation of these dressings lies beyond the scope of the present work.

\vskip 4pt
Mellin space provides a convenient representation of gauge boson and graviton propagators, including all longitudinal components. This enables the computation of their full contribution to (EA)dS boundary correlators, as well as the study of Ward–Takahashi identities and gauge invariance.

\vskip 4pt
Late–time correlators of gauge bosons and gravitons can exhibit IR divergences for certain falloffs and boundary dimensions. In such cases one should establish an appropriate renormalisation procedure. It would be interesting to combine the results of the present work with the renormalisation analysis of \cite{Bzowski:2023nef}, extending it to spinning correlators.

\vskip 4pt
We hope the results presented here provide a useful foundation for these directions, and more generally serve as a practical tool for future investigations of cosmological correlators in gauge theory and gravity.

\section*{Acknowledgments}

This research was supported by the European Union (ERC grant ``HoloBoot'', project number 101125112),\footnote{Views and opinions expressed are however those of the author(s) only and do not necessarily reflect those of the European Union or the European Research Council. Neither the European Union nor the granting authority can be held responsible for them.} by the MUR-PRIN grant No. PRIN2022BP52A (European Union - Next Generation EU) and by the INFN initiative STEFI.

\newpage

\begin{appendix}

\section{Propagators}
\label{app::props}

In this appendix we compile various technical details regarding the Mellin space representation of bulk-to-bulk propagators and their relation to other representations available in the literature.

\subsection{Mellin transform}
\label{app::MTprops}

In this section we review the derivation of the Mellin transform \eqref{EAdS scalar prop Mellin bubu} of the EAdS bulk-to-bulk propagator for scalar fields given in  section 4.7 of \cite{Sleight:2019hfp} and more recently \cite{Sleight:2021plv} in appendix A.1.

\vskip 4pt
It is convenient to start from the harmonic function decomposition of the EAdS bulk-to-bulk propagator, which for the normalisable boundary condition $\Delta_+$ reads \cite{Penedones:2010ue}
\begin{equation}\label{dricheads}
    G^{\text{AdS}}_{\Delta^+}\left(x;{\bar x}\right)=\int^{+\infty}_{-\infty}\frac{{\rm d}\nu}{\nu^2+\left(\Delta_+-\frac{d}{2}\right)^2}\,\Omega^{\text{AdS}}_{\nu}\left(x;{\bar x}\right),
\end{equation}
where $\Omega^{\text{AdS}}_{\nu}$ is the scalar harmonic function, which admits the following (``plane-waves" or ``split") representation \cite{Bros:1995js,Penedones:2010ue}
\begin{equation}\label{splitrep}
    \Omega^{\text{AdS}}_{\nu}\left(z,{\bf x};{\bar z},{\bar {\bf x}}\right) = \frac{\nu^2}{\pi} \int {\rm d}^d{\bf y}\, K^{\text{AdS}}_{\frac{d}{2}+i\nu}\left(z,{\bf x};{\bf y}\right)K^{\text{AdS}}_{\frac{d}{2}-i\nu}\left({\bar z},{\bar {\bf x}};{\bf y}\right),
\end{equation}
which is a product of two bulk-to-boundary propagators with scaling dimensions $\frac{d}{2}\pm i\nu$ integrated over their common boundary point ${\bf y}$. In Fourier space \eqref{FT} this factorises as
\begin{equation}\label{splitrep}
    \Omega^{\text{AdS}}_{\nu}\left(z,{\bar z},{\bf k}\right) = \frac{\nu^2}{\pi} K^{\text{AdS}}_{\frac{d}{2}+i\nu}\left(z,{\bf k}\right)K^{\text{AdS}}_{\frac{d}{2}-i\nu}\left({\bar z}, {\bf k}\right).
\end{equation}
The bulk-to-boundary propagator is a modified Bessel function of the second kind:
 \begin{align}
   K^{\text{AdS}}_{\frac{d}{2}+i\nu}\left(z;{\bf k}\right)= \left(\frac{k}{2}\right)^{i\nu}\frac{z^{\frac{d}{2}}}{\Gamma(i\nu+1)}K_{i\nu}(kz).
\end{align} 
Inserting the Mellin representation \eqref{Bessels} of the Bessel function and taking the Mellin transform one obtains:
\begin{align}
G^{\text{AdS}}_{\Delta^+}\left(u,{\bar u};{\bf k}\right)&= \int^\infty_0 \frac{{\rm d}z}{z} \frac{{\rm d}{\bar z}}{\bar z}\, G^{\text{AdS}}_{\frac{d}{2}+i\nu}(z,\bar{z};{\bf  k}) z^{2u-\frac{d}{2}} {\bar z}^{2{\bar u}-\frac{d}{2}}\\ \nonumber 
    &=\int^{+\infty}_{-\infty}\frac{{\rm d}\nu}{\nu^2+\left(\Delta_+-\frac{d}{2}\right)^2} \frac{\Gamma\left(u+\tfrac{i\nu}{2}\right)\Gamma\left(u-\tfrac{i\nu}{2}\right)\Gamma\left({\bar u}+\tfrac{i\nu}{2}\right)\Gamma\left({\bar u}-\tfrac{i\nu}{2}\right)}{16\pi\Gamma\left(+i\nu\right)\Gamma\left(-i\nu\right)}\left(\frac{k}{2}\right)^{-2\left(u+{\bar u}\right)}.
\end{align}
The integral over $\nu$ is of the same form as those encountered in \cite{Sleight:2018ryu}, where in particular it was shown that:
\begin{multline}\label{specident}
    \int_{-\infty}^\infty {\rm d}\nu \frac{\Gamma \left(a_1+\frac{i \nu }{2}\right)\Gamma \left(a_1-\frac{i \nu }{2}\right)\Gamma \left(a_2+\frac{i \nu }{2}\right)\Gamma \left(a_2-\frac{i \nu }{2}\right)\Gamma \left(a_3+\frac{i \nu }{2}\right)\Gamma \left(a_3-\frac{i \nu }{2}\right) }{\Gamma (-i \nu ) \Gamma (i \nu ) \Gamma \left(a_4+\frac{i \nu }{2}+1\right) \Gamma \left(a_4-\frac{i \nu }{2}+1\right)}\\=\frac{8 \pi  \Gamma (a_1+a_2) \Gamma (a_1+a_3) \Gamma (a_2+a_3) \Gamma (-a_1-a_2-a_3+a_4+1)}{\Gamma (1-a_1+a_4) \Gamma (1-a_2+a_4) \Gamma (1-a_3+a_4)}.
\end{multline}
From this is follows that
\begin{multline}
    \int^{+\infty}_{-\infty}\frac{{\rm d}\nu}{\nu^2+\left(\Delta_+-\frac{d}{2}\right)^2}\,\frac{\Gamma(u+\tfrac{i\nu}2)\Gamma(u-\tfrac{i\nu}2)\Gamma({\bar u}+\tfrac{i\nu}2)\Gamma({\bar u}-\tfrac{i\nu}2)}{\Gamma (i \nu) \Gamma (-i \nu )}\\
    =\frac{2 \pi ^2 \csc (\pi  (u+\bar{u}))  \Gamma \left(u+\frac{1}{2}\left(\Delta_+-\frac{d}{2}\right)\right)\Gamma \left(\bar{u}+\frac{1}{2}\left(\Delta_+-\frac{d}{2}\right)\right)}{\Gamma \left(1-u+\frac{1}{2}\left(\Delta_+-\frac{d}{2}\right)\right) \Gamma \left(1-\bar{u}+\frac{1}{2}\left(\Delta_+-\frac{d}{2}\right)\right)}\,.
\end{multline}
Replacing $\Delta_+ = \frac{d}{2}+i\nu$ and using
\begin{equation}
    \frac{1}{\Gamma\left(1-u+\tfrac{i\nu}{2}\right)\Gamma\left(u-\tfrac{i\nu}{2}\right)} = \frac{1}{\pi} \sin\left(\pi\left(u-\tfrac{i\nu}{2}\right)\right),
\end{equation}
one recovers the expression \eqref{EAdS scalar prop Mellin bubu}:
\begin{equation}\label{bubueadsmtscapp}
G^{\text{AdS}}_{\frac{d}{2}+i\nu}\left(u,{\bar u};{\bf k}\right)=\frac1{16\pi} \csc(\pi(u+\bar{u})) \omega_{\nu}(u,\bar{u})\Gamma(u\pm \tfrac{i\nu}{2})\Gamma(\bar{u}\pm \tfrac{i\nu}{2})\left(\frac{k}{2}\right)^{-2 u-2 \bar{u}}. 
\end{equation}
Notice that, while we started from the propagator with normalisable $\Delta_+$ boundary condition \eqref{dricheads}, the expression \eqref{bubueadsmtscapp} is an analytic function of $\nu$ and therefore valid for both normalisable and non-normalisable boundary conditions $\Delta_\pm$.

\vskip 4pt
Another approach is to start from the standard representation \eqref{standard bubu sc eads} of the bulk-to-bulk propagator in Fourier space which is a sum of ordered terms in the bulk coordinate. This approach was taken in appendix A.2 of \cite{Sleight:2021plv} to determine the Mellin transform of the dS Schwinger-Keldysh propagators from their analogous representation \eqref{SK mode functions} in terms of the mode functions.

\subsection{Contour choice}
\label{app::mellincontour}

In Mellin space it is important to keep track of the integration contour for the various
terms. Basic algebraic manipulations of a given expression in Mellin space should only be
performed if the various terms share the same Mellin integration contour.

\vskip 4pt
This applies in particular to the Mellin-space form of the gauge-boson
\eqref{Spin 1 bubu Mellin} and graviton \eqref{Spin 2 bubu Mellin} propagators. In these
cases the different tensor structures are accompanied by cosecant functions with shifted
arguments. Owing to the prescription \eqref{csc} for the cosecant poles, these terms do
not initially share the same Mellin integration contour. This is discussed in more detail in
the following.

\paragraph{Gauge boson propagator.}

For the gauge-boson propagator \eqref{Spin 1 bubu Mellin} we have
\begin{multline}\label{Spin 1 bubu Mellin A}
    G^{\text{AdS}\,\Delta_A}_{ij}(u,\bar{u};{\bf k})
    =
    \frac1{16\pi}
    \left[
        \delta_{ij}\csc(\pi(u+\bar{u}))
        +
        \frac{k_i k_j}{k^2}\,
        \csc(\pi(u+\bar{u}+1))
    \right]
    \\
    \times
    \omega_{\Delta_A}(u,\bar u)
    \Gamma\left(
        u\pm\frac12\left(\Delta_A-\frac d2\right)
    \right)
    \Gamma\left(
        \bar u\pm\frac12\left(\Delta_A-\frac d2\right)
    \right)
    \left(\frac{k}{2}\right)^{-2u-2\bar u}.
\end{multline}
Here $\Delta_A$ is to be set equal to either $\Delta_A^+=d-1$ or
$\Delta_A^-=1$. The terms proportional to $\delta_{ij}$ and $k_i k_j$ do not share the
same Mellin integration contour owing to the prescription \eqref{csc} for the cosecant
poles.

The integration contour for the term proportional to $\delta_{ij}$ separates the poles
\begin{equation}
    u=-\bar u-m,
    \qquad
    u=-\bar u+1+m,
    \qquad
    m\in\mathbb{N}_0,
\end{equation}
while the integration contour for the term proportional to $k_i k_j$ separates the poles
\begin{equation}
    u=-\bar u-1-m,
    \qquad
    u=-\bar u+m,
    \qquad
    m\in\mathbb{N}_0.
\end{equation}
Thus the term proportional to $\delta_{ij}$ naturally has contour
\begin{equation}
    -\operatorname{Re}[\bar u]
    <
    \operatorname{Re}[u]
    <
    1-\operatorname{Re}[\bar u],
\end{equation}
whereas the term proportional to $k_i k_j$ naturally has contour
\begin{equation}
    -1-\operatorname{Re}[\bar u]
    <
    \operatorname{Re}[u]
    <
    -\operatorname{Re}[\bar u].
\end{equation}
To combine both terms under the same Mellin integration contour, we shift the contour of
the $k_i k_j$ term to
\begin{equation}
    -\operatorname{Re}[\bar u]
    <
    \operatorname{Re}[u]
    <
    1-\operatorname{Re}[\bar u].
\end{equation}
In doing so, we cross the pole at $u=-\bar u$. The corresponding residue must therefore
be added, with the sign fixed by the orientation of the contour deformation.

Equivalently, writing
\begin{equation}
    a_A=\frac12\left(\Delta_A-\frac d2\right),
\end{equation}
the contribution from the crossed pole is obtained from the residue of
\begin{equation}
    \frac{
        \Gamma(u+a_A)\Gamma(\bar u+a_A)
    }{
        \Gamma(1-u+a_A)\Gamma(1-\bar u+a_A)
    }
    \left(\frac{k}{2}\right)^{-2(u+\bar u)}
    \frac{k_i k_j}{k^2}
    \pi\,\csc\left(\pi(u+\bar u+1)\right)
\end{equation}
at $u=-\bar u$. The contour around the crossed pole is clockwise, which gives the
corresponding residue an extra minus sign. This shifting of the contour is illustrated in
the figure below, where the real axis has been translated so that the crossed pole
$u=-\bar u$ is located at the origin.

\begin{equation*}
\begin{tikzpicture}
    \draw[->,line width=0.03cm] (-3,0) -- (3,0);
    \draw[->,line width=0.03cm] (0,-2) -- (0,2);
    \draw[->] (-0.8,-1.5) -- (-0.8,1);
    \draw (-0.8,1) -- (-0.8,1.5);
    \node at (0.35,0.3) {$-\bar u$};
    \node at (1,0) {$\bullet$};
    \node at (2,0) {$\bullet$};
    \node at (0,0) {$\bullet$};
    \node at (-1,0) {$\times$};
    \node at (-2,0) {$\times$};
    \node at (-2,-0.3) {$-2$};
    \node at (-1,-0.3) {$-1$};
    \node at (1,-0.3) {$1$};
    \node at (2,-0.3) {$2$};
    \node at (-3.3,0) {$\cdots$};
    \node at (3.5,0) {$\operatorname{Re}[u]$};
    \node at (0,2.5) {$\operatorname{Im}[u]$};
\end{tikzpicture}
\qquad
\begin{tikzpicture}
    \draw[->,line width=0.03cm] (-3,0) -- (3,0);
    \draw[->,line width=0.03cm] (0,-2) -- (0,2);
    \draw[->] (0.8,-1.5) -- (0.8,1);
    \draw (0.8,1) -- (0.8,1.5);
    \draw[
        decoration={markings, mark=at position 0.625 with {\arrow{<}}},
        postaction={decorate}
    ]
    (0,0) circle (0.7);
    \node at (0.35,0.3) {$-\bar u$};
    \node at (1,0) {$\bullet$};
    \node at (2,0) {$\bullet$};
    \node at (0,0) {$\bullet$};
    \node at (-1,0) {$\times$};
    \node at (-2,0) {$\times$};
    \node at (-2,-0.3) {$-2$};
    \node at (-1,-0.3) {$-1$};
    \node at (1,-0.3) {$1$};
    \node at (2,-0.3) {$2$};
    \node at (-3.3,0) {$\cdots$};
    \node at (3.5,0) {$\operatorname{Re}[u]$};
    \node at (0,2.5) {$\operatorname{Im}[u]$};
\end{tikzpicture}
\end{equation*}

\paragraph{Graviton propagator.}

Likewise, for the Mellin-space form \eqref{Spin 2 bubu Mellin} of the graviton propagator,
\begin{multline}\label{Spin 2 bubu Mellin A}
   G^{\text{AdS}\,\Delta_h}_{i_1i_2;j_1j_2}(u,\bar{u};{\bf k})
   =
   \frac1{16\pi}
   \Big[
        P^{(0)}_{i_1i_2;j_1j_2}\csc(\pi(u+\bar{u}))
        +
        P^{(1)}_{i_1i_2;j_1j_2}
        \csc(\pi(u+\bar{u}+1))k^{-2}
        \\
        +
        P^{(2)}_{i_1i_2;j_1j_2}
        \csc(\pi(u+\bar{u}+2))k^{-4}
   \Big]
   \\
   \times
   \omega_{\Delta_h}(u,\bar u)
   \Gamma\left(
        u\pm\frac12\left(\Delta_h-\frac d2\right)
   \right)
   \Gamma\left(
        \bar u\pm\frac12\left(\Delta_h-\frac d2\right)
   \right)
   \left(\frac{k}{2}\right)^{-2u-2\bar u}.
\end{multline}
Here $\Delta_h$ is to be set equal to either $\Delta_h^+=d$ or $\Delta_h^-=0$.
Again, the different tensor structures do not share the same Mellin integration contour.

The term proportional to $P^{(0)}_{i_1i_2;j_1j_2}$ has poles
\begin{equation}
    u=-\bar u-m,
    \qquad
    u=-\bar u+1+m,
    \qquad
    m\in\mathbb{N}_0.
\end{equation}
The term proportional to $P^{(1)}_{i_1i_2;j_1j_2}$ has poles
\begin{equation}
    u=-\bar u-1-m,
    \qquad
    u=-\bar u+m,
    \qquad
    m\in\mathbb{N}_0.
\end{equation}
Finally, the term proportional to $P^{(2)}_{i_1i_2;j_1j_2}$ has poles
\begin{equation}
    u=-\bar u-2-m,
    \qquad
    u=-\bar u-1+m,
    \qquad
    m\in\mathbb{N}_0.
\end{equation}
To combine all terms under the same Mellin integration contour, we shift the contours to
\begin{equation}
    -\operatorname{Re}[\bar u]
    <
    \operatorname{Re}[u]
    <
    1-\operatorname{Re}[\bar u].
\end{equation}
This crosses the pole at $u=-\bar u$ for the term proportional to
$P^{(1)}_{i_1i_2;j_1j_2}$, and the poles at $u=-\bar u-1$ and $u=-\bar u$ for the term
proportional to $P^{(2)}_{i_1i_2;j_1j_2}$. The corresponding residues must therefore be
added, with signs fixed by the orientation of the contour deformation.

This shifting of the contour is illustrated in the figure below, again with the real axis
translated so that the pole $u=-\bar u$ is located at the origin.

\begin{equation*}
\begin{tikzpicture}
    \draw[->,line width=0.03cm] (-4,0) -- (3,0);
    \draw[->,line width=0.03cm] (0,-2) -- (0,2);
    \draw[->] (0.8,-1.5) -- (0.8,1);
    \draw (0.8,1) -- (0.8,1.5);
    \draw[
        decoration={markings, mark=at position 0.625 with {\arrow{<}}},
        postaction={decorate}
    ]
    (0,0) circle (0.25);
    \draw[
        decoration={markings, mark=at position 0.625 with {\arrow{<}}},
        postaction={decorate}
    ]
    (-1,0) circle (0.25);
    \node at (0.35,0.35) {$-\bar u$};
    \node at (1,0) {$\bullet$};
    \node at (2,0) {$\bullet$};
    \node at (0,0) {$\bullet$};
    \node at (-1,0) {$\bullet$};
    \node at (-2,0) {$\times$};
    \node at (-3,0) {$\times$};
    \node at (-3,-0.35) {$-3$};
    \node at (-2,-0.35) {$-2$};
    \node at (-1,-0.35) {$-1$};
    \node at (1,-0.35) {$1$};
    \node at (2,-0.35) {$2$};
    \node at (-4.3,0) {$\cdots$};
    \node at (3.5,0) {$\operatorname{Re}[u]$};
    \node at (0,2.5) {$\operatorname{Im}[u]$};
\end{tikzpicture}
\end{equation*}

\subsection{Comparison with other representations}
\label{app::raju}

In this appendix we compare the Mellin-space representation of the massive scalar, gauge
boson and graviton propagators considered in this work with the representations given in
\cite{Liu:1998ty,Muck:1998rr,Bertola:2000mx,Bertola:2000pp,Raju:2010by,Raju:2011mp}.

\paragraph{Massive scalar.}

The bulk-to-bulk propagator for a scalar field of mass
$m^2=-\Delta_+\Delta_-$ in AdS admits the following representation in Fourier space:
\begin{equation}
   G^{\text{AdS}}_{\frac{d}{2}+i\nu}(z,\bar z;{\bf k})
   =
   \int_0^\infty
   \frac{{\rm d}p^2}{2}
   \frac{
        z^{\frac{d}{2}}
        J_{i\nu}(pz)
        J_{i\nu}(p\bar z)
        \bar z^{\frac{d}{2}}
   }
   {p^2+k^2}.
\end{equation}
The Mellin-space representation \eqref{EAdS scalar prop Mellin bubu} follows simply by
employing the Mellin representation \eqref{Bessels} of the Bessel functions. This gives
\begin{subequations}
\begin{align}
  G^{\text{AdS}}_{\frac{d}{2}+i\nu}(u,\bar{u};{\bf k})
  &=
  \int^\infty_0
  \frac{{\rm d}z}{z}
  \frac{{\rm d}{\bar z}}{{\bar z}}\,
  G^{\text{AdS}}_{\frac{d}{2}+i\nu}(z,\bar{z};{\bf k})
  z^{2u-\frac{d}{2}}
  {\bar z}^{2{\bar u}-\frac{d}{2}}
  \\
  &=
  \frac{1}{8}
  \int_0^\infty
  \frac{{\rm d}p^2}{p^2+k^2}
  \frac{
      \Gamma\left(u+\frac{i\nu}{2}\right)
      \Gamma\left(\bar u+\frac{i\nu}{2}\right)
  }{
      \Gamma\left(1-u+\frac{i\nu}{2}\right)
      \Gamma\left(1-\bar u+\frac{i\nu}{2}\right)
  }
  \left(\frac{p}{2}\right)^{-2(u+\bar u)} .
\end{align}
\end{subequations}
The integral in $p^2$ can be evaluated using
\begin{equation}\label{p2intcsc}
    \int_0^\infty
    {\rm d}p^2
    \frac{p^{-2(u+\bar{u}+n)}}{p^2+k^2}
    =
    \pi
    \csc\left(\pi(u+\bar u+n)\right)
    k^{-2(u+\bar u+n)},
    \qquad
    0<\operatorname{Re}[u+\bar u]+n<1 .
\end{equation}
It then follows, using the reflection formula for the gamma function, that
\begin{multline}
   G^{\text{AdS}}_{\frac{d}{2}+i\nu}(u,\bar{u};{\bf k})
   =
   \frac{1}{16\pi}
   \csc\left(\pi(u+\bar{u})\right)
   \omega_\nu(u,\bar u)
   \Gamma\left(u\pm\frac{i\nu}{2}\right)
   \Gamma\left(\bar u\pm\frac{i\nu}{2}\right)
   \\
   \times
   \left(\frac{k}{2}\right)^{-2(u+\bar u)} ,
\end{multline}
which recovers the Mellin-space representation
\eqref{EAdS scalar prop Mellin bubu}.

\paragraph{Gauge boson.}

An analogous representation for the AdS gauge-boson propagator in axial gauge is
\begin{equation}
   G^{\text{AdS}\,\Delta_A}_{ij}(z,\bar z;{\bf k})
   =
   \int_0^\infty
   \frac{{\rm d}p^2}{2}
   \frac{
        z^{\frac{d}{2}-1}
        J_{\Delta_A-\frac d2}(pz)
        J_{\Delta_A-\frac d2}(p\bar z)
        \bar z^{\frac{d}{2}-1}
   }
   {p^2+k^2}
   \mathcal{T}_{ij},
\end{equation}
where
\begin{equation}
    \mathcal{T}_{ij}
    =
    \delta_{ij}
    +
    \frac{k_i k_j}{p^2}.
\end{equation}
On the pole $p^2=-k^2$, this reduces to the transverse projector
$\pi_{ij}=\delta_{ij}-k_i k_j/k^2$.

As before, the Mellin-space representation can be recovered using the Mellin
representation \eqref{Bessels} of the Bessel functions. Writing
\begin{equation}
    a_A=\frac12\left(\Delta_A-\frac d2\right),
\end{equation}
one obtains
\begin{align}
   G^{\text{AdS}\,\Delta_A}_{ij}(u,\bar u;{\bf k})
   &=
   \int^\infty_0
   \frac{{\rm d}z}{z}
   \frac{{\rm d}{\bar z}}{\bar z}\,
   G^{\text{AdS}\,\Delta_A}_{ij}(z,\bar z;{\bf k})
   z^{2u-\frac{d}{2}+1}
   {\bar z}^{2\bar u-\frac{d}{2}+1}
   \nonumber
   \\
   &=
   \frac{1}{8}
   \int_0^\infty
   \frac{{\rm d}p^2}{p^2+k^2}
   \frac{
      \Gamma(u+a_A)
      \Gamma(\bar u+a_A)
   }{
      \Gamma(1-u+a_A)
      \Gamma(1-\bar u+a_A)
   }
   \left(\frac{p}{2}\right)^{-2(u+\bar u)}
   \left(
        \delta_{ij}
        +
        \frac{k_i k_j}{p^2}
   \right).
\end{align}
The two terms are then evaluated using \eqref{p2intcsc} with $n=0$ and $n=1$,
respectively. Using the reflection formula, this gives
\begin{multline}
   G^{\text{AdS}\,\Delta_A}_{ij}(u,\bar u;{\bf k})
   =
   \frac1{16\pi}
   \left[
        \delta_{ij}
        \csc\left(\pi(u+\bar u)\right)
        +
        \frac{k_i k_j}{k^2}
        \csc\left(\pi(u+\bar u+1)\right)
   \right]
   \\
   \times
   \omega_{\Delta_A}(u,\bar u)
   \Gamma\left(
        u\pm\frac12\left(\Delta_A-\frac d2\right)
   \right)
   \Gamma\left(
        \bar u\pm\frac12\left(\Delta_A-\frac d2\right)
   \right)
   \left(\frac{k}{2}\right)^{-2u-2\bar u}.
\end{multline}
This recovers the Mellin-space representation \eqref{Spin 1 bubu Mellin}. The two
possible gauge-field falloffs are
\begin{equation}
    \Delta_A^+=d-1,
    \qquad
    \Delta_A^-=1.
\end{equation}

\paragraph{Graviton.}

Likewise, for the AdS graviton propagator we have
\begin{multline}
  G^{\text{AdS}\,\Delta_h}_{i_1i_2;j_1j_2}(z,\bar z;{\bf k})
  =
  \int_0^\infty
  \frac{{\rm d}p^2}{2}
  \frac{
        z^{\frac{d}{2}-2}
        J_{\Delta_h-\frac d2}(pz)
        J_{\Delta_h-\frac d2}(p\bar z)
        \bar z^{\frac{d}{2}-2}
  }
  {p^2+k^2}
  \\
  \times
  \frac{1}{2}
  \left(
      \mathcal{T}_{i_1j_1}\mathcal{T}_{i_2j_2}
      +
      \mathcal{T}_{i_1j_2}\mathcal{T}_{i_2j_1}
      -
      \frac{2}{d-1}
      \mathcal{T}_{i_1i_2}\mathcal{T}_{j_1j_2}
  \right),
\end{multline}
where again
\begin{equation}
    \mathcal{T}_{ij}
    =
    \delta_{ij}
    +
    \frac{k_i k_j}{p^2}.
\end{equation}
Writing
\begin{equation}
    a_h=\frac12\left(\Delta_h-\frac d2\right),
\end{equation}
the Mellin transform is
\begin{align}
     G^{\text{AdS}\,\Delta_h}_{i_1i_2;j_1j_2}(u,\bar u;{\bf k})
     &=
     \int^\infty_0
     \frac{{\rm d}z}{z}
     \frac{{\rm d}{\bar z}}{\bar z}\,
     G^{\text{AdS}\,\Delta_h}_{i_1i_2;j_1j_2}(z,\bar z;{\bf k})
     z^{2u-\frac{d}{2}+2}
     {\bar z}^{2\bar u-\frac{d}{2}+2}
     \nonumber
     \\
     &=
     \frac{1}{8}
     \int_0^\infty
     \frac{{\rm d}p^2}{p^2+k^2}
     \frac{
        \Gamma(u+a_h)
        \Gamma(\bar u+a_h)
     }{
        \Gamma(1-u+a_h)
        \Gamma(1-\bar u+a_h)
     }
     \left(\frac{p}{2}\right)^{-2(u+\bar u)}
     \nonumber
     \\
     &\quad\times
     \frac{1}{2}
     \left[
        \left(
            \delta_{i_1j_1}
            +
            \frac{k_{i_1}k_{j_1}}{p^2}
        \right)
        \left(
            \delta_{i_2j_2}
            +
            \frac{k_{i_2}k_{j_2}}{p^2}
        \right)
        +
        (j_1\leftrightarrow j_2)
        \right.
        \nonumber
        \\
        &\hspace{2.8cm}
        \left.
        -
        \frac{2}{d-1}
        \left(
            \delta_{i_1i_2}
            +
            \frac{k_{i_1}k_{i_2}}{p^2}
        \right)
        \left(
            \delta_{j_1j_2}
            +
            \frac{k_{j_1}k_{j_2}}{p^2}
        \right)
     \right].
\end{align}
Collecting the tensor structures with the same power of $p^{-2}$ and evaluating the
$p^2$ integrals using \eqref{p2intcsc}, one obtains
\begin{multline}
   G^{\text{AdS}\,\Delta_h}_{i_1i_2;j_1j_2}(u,\bar u;{\bf k})
   =
   \frac1{16\pi}
   \Big[
        P^{(0)}_{i_1i_2;j_1j_2}
        \csc\left(\pi(u+\bar u)\right)
        +
        P^{(1)}_{i_1i_2;j_1j_2}
        \csc\left(\pi(u+\bar u+1)\right)
        k^{-2}
        \\
        +
        P^{(2)}_{i_1i_2;j_1j_2}
        \csc\left(\pi(u+\bar u+2)\right)
        k^{-4}
   \Big]
   \\
   \times
   \omega_{\Delta_h}(u,\bar u)
   \Gamma\left(
        u\pm\frac12\left(\Delta_h-\frac d2\right)
   \right)
   \Gamma\left(
        \bar u\pm\frac12\left(\Delta_h-\frac d2\right)
   \right)
   \left(\frac{k}{2}\right)^{-2u-2\bar u}.
\end{multline}
This recovers the Mellin-space representation \eqref{Spin 2 bubu Mellin}. The two
possible graviton falloffs are
\begin{equation}
    \Delta_h^+=d,
    \qquad
    \Delta_h^-=0.
\end{equation}

\end{appendix}
\bibliographystyle{JHEP}
\bibliography{refs}

\end{document}